%% file: article.tex
\documentclass[final]{siamart1116} 

\newcommand{\TheTitle}{Extreme-Scale Block-Structured Adaptive Mesh Refinement}
\newcommand{\TheAbbreviatedTitle}{Extreme-Scale Block-Structured Adaptive Mesh Refinement}

\newcommand{\TheAuthors}{F. Schornbaum and U. R{\"u}de}
\newcommand{\TheKeywords}{adaptive mesh refinement, dynamic load balancing, supercomputing, scalable parallel algorithms, parallel performance, lattice Boltzmann method, AMR, HPC, LBM}

\usepackage{amssymb}
\usepackage{MnSymbol}
\usepackage{xfrac}
\usepackage{siunitx}
\usepackage[utf8]{inputenc}
\usepackage{setspace}

\usepackage[ruled,lined,linesnumbered,algo2e]{algo2e}
\renewcommand\AlCapFnt{\footnotesize\normalfont\scshape}
\SetAlgoCaptionSeparator{}
\SetAlgoInsideSkip{medskip}
\SetNlSty{}{}{}
\SetAlFnt{\fontfamily{lmtt}\fontseries{l}\selectfont}

\SetKwSty{mykwfont}

\SetFuncSty{scshape}
\SetFuncArgSty{myttfont}
\SetProgSty{scshape}
\SetArgSty{myttfont}
\SetProcNameSty{myttfont}
\SetProcArgSty{myttfont}

\SetCommentSty{mycommentfont}
\crefname{algocf}{Algorithm}{Algorithms}
\Crefname{algocf}{Algorithm}{Algorithms}
\newcommand{\alineref}[1]{\hyperref[#1]{Line~\ref*{#1}}}
\newcommand{\alinerefpl}[1]{\hyperref[#1]{Lines~\ref*{#1}}}

\usepackage[normalem]{ulem}
\usepackage{booktabs}
\usepackage{multirow}
\usepackage{relsize}

\usepackage{array}
\newcolumntype{L}[1]{>{\raggedright\let\newline\\\arraybackslash\hspace{0pt}}m{#1}}
\newcolumntype{C}[1]{>{\centering\let\newline\\\arraybackslash\hspace{0pt}}m{#1}}
\newcolumntype{R}[1]{>{\raggedleft\let\newline\\\arraybackslash\hspace{0pt}}m{#1}}

\usepackage[acronym,nogroupskip,nomain,indexonlyfirst,toc]{glossaries}
\glsdisablehyper
\setacronymstyle{long-short}
\newacronym{hpc}{HPC}{high performance computing}
\newacronym{lbm}{LBM}{lattice Boltzmann method}
\newacronym{cfd}{CFD}{computational fluid dynamics}
\newacronym{amr}{AMR}{adaptive mesh refinement}
\newacronym{samr}{SAMR}{block-structured adaptive mesh refinement}
\newacronym{d3q19}{D3Q19}{three-dimensional lattice model with 19 directions (corners missing)}
\newacronym{d3q27}{D3Q27}{three-dimensional lattice model with 27 directions}
\newacronym{mpi}{MPI}{message passing interface}
\newacronym{sfc}{SFC}{space filling curve}
\newacronym{simd}{SIMD}{single instruction/multiple data}
\newacronym{cpu}{CPU}{central processing unit}
\newacronym{pdf}{PDF}{particle distribution function}
\newacronym{trt}{TRT}{two-relaxation-time}
\newacronym{id}{ID}{identifier}
\newacronym{aabb}{AABB}{axis-aligned bounding box}

\newcommand{\dthqnt}{\glsdisp{d3q19}{\glsentryshort{d3q19}}}
\newcommand{\dthqts}{\glsdisp{d3q27}{\glsentryshort{d3q27}}}
\newcommand{\mpi}{\glsdisp{mpi}{\glsentryshort{mpi}}}
\newcommand{\cpu}{\glsdisp{cpu}{\glsentryshort{cpu}}}

\newcommand{\Walberla}{\textsc{waLBerla}}

\newcommand\CPP{C\nolinebreak\hspace{-.05em}\raisebox{.4ex}{\relsize{-3}{\textbf{+}}}\nolinebreak\hspace{-.10em}\raisebox{.4ex}{\relsize{-3}{\textbf{+}}}}
\newcommand\CPPE{C\nolinebreak\hspace{-.05em}\raisebox{.4ex}{\relsize{-3}{\textbf{+}}}\nolinebreak\hspace{-.10em}\raisebox{.4ex}{\relsize{-3}{\textbf{+}}}\nolinebreak\hspace{-.05em}11}
\newcommand\CPPF{C\nolinebreak\hspace{-.05em}\raisebox{.4ex}{\relsize{-3}{\textbf{+}}}\nolinebreak\hspace{-.10em}\raisebox{.4ex}{\relsize{-3}{\textbf{+}}}\nolinebreak\hspace{-.05em}14}

\newcommand\CharmPP{Charm\nolinebreak\hspace{-.05em}\raisebox{.4ex}{\relsize{-3}{\textbf{+}}}\nolinebreak\hspace{-.10em}\raisebox{.4ex}{\relsize{-3}{\textbf{+}}}}

\newcommand{\code}[1]{{\upshape\ttfamily #1}}

\title{\TheTitle{}}

\author{
  Florian Schornbaum\thanks{Chair for System Simulation, Friedrich-Alexander-Universit\"at Erlangen-N\"urnberg, Erlangen, Germany
  (\email{florian.schornbaum@fau.de, ulrich.ruede@fau.de}).}
  \and
  Ulrich R\"ude\,\footnotemark[1] \thanks{Parallel Algorithms Project, CERFACS, Toulouse, France.}
}

\crefname{section}{Section}{Sections}
\Crefname{section}{Section}{Sections}

\crefname{subsection}{Section}{Sections}
\Crefname{subsection}{Section}{Sections}

\crefname{subsubsection}{Section}{Sections}
\Crefname{subsubsection}{Section}{Sections}

\newcommand{\nocontentsline}[3]{}
\newcommand{\tocless}[2]{\bgroup\let\addcontentsline=\nocontentsline#1{#2}\egroup}

\headers{\TheAbbreviatedTitle}{\TheAuthors}

\ifpdf
\hypersetup{
    pdfpagemode=UseThumbs,
    pdfstartview={Fit},              
    pdftitle={\TheTitle{}},          
    pdfauthor={\TheAuthors{}},       
    pdfsubject={\TheTitle{}},        
    pdfkeywords={\TheKeywords{}}     
}
\fi

\begin{document}

\maketitle

\begin{abstract}
In this article,
we present a novel approach for
block-structured adaptive mesh refinement (AMR) that is suitable for extreme-scale parallelism.
All data structures are designed such that the
size of the meta data in each distributed processor memory
remains bounded independent of the processor number.
In all stages of the AMR process, we use only distributed algorithms.
No central resources such as a master process or replicated data are employed,
so that an unlimited scalability can be achieved.
For the dynamic load balancing in particular,
we propose to exploit the hierarchical nature of the block-structured domain partitioning by
creating a lightweight, temporary copy of the core data structure.
This copy acts as a local and fully distributed proxy data structure.
It does not contain simulation data,
but only provides topological information about the domain partitioning into blocks.
Ultimately, this approach enables an inexpensive, local, diffusion-based dynamic load balancing scheme.

We demonstrate the excellent performance and the full scalability of 
our new AMR implementation for two architecturally different petascale supercomputers.
Benchmarks on an IBM Blue Gene/Q system with a mesh containing 3.7 trillion unknowns distributed to 458,752 processes
confirm the applicability for future extreme-scale parallel machines.
The algorithms proposed in this article operate on
blocks that result from the domain partitioning.
This concept and its realization support the storage of arbitrary data.
In consequence,
the software framework can be used for
different simulation methods, including mesh-based and meshless methods.
In this article, we demonstrate fluid simulations 
based on the lattice Boltzmann method.
\end{abstract}



\section{Introduction}\label{sec:intro}
With the availability of modern computers and the continuing increase in their computational performance,
the simulation of physical phenomena plays an important role in many areas of research.
These simulations in, e.g., fluid dynamics, mechanics, chemistry,
or astronomy often require a large amount of computational resources
and therefore rely on massively parallel execution on state-of-the-art supercomputers.

\subsection{Adaptive Mesh Refinement}
If only parts of the simulation domain require high resolution, many advanced models rely on grid refinement
in order to focus the computational resources in those regions where a high resolution is necessary.
For time dependent, instationary simulations and in particular for many
fluid dynamics simulations of complex flows,
a priori static grid refinement cannot capture the inherently dynamic behavior.
In such situations, \gls{amr} must be used to repeatedly adapt the grid resolution to the current state of the simulation.
For \gls{amr} to work efficiently on distributed parallel systems, 
the underlying data structures must support dynamic modifications to the grid,
migration of data between processes, and dynamic load balancing.

\begin{figure}[tp]
  \centering
  \includegraphics[width=0.95\textwidth]{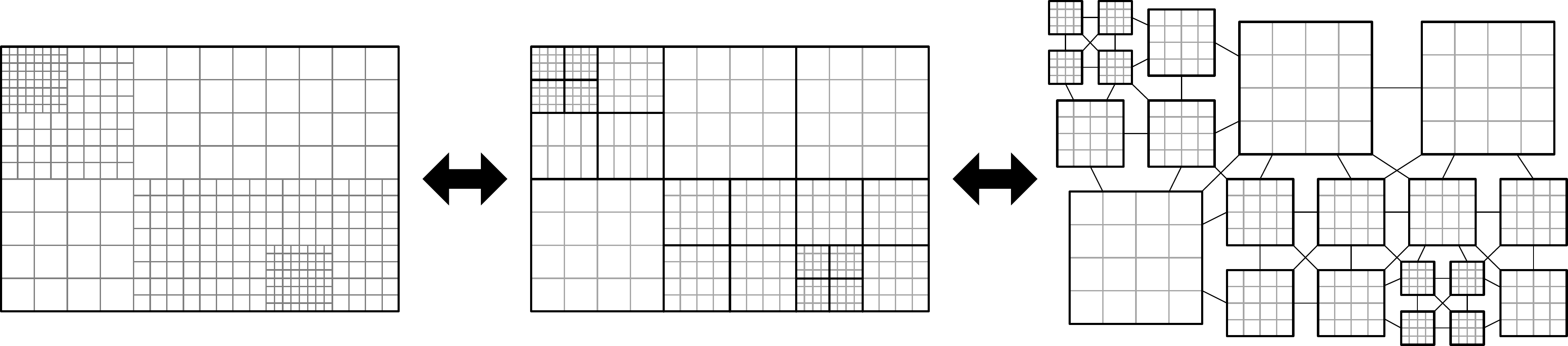}
  \caption{The figure on the left outlines the entire global mesh of
  an exemplary simulation that uses local mesh refinement in the
  upper left and the lower right corner of the domain.
  The figure in the middle shows the corresponding partitioning of the
  simulation domain into blocks, where each block contains a
  uniform Cartesian grid.
  The underlying data structure does not contain global information
  about all blocks of the simulation,
  but for each block, it stores a reference to all neighbor blocks
  visualized in the figure on the right.
  In parallel simulations, these blocks are distributed to
  all available processes as outlined in \cref{fig:dynamic:proc}.
  This kind of a block-structured domain partitioning serves as the
  basis of the \gls{amr} procedure presented in this article.
  For more details about the domain partitioning concepts and
  their parallel implementation, we refer to~\cite{schornbaum16}.
  Although the code fully supports 3D and all simulations in
  \cref{sec:bench} are performed in 3D,
  this illustration as well as all later illustrations are in 2D in 
  order to facilitate the schematic visualization of the methods.}
  \label{fig:intro}
\end{figure}

In this article, we present a novel approach for a block-structured domain partitioning 
that supports fully scalable \gls{amr} on massively parallel systems.
We build here on the parallelization concepts,
data structures, algorithms, and computational models introduced in~\cite{schornbaum16}.
The hierarchical approach of~\cite{schornbaum16} consists of a
distributed forest of octrees-like domain partitioning into blocks similar to~\cite{Bordner:2012,forestclaw},
with each block representing a container for arbitrary simulation data.
When the blocks are used to store uniform Cartesian grids,
this leads to a piecewise uniform, globally nonuniform mesh as outlined in \cref{fig:intro}.
The data structure that manages the block partitioning requires, and thus enforces,
2:1 balance between neighboring blocks.
The 2:1 balance constraint
requires the levels of neighboring blocks to differ by at most 1.
This has, for example, been used in~\cite{schornbaum16} to construct a parallelization scheme for 
the \gls{lbm}~\cite{Aidun2010,Chen98} on nonuniform grids.
The corresponding implementation shows
excellent and scalable performance for two architecturally different petascale supercomputers.
Weak scalability with up to 1.8 million threads and close to one trillion grid cells is demonstrated.
In a strong scaling scenario, the implementation reaches 
one thousand time steps per second for 8.5 million lattice cells.

In the present article,
we develop algorithms that enable scalable, parallel \gls{amr} functionality by
operating on the aforementioned distributed data structures.
This requires algorithms that are responsible for
dynamically adapting the domain partitioning, performing dynamic load balancing, and migrating data between processes.
All concepts and algorithms described in this article are, however,
only based on the blocks that result from domain partitioning.
Since the blocks act as containers and
since the data that is stored inside these containers is kept conceptually distinct,
our new implementation of \gls{amr} does not only apply to
simulations that rely on an underlying mesh,
but can, for example, also be employed for particle-based methods.
In particular,
the implementation can also be used to load balance
multibody simulations and granular flow scenarios
such as in~\cite{Preclik2015}.
We emphasize here that this generality of the distributed data structures and algorithms
is key to reaching
fully scalable, adaptive multiphysics simulations that require the coupling
of different solvers and different simulation methods
in a massively parallel environment~\cite{Bartuschat:2015:CP,Gotz2010}.

\subsection{Related Work}\label{sec:intro:related}
Software frameworks for \gls{samr} have been available for the last three decades.
Recently, many \gls{samr} codes have been compared in terms of their design, capabilities, and limitations in~\cite{Dubey20143217}.
All codes covered in this survey can run on large-scale parallel systems,
are written in C/\CPP{} or Fortran, and are publicly available.
Moreover, almost all these software packages can, among other approaches, make use of \glspl{sfc} during load balancing.
Some of the \gls{samr} codes are focused on specific applications and methods,
while others are more generic and provide the building blocks for a larger variety of computational models.
The codes also differ in the extent to which their underlying data structures require the redundant replication and
synchronization of meta data among all processes.
Meta data that increases with the size of the simulation is often an issue on large-scale parallel systems,
and eliminating this need for global meta data replication is a challenge that all \gls{samr} codes are facing.

Both BoxLib~\cite{boxlib} and Chombo~\cite{chombo}, with Chombo being a fork of BoxLib that started in 1998,
are general \gls{samr} frameworks that are not tied to a specific application.
Both, however, rely on a patch-based \gls{amr} implementation that
is subject to redundant replication of meta data across all processes.
Another generic framework for \gls{samr} is Cactus~\cite{cactusweb,cactus} with mesh refinement support through Carpet~\cite{carpetweb,carpet}.
According to~\cite{Dubey20143217},
FLASH~\cite{flash,flashweb} with \gls{amr} capabilities provided by the octree-based PARAMESH package~\cite{paramesh,parameshweb}
was among the first \gls{samr} codes to eliminate redundant meta data replication.
Besides FLASH, the authors of~\cite{Dubey20143217} conclude that Uintah~\cite{uintah,uintahweb} has gone the farthest in overcoming the limitations
of replicating meta data and adapting the software basis to current and future architectures.
Uintah employs task-based parallelism with dynamic task scheduling similar to \CharmPP{}~\cite{charmpp}.
Recently, however, the developers behind Enzo~\cite{enzo,enzoweb}, an \gls{samr} code mainy focused on astrophysical and cosmological simulations,
have also analyzed that the increasing memory consumption due to data replication is the cause of Enzo's most significant scaling bottlenecks~\cite{Bordner:2012}.
A subsequent redesign of the software basis resulted in Enzo-P/Cello~\cite{Bordner:2012,enzopweb}.
Enzo-P/Cello uses a domain partitioning based on a fully distributed forest of octrees
similar to the methods studied in this article.
Moreover, Enzo-P/Cello is build on top of \CharmPP{} and therefore program flow also follows a task-based parallelism model similar to Uintah.
The idea of first partitioning the simulation domain into blocks using an octree approach
and later creating Cartesian meshes inside these blocks was recently also adopted by a new software project: ForestClaw~\cite{forestclaw}.
ForestClaw uses the p4est library~\cite{p4est} for domain partitioning, a library that already demonstrated scalability to massively parallel systems
and, being based on a distributed forest of octrees implementation, also shares similarities with our approach.
For more details on many of these \gls{samr} codes, we refer to the survey in~\cite{Dubey20143217}.

Specifically for the \gls{lbm}, several \gls{amr} approaches have been published.
\cite{Toelke2006820}~employs octrees on a cell level in order to realize cell-based \gls{amr}.
\cite{Yu20096456}~describes an \gls{samr} implementation based on the PARAMESH package.
Other \gls{samr} approaches for the \gls{lbm} are presented in~\cite{Neumann2013}, which makes use of the Peano framework~\cite{Bungartz2010} that is based on a generalized spacetree concept,
and \cite{Fakhari2014}, which relies on an octree-like domain partitioning.
\cite{Fakhari2014},~however, solely focuses on the \gls{amr} methodology and does not provide information about parallelization capabilities of the underlying code.
Recently, the \gls{amr} scheme of~\cite{Fakhari2014} was further extended to support two-phase flow in~\cite{Fakhari2016}.
First results of a cell-based \gls{amr} implementation build on top of the p4est library are presented in~\cite{Lahnert2016}.
This implementation, however, does not yet support levelwise load balancing
as it is necessary for balanced simulations in the context of the \gls{amr} schemes for the \gls{lbm}.
A priori static grid refinement in parallel simulations based on the \gls{lbm} is studied in~\cite{Freudiger08,Hasert2014784,Schoenherr20113730}.
To our best knowledge, other popular simulation codes based on the \gls{lbm} such as Palabos~\cite{Lagrava20124808,palabos},
OpenLB~\cite{Fietz2012,heuveline2007,openlb}, LB3D~\cite{groen2011,lb3d},
HemeLB~\cite{Groen2013412}, HARVEY~\cite{Randles2013}, or LUDWIG~\cite{Desplat2001273} are at a state that they either do not support grid refinement,
only support grid refinement for 2D problems, or have not yet demonstrated large-scale simulations on nonuniform grids.
The implementation for the \gls{lbm} on nonuniform grids that we are using~\cite{schornbaum16} consists of a distributed memory parallelization
of the algorithm described in~\cite{Rohde2006} combined with the interpolation scheme suggested by~\cite{Chen2006}.
For more information about various grid refinement approaches for the \gls{lbm}, we refer to our summary of related work in~\cite{schornbaum16}.

\subsection{Contribution and Outline}\label{sec:intro:outline}
The contribution of this article is a pipeline for \gls{samr} that relies on a temporary, lightweight, shallow copy of the core data structure that
acts as a proxy and only contains topological information without any additional computational data.
The data structure imposes as few restrictions as possible on the dynamic load balancing algorithm,
thus enabling a wide range of different balancing strategies, including load balancing implementations that rely on \glspl{sfc} and graph-based balancing schemes.
Particularly, this lightweight proxy data structure allows for inexpensive iterative balancing schemes that make use of fully distributed diffusion-based algorithms.
Besides support for distributed algorithms, the data structures themselves are stored distributedly.
Each process only stores data and meta data for process-local blocks,
including information about direct neighbor blocks in the form of
block \gls{id} and process rank pairs.
No data or meta data is stored about the blocks that are located on
other processes~\cite{schornbaum16}.
Consequently, if a fixed number of grid cells are assigned to each process,
the per process memory requirement of a simulation remains unchanged and constant,
independent of the total number of active processes.
Distributed storage of all data structures is the foundation for scalability to extreme-scale supercomputers.
As such, the implementation presented in this article is an example for a state-of-the-art \gls{samr} code.
Moreover, to the best knowledge of the authors,
the total number of cells that can be handled
and the overall performance achieved
significantly exceed the performance data that has been
published for the \gls{lbm} on nonuniform grids to date~\cite{Freudiger08,Hasert2014784,Lahnert2016,FLD:FLD2469,Neumann2013,Schoenherr20113730,Yu20096456}.

The remainder of this article is organized as follows.
\Cref{sec:dynamic} contains a detailed description of our \gls{samr} pipeline, from marking blocks for refinement and coarsening to load balancing
the new domain partitioning and finally migrating the simulation data between processes.
Special focus is put on the design and realization of a lightweight proxy data structure
and its implications especially on the dynamic load balancing algorithm.
Consequently, the section also provides a detailed discussion about the applicability, the advantages, and the disadvantages of and the differences between
a dynamic load balancing scheme based on \glspl{sfc} and
an algorithm that is based on a fully distributed diffusion approach.
In \cref{sec:bench}, we present several benchmarks that demonstrate the performance and scalability of our \gls{samr} approach on two petascale supercomputers.
We conclude the article in \cref{sec:conclusion}.

\section{Dynamic domain repartitioning}\label{sec:dynamic}
Our \gls{amr} algorithms are built on a forest of octrees-like domain partitioning into blocks.
The resulting tree structure, however, is not stored explicitly,
but it is implicitly defined by a unique identification scheme for all blocks.
Additionally, every block is aware of all of its spatially adjacent neighbor blocks,
effectively creating a distributed adjacency graph for all blocks (see \cref{fig:intro}).
Consequently,
the new software framework supports the implementation of algorithms that
operate on the tree-like space decomposition
as well as on the distributed graph representation.
All concepts, algorithms, and data structures presented in this article are implemented in the open source software framework \Walberla{}~\cite{Godenschwager2013,walberla}.
The entire code is written in \CPP{}\footnote{including the subset of features of \CPPE{} and \CPPF{} that are supported by all major \CPP{} compilers
(GCC, Clang, Intel, Microsoft Visual \CPP{}, IBM XL C/\CPP{})}
and besides parallelization with only OpenMP for shared memory systems or only with \mpi{}
for distributed memory, it also supports hybrid parallel execution where multiple OpenMP threads are executed per \mpi{} process.
Moreover, the framework does not impose any constraints on the algorithms concerning program flow.
As a result, methods that require more time steps on finer levels as well as methods that perform one synchronous time step on all levels
can be implemented on top of the underlying data structures.
For more details about our domain partitioning and parallelization concepts and
the specifics of the data structures, we refer to section 3 of~\cite{schornbaum16}.

In the following five subsections, we outline the different steps of our \gls{samr} process that are necessary
to dynamically adapt the domain partitioning and rebalance and redistribute the simulation data.
The implementation of these algorithms follows the open/closed software design principle that states that
``software entities (classes, modules, functions, etc.) should be open for extension, but closed for modification''~\cite{tocp1996,oosc1988}.
Consequently, key components of the algorithms are customizable and extensible via user-provided callback functions.
These callbacks are fundamental to our software architecture.
They allow to adapt the core algorithms and data structures to the specific
requirements of the underlying simulation without any need to modify source code in our \gls{amr} framework.

When blocks are exchanged between processes, for example, the framework does not perform
the serialization of the block data since this data can be of arbitrary type.
Simulations are allowed to define their own \CPP{} classes and
add instances of these classes as data to the blocks.
As a result, it may be impossible for the framework to know how to serialize
the concrete block data.
Therefore, when data is added to the blocks,
also corresponding serialization functions must be registered.
More precisely, any callable object that adheres to a certain signature can be registered.
These callable objects can be C-style function pointers,
instances of classes that overload \code{operator()}, or lambda expressions.
Internally, they are bound to \code{std::function} objects\footnote{
For compatibility with compilers that do not yet fully support \CPPE{},
\code{boost::function} from the boost library~\cite{boost}
can be used instead of \code{std::function}.}.
The code in the framework that manages the exchange of blocks
uses these callable objects for the serialization of the data to a byte stream.
This principle of having callback functions which are registered at simulation setup
in order to be later executed by specific parts of the framework
is also employed in various essential parts of the \gls{amr} pipeline.
These callbacks make the framework flexible and extensible. 
They are, for example, used in order to decide which blocks need to be refined
(see \cref{sec:dynamic:ref})
and in order to represent the concrete dynamic load balancing algorithm
(see \cref{sec:dynamic:lb}).

\subsection{Four-step procedure}\label{sec:dynamic:proc}
The entire \gls{amr} pipeline consists of four distinct steps.
In the first step, blocks are marked for refinement and coarsening.
In the second step, this block-level refinement information is used in order to create 
a second, lightweight, proxy data structure that only contains this new topological information but no simulation data.
The proxy blocks can be assigned weights that correspond to the expected workload generated by the actual simulation data.
In the third step of the \gls{amr} procedure, the proxy data structure is then load balanced and
proxy blocks are redistributed between all available processes.
In the fourth and final step, the actual, still unmodified data structure containing the simulation data is adapted according to
the state of the proxy data structure, including refinement/coarsening and redistribution of the simulation data.

This four-step \gls{amr} procedure is outlined in \cref{algo:dynamic:proc}.
As indicated by \cref{algo:dynamic:proc},
the reevaluation of block weights that potentially results in a redistribution of all blocks can
be triggered without the need of block-level refinement or coarsening,
meaning the entire pipeline can be forced to be executed without any blocks being marked for refinement or coarsening.
Moreover, the implementation allows multiple \gls{amr} cycles before the
simulation resumes.
As a result, blocks can be split/merged multiple times during one dynamic repartitioning phase.

\begin{algorithm2e}[tp]
\begin{spacing}{0.8}
\caption{\footnotesize\itshape
Program flow of the \gls{amr} scheme.
The entire \gls{amr} pipeline consists of four distinct steps.
First, blocks are marked for refinement/coarsening in \alineref{algo:dynamic:proc:blr1}.
Then, a proxy data structure representing the new domain partitioning is created in \alineref{algo:dynamic:proc:pi} and
subsequently balanced in \alineref{algo:dynamic:proc:dlb}.
Finally, the actual simulation data is adapted and redistributed in \alineref{algo:dynamic:proc:dm}.
Illustrations corresponding to these four steps are presented in \cref{fig:dynamic:ref,fig:dynamic:proxy,fig:dynamic:lb,fig:dynamic:data}.
}\label{algo:dynamic:proc}
\end{spacing}
\footnotesize
\DontPrintSemicolon
\SetKwFunction{BLR}{BlockLevelRefinement}
\SetKwFunction{PI}{ProxyInitialization}
\SetKwFunction{DLB}{DynamicLoadBalancing}
\SetKwFunction{DM}{DataMigration}
\SetKw{or}{or}
\SetKwProg{Fn}{Function}{}{end}
\Fn{DynamicRepartitioning}
{
call \BLR\tcc*[r]{blocks are marked for refinement/coarsening}\label{algo:dynamic:proc:blr1}
\tcc*[r]{(by a user-provided callback and the framework that enforces 2:1 balance)}
\While{blocks marked for refinement/coarsening exist \or block weights must be reevaluated and blocks must be redistributed}{
call \PI\tcc*[r]{construction of the proxy data structure}\label{algo:dynamic:proc:pi}
call \DLB\tcc*[r]{redistribution of proxy blocks}\label{algo:dynamic:proc:dlb}
call \DM\tcc*[r]{migration and refinement/coarsening of the ...}\label{algo:dynamic:proc:dm}
\tcc*[r]{... actual simulation data according to the state of the proxy structure}
\If{multiple \gls{amr} cycles are allowed}{
call \BLR\tcc*[r]{same as \alineref{algo:dynamic:proc:blr1}}\label{algo:dynamic:proc:blr2}
}
}
}
\end{algorithm2e}

\begin{figure}[tp]
  \centering
  \includegraphics[width=0.95\textwidth]{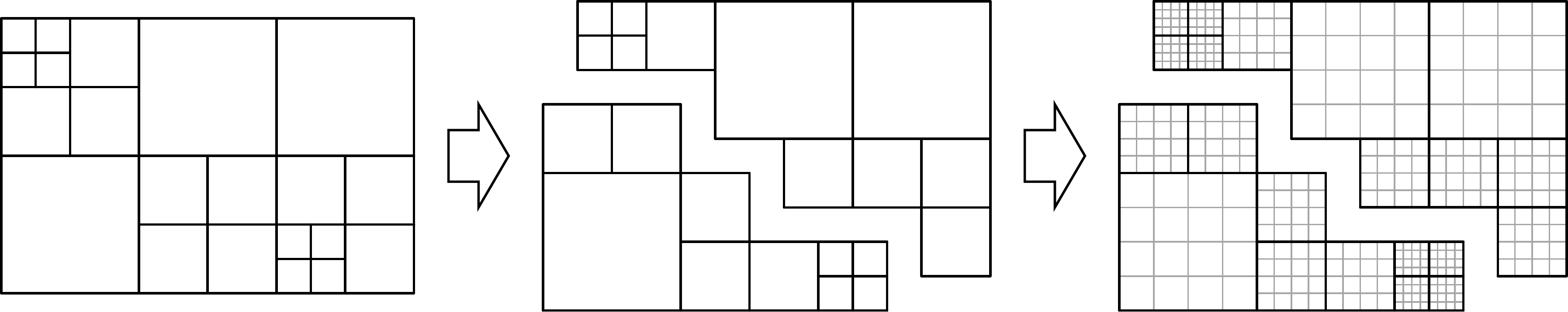}
  \caption{Example domain partitioning for a simulation
  with blocks containing 16 cells ($4 \times 4$) each.
  These blocks are distributed among two processes
  as indicated by the separation of the data into two parts
  in the figures in the middle and on the right.
  This example partitioning illustrates the initial state of a simulation before \gls{amr} is triggered.
  The \gls{amr} pipeline then
  starts with distributed block-level refinement/coarsening in \cref{fig:dynamic:ref},
  continues with the creation and load balancing of a lightweight proxy data structure in \cref{fig:dynamic:proxy,fig:dynamic:lb},
  and ends with the actual migration and refinement/coarsening of the grid data in \cref{fig:dynamic:data}.}
  \label{fig:dynamic:proc}
\end{figure}

The four steps of our \gls{amr} scheme are discussed in more detail in the following four subsections.
The entire \gls{amr} pipeline is also illustrated in \cref{fig:dynamic:ref,fig:dynamic:proxy,fig:dynamic:lb,fig:dynamic:data}
starting with an example initial domain partitioning outlined in \cref{fig:dynamic:proc}.

\subsection{Distributed block-level refinement}\label{sec:dynamic:ref}
The objective of the block-level refinement and coarsening phase is to assign a target level to each block such that
\begin{equation*}
\ell_{\text{target}} \in \{\ell_{\text{current}} - 1, \ell_{\text{current}}, \ell_{\text{current}} + 1\} .
\end{equation*}
In order to assign target levels to all blocks, the block-level refinement and coarsening phase is divided into two steps.
First, an application-dependent callback function is evaluated on every process.
The task of this function is to assign a target level to every block handled by the process.
As such, initially marking blocks for refinement or coarsening is a perfectly
distributed operation that can be executed in parallel on all processes.

These new target levels, as set by an application-dependent callback, 
might violate the 2:1 balance constraint of the domain partitioning.
Consequently, after the application-dependent callback was evaluated,
the framework guarantees 2:1 balance by first accepting all blocks marked for refinement and subsequently
potentially forcing additional blocks to split in order to maintain 2:1 balance.
Afterwards, blocks marked for coarsening are accepted for merging if, and only if,
all small blocks that merge to the same large block are marked for coarsening
and 2:1 balance is not violated.
Consequently, blocks marked for refinement by the application-dependent callback are guaranteed to be split,
whereas blocks marked for coarsening are only merged into a larger block if
they can be grouped together according to the octree structure.

This process of guaranteeing 2:1 balance can be achieved by exchanging the target levels of all process-local blocks with all neighboring processes.
Afterwards, this information can be used to check if process-local blocks must change their target levels,
i.e., check if they must be forced to split or are allowed to merge,
due to the state of neighboring blocks.
This process of exchanging block target levels with neighbors must be repeated multiple times.
The number of times every process must communicate with all of its neighbors is, however, limited by the depth of the forest of octrees,
i.e., the number of levels available in the block partitioning.
Consequently, the runtime of this first stage in the \gls{amr} pipeline scales linearly with
the number of levels in use,
but its complexity is constant with regard to the total number of processes.
For the rest of this article,
if the complexity of an algorithm is constant with regard to the number of processes,
we refer to that algorithm (and the algorithm's runtime)
as being independent of the number of processes.

\begin{figure}[tp]
  \centering
  \resizebox{0.95\textwidth}{!}{\input{dyn1_v2_.tex}}
  \caption{Starting from the domain partitioning outlined in \cref{fig:dynamic:proc},
  during the distributed block-level refinement/coarsening phase, an application-dependent callback function determines which blocks must be split and
  which blocks may be merged (figure on the bottom left).
  Blocks are marked for coarsening independent of their neighbors.
  After the evaluation of the callback function, all blocks marked for refinement are accepted (figure (1) on the right) and
  additional blocks are registered for refinement by an iterative process in order to maintain 2:1 balance (2).
  Finally, another iterative procedure accepts blocks for coarsening if all small blocks that merge to the same large block are marked for coarsening and
  if 2:1 balance is not violated (figures (3) and (4)).}
  \label{fig:dynamic:ref}
\end{figure}
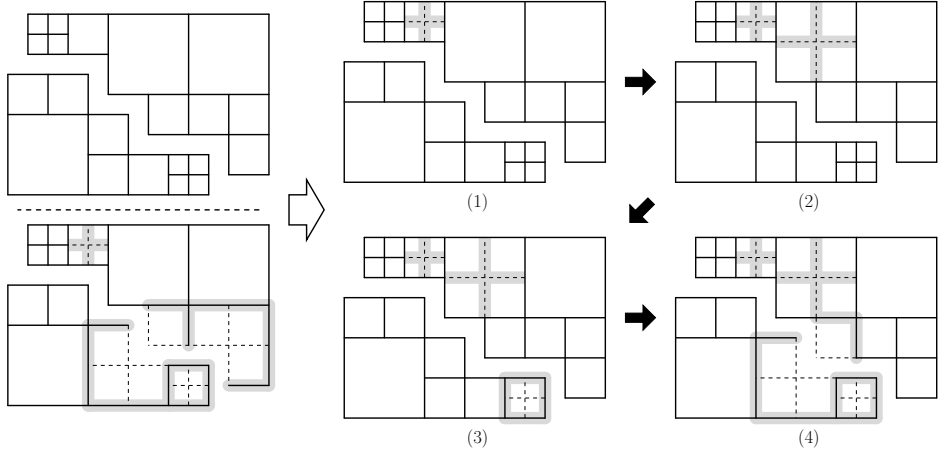

The block-level refinement and coarsening phase is illustrated in \cref{fig:dynamic:ref}.
The iterative process of
evaluating block neighborhoods multiple times
means that the medium sized blocks in \cref{fig:dynamic:ref}(4) are only accepted for coarsening after their smaller neighbor blocks 
were accepted for coarsening in \cref{fig:dynamic:ref}(3).

Since perfectly distributed algorithms and data structures require that
every process possesses only local, but no global knowledge,
every process must assume that on distant, non-neighbor processes blocks are marked to be split or to be merged.
Consequently, all processes must continue with the next step in the four-step \gls{amr} procedure,
even if there are no changes to the block partitioning.
The actual implementation of the block-level refinement and coarsening phase therefore uses two global reductions of a boolean variable as a means of optimizing the execution time as follows:
Immediately after the application-dependent callback is executed,
the first reduction can be used to abort the entire \gls{amr} procedure early
if no blocks have been marked for refinement or coarsening.
Even if some blocks are marked for coarsening, they all might violate the requirements
that are necessary for merging.
Therefore, a second reduction at the end of the block-level refinement 
and coarsening phase provides another opportunity
to terminate the \gls{amr} process early.
On current petascale supercomputers,
the benefit of aborting the \gls{amr} algorithm
often outweighs the costs of the two global reductions.

\subsection{Proxy data structure}\label{sec:dynamic:proxy}
The \gls{amr} implementation in this article is characterized by using a light-weight
proxy data structure to manage load balancing and the dynamic data redistribution in an efficient way.
Consequently,
the first phase of the \gls{amr} procedure outlined in the previous section
only assigns target levels to all blocks,
but it does not yet apply any changes to the block partitioning.
During the second step of the \gls{amr} procedure,
these target levels are used in order to create a second block partitioning that
conforms with the new topology as defined by the target levels.
This second data structure acts as a proxy for the actual, still unmodified simulation data structure.
For the rest of this article,
we will use the term {\em proxy data structure} as opposed to the primary {\em actual data structure}.
Similarly, we distinguish between {\em proxy blocks} and {\em actual blocks}.

The proxy data structure only manages the process association and the connectivity information of all of its blocks,
but it does not store any simulation data.
Furthermore, during creation of the proxy data structure,
links are established between all proxy blocks and their corresponding actual blocks.
Consequently, every proxy block always knows the process where its corresponding actual block is located, and vice versa.
Particularly during the third step of the \gls{amr} procedure when proxy blocks might migrate to different processes,
maintaining these bilateral links is vital for the success of the entire \gls{amr} scheme.
Typically, these links are represented by a {\em target process} that is stored for each actual block.
This target process is the process owning the corresponding proxy block.
Additionally, there is a {\em source process} stored for each proxy block.
Analogously, this is the process on which the corresponding actual block is located on.
If an actual block corresponds to eight smaller proxy blocks
due to being marked for refinement,
the actual block stores eight distinct target processes, one for each proxy block.
Similarly, if eight actual blocks correspond to one larger proxy block due to all actual blocks being marked for coarsening,
the proxy block stores eight source processes.
For various figures in this article,
these eight-to-one relationships in the 3D implementation
correspond to four-to-one relationships in the 2D illustrations.

\begin{figure}[tp]
  \centering
  \includegraphics[width=0.95\textwidth]{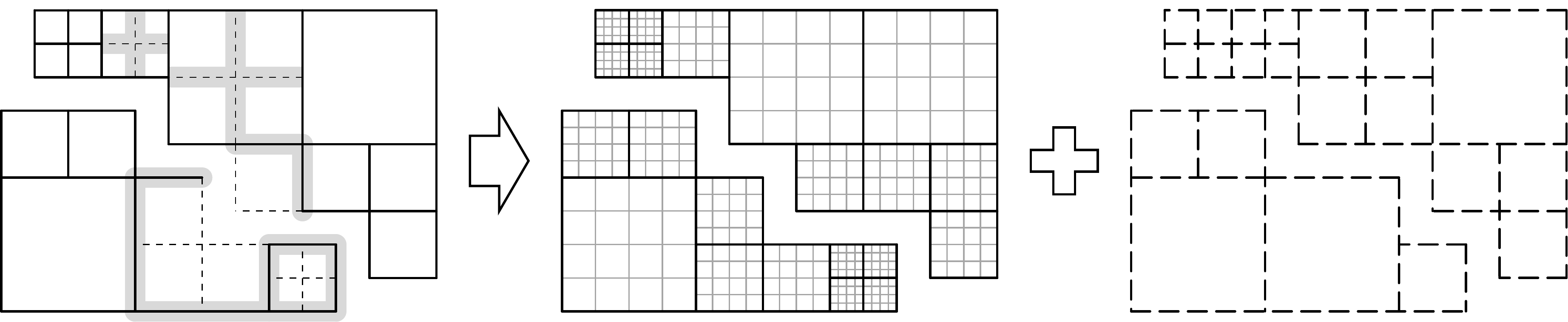}
  \caption{Using the results from the block-level refinement phase outlined in \cref{fig:dynamic:ref},
  a lightweight proxy data structure that conforms with the new topology is created.
  As a result, every process stores the current, unmodified
  data structure that maintains the simulation data (figure in the middle)
  as well as the temporarily created proxy data structure that does not store any simulation data (figure on the right).
  Additionally, every block in each of these two data structures maintains a link/links (which are not visualized in the illustration)
  to the corresponding block(s) in the other data structure.
  Initially, the proxy data structure is in an unbalanced state.
  In this example, most of the blocks, including all of the smallest blocks, are initially assigned to the same process.
  Therefore, load balancing the proxy data structure is the next step in the \gls{amr} pipeline (see \cref{fig:dynamic:lb}).}
  \label{fig:dynamic:proxy}
\end{figure}

The creation of the proxy data structure is outlined in \cref{fig:dynamic:proxy}.
As illustrated by \cref{fig:dynamic:proxy},
for multiple blocks to be merged into one larger block,
the smaller blocks do not have to be located on the same process.
Also, the creation of all proxy blocks, including the initialization of target and source processes for proxy and actual blocks, is a process-local operation.
Only when setting up the connectivity information for the proxy blocks,
communication with neighboring processes is required.
Consequently, the runtime of this second step of the \gls{amr} procedure, the creation of the proxy data structure,
is, just as the block-level refinement and coarsening phase, also independent of the total number of processes.

\subsection{Dynamic load balancing}\label{sec:dynamic:lb}
The third step in the \gls{amr} scheme is the dynamic load balancing phase.
Here, the goal is to redistribute the proxy blocks among all processes
such that the proxy data structure is in balance.
Just like the block-level refinement phase (see \cref{sec:dynamic:ref}),
the dynamic load balancing procedure is also divided into two parts.
First, a simulation-dependent, possibly user-provided callback function determines a target process for every proxy block.
This callback function represents the actual, customizable load balancing algorithm.
Once this callback is finished and returns,
the framework part takes over execution and sends
proxy blocks to different processes according to their just assigned target processes.
During this migration process, the framework also maintains the bilateral links between proxy blocks and actual blocks.

To be exact,
the callback function of the load balancing stage must perform three distinct tasks.
It must assign a target process to every proxy block,
it must notify every process about the proxy blocks that are expected to arrive from other processes,
and it must return whether or not another execution of the dynamic load balancing procedure
must be performed immediately after the migration of the proxy blocks.
Requesting another execution of the entire dynamic load balancing step enables iterative balancing schemes, as we will use below.
Executing the dynamic load balancing procedure multiple times
means proxy blocks are also exchanged multiple times.
Transferring a proxy block to another process, however, only requires to send a few bytes of data
(block \gls{id}, the source process of its corresponding actual block, and the block \glspl{id} of its neighbors).
As a result, the proxy transfers are inexpensive communication operations.

The implementation is customizable to
the requirements of the underlying simulation and the user-defined load balancing algorithm.
It is possible to augment each proxy block with additional data that is
also transfered when proxy blocks migrate to other processes.
The additional proxy data can, for example, be used to encode the workload/weight of a block
as it is used in the load balancing algorithm.
Generally, the extensibility of the proxy block data is similar to
the extensibility of the simulation block data. 

\begin{figure}[tp]
  \centering
  \includegraphics[width=0.95\textwidth]{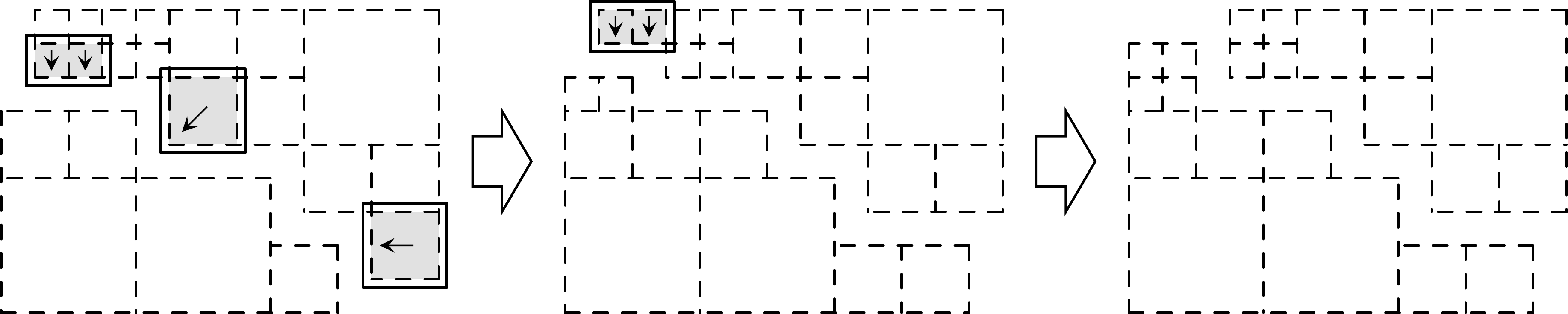}
  \caption{During dynamic load balancing, an application-dependent callback function determines a target process for every proxy block.
  The framework then performs the migration of proxy blocks between processes,
  including an update of all links between these proxy blocks and their
  counterparts in the simulation data structure.
  As illustrated in the figure,
  this process of first determining target processes for the proxy blocks and then migrating them accordingly
  can be repeated multiple times, enabling iterative, diffusion-based load balancing schemes.
  In this example, balance is achieved after two steps.
  Nevertheless, globally load balancing the proxy data structure in one step is also possible.}
  \label{fig:dynamic:lb}
\end{figure}

The dynamic load balancing procedure is illustrated in \cref{fig:dynamic:lb}.
The migration of proxy blocks, including the preservation
of all links between proxy blocks and actual blocks,
only requires point-to-point communication between different process pairs.
As a result,
the runtime of the framework part of the dynamic load balancing procedure is independent of the total number of processes.
Consequently, the runtime and scalability of the entire dynamic load balancing procedure is mainly
determined by the runtime and scalability of the callback function.
Ultimately, for large-scale simulations,
the chosen dynamic load balancing algorithm determines
the runtime and scalability of the entire \gls{amr} scheme.

The following two subsections present two different dynamic load balancing approaches
as they are currently provided by the framework. 
\Cref{sec:dynamic:lb:sfc} outlines a balancing scheme based on \glspl{sfc} that requires global data synchronization among all processes,
whereas \cref{sec:dynamic:lb:diffusion} presents a balancing algorithm built on a fully distributed,
local diffusion-based redistribution scheme.
The runtime of the latter is independent of the total number of processes.
Since the dynamic load balancing algorithms only operate on proxy blocks,
the terms ``proxy blocks'' and ``blocks'' are used synonymously in the following two subsections.

Both implementations also provide the ability to balance the blocks per level.
For \gls{lbm}-based simulations on nonuniform grids
as they are used by all benchmarks and the example application in \cref{sec:bench},
levelwise balancing of the blocks is essential for achieving good performance.
Only blocks balanced per level 
perfectly match the structure of the underlying algorithm~\cite{schornbaum16}.
Consequently,
providing the ability to balance the blocks per level is a necessary requirement
for any balancing algorithm that is used for simulations based on the \gls{lbm}.

The \gls{sfc} as well as the diffusion-based approach presented in the
next two subsections are both injected into the \gls{samr} framework as callback functions
that adhere to the requirements for
load balancing algorithms for the proxy data structure
described at the beginning of this section. 
Since the actual load balancing algorithm is injected into the \gls{amr} pipeline
as a callback function,
other dynamic load balancing strategies can be incorporated
without the need to modify framework code.
Consequently,
future work that builds on the \gls{samr} framework presented in this article 
will investigate the applicability of dynamic load balancing algorithms
provided by other, specialized libraries like ParMETIS~\cite{parmetisweb,parmetis},
Zoltan~\cite{zoltan,zoltanweb}, or PT-Scotch~\cite{ptscotch,ptscotchweb}.

\subsubsection{Space filling curves}\label{sec:dynamic:lb:sfc}
Many \gls{samr} codes employ \glspl{sfc} for load balancing.
In general, \glspl{sfc} map multidimensional data to one dimension while preserving good spatial locality of the data.
As such, they can be used to define a global ordering for all the octants of an octree.
Consequently, \glspl{sfc} allow to construct an ordered list of all the blocks stored in our data structure.
This list can be used for load balancing by
first dividing the list into as many pieces of consecutive blocks as there are processes and
then assigning one of these pieces to each process.
For this approach to work on homogeneous systems, the sum of the weights of all blocks in each piece must be identical.
Consequently, if all blocks have the same weight, i.e, generate the same workload,
each piece must consist of the same number of blocks.
The generalization to a forest of octrees
where each root block corresponds to an octree~\cite{schornbaum16}
can be realized by defining a global ordering for all root blocks~\cite{p4est}.

The current version of our code features \gls{sfc}-based
load balancing routines that make use of either
Morton~\cite{morton} or Hilbert~\cite{hilbert} order.
With Hilbert order, consecutive blocks are always connected via faces,
whereas with Morton order, several consecutive blocks are only connected via edges or corners,
with some consecutive blocks not being connected to each other at all.
Hilbert order, therefore, results in
better spatial locality for blocks assigned to the same process than Morton order.
Both curves can be constructed by depth-first tree traversal.
For Morton order, child nodes are always visited in the same order while descending
the tree, whereas for Hilbert order, the order in which child nodes are traversed
depends on their position within the tree.
For the construction of the Hilbert curve,
lookup tables exist that specify the exact traversal order~\cite{hilbertconstruction}.
Consequently, the construction of a \gls{sfc} based on Hilbert instead of Morton order
only results in an insignificant overhead.
Since in our implementation, the block \glspl{id},
which uniquely identify all blocks within the distributed data structure,
can be represented as numbers that are related to the Morton order similar to the identification scheme of~\cite{p4est},
simply sorting all blocks by \gls{id} also results in a Morton ordering of the blocks.

Regardless of whether Morton or Hilbert order is used during the \gls{sfc}-based load balancing phase,
the construction of the curve is built on a global information exchange among all processes
as it is also used by other established software~\cite{p4est}.
This global data synchronization is usually best
realized with an \code{allgather} operation.
If all blocks share the same weight and if all blocks are treated equally regardless of their level,
then globally synchronizing the number of blocks stored on each process is enough to determine,
locally on every process,
where blocks need to migrate in order to achieve a balanced redistribution along the curve.
Consequently, this approach results in the synchronization and global replication of one number (typically 1 byte is enough) per process.

\begin{figure}[tbp]
  \centering
  \resizebox{0.95\textwidth}{!}{\input{sfc_.tex}}
  \caption{Example of a \gls{sfc}-based distribution to 4 processes
  (as indicated by the small numbers inside the blocks).
  In 1.1 and 1.2, the blocks are distributed to the 4 processes
  following the global ordering as given by the curve.
  If blocks split/merge due to refinement/coarsening (indicated by the gray boxes in 1.1),
  the new blocks always align themselves correctly (1.2):
  Blocks along the new curve will always still be in order,
  but rebalance may be required.
  Since all blocks are still in order, this rebalancing operation is cheap and only requires
  knowledge about the number of blocks on every process.
  If, however, blocks on different levels must be balanced separately (as is illustrated in 2.1 and 2.2),
  new blocks that are the result of refinement or coarsening may fall out of line (2.2):
  The lists of blocks for every level are not in order anymore.
  In contrast to 1.2, rebalancing 2.2 also requires a complete reordering,
  which is a far more expensive operation in terms of
  the amount of information that must be synchronized between all processes.}
  \label{fig:dynamic:sfcbalanceorder}
\end{figure}
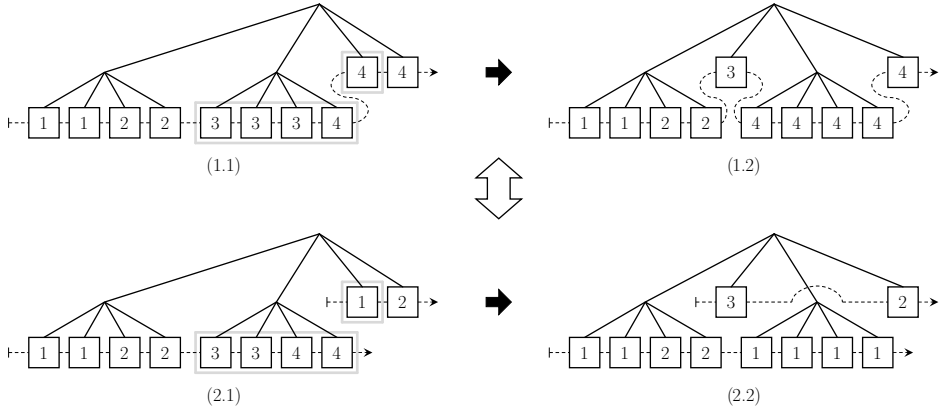

If the blocks must be balanced per level,
the \gls{sfc} is used to construct one list of blocks for every level.
Load balance is then achieved by distributing each list separately.
During dynamic refinement and coarsening,
this per-level ordering is mixed up (see \cref{fig:dynamic:sfcbalanceorder}).
As a result,
restoring the order for every level and the subsequent rebalancing
is not cheap anymore and requires an \code{allgather} of all block \glspl{id} (typically 4 to 8 bytes per block).
Knowing the \gls{id} of every block then allows each process to, locally, reconstruct and redistribute all \gls{sfc}-based lists of blocks, sorted by level.

If, additionally, blocks are allowed to have individual weights,
the amount of data that must be synchronized increases further by 1 to 4 bytes for every block (cf.\ \cref{tab:dynamic:sfcdata}).
Ultimately, \gls{sfc}-based dynamic load balancing requires global synchronization regardless of the requirements of the underlying simulation.
The communication time as well as the memory consumption per process
therefore increases linearly with the number of processes.
If all blocks share the same weight and if per-level balancing is not necessary,
\gls{sfc}-based dynamic load balancing may, however, still be feasible with several millions of processes.

\begin{table}[tp]
  \centering
  \caption{Typical amount of data that must be globally replicated and
  synchronized to all processes when using \gls{sfc}-based dynamic load balancing.}
  \label{tab:dynamic:sfcdata}
  \footnotesize
  \setlength{\extrarowheight}{4pt}
   \begin{tabular}{cc|c|c|}
   & \multicolumn{1}{c}{} & \multicolumn{2}{c}{blocks must be balanced per level}\\
   & \multicolumn{1}{c}{} & \multicolumn{1}{c}{no}  & \multicolumn{1}{c}{yes} \\[2pt] \cline{3-4}
   \multirow{2}*{every block has the same weight}  & yes & 1 byte per process & 4-8 bytes per block \\[2pt] \cline{3-4}
   & no & 1-4 bytes per block & 5-12 bytes per block \\[2pt] \cline{3-4}
   \end{tabular}
\end{table}

\subsubsection{Diffusion-based approach}\label{sec:dynamic:lb:diffusion}
The idea of diffusion-based load balancing schemes is to use a process motivated by physical diffusion
in order to determine, for every process $x$,
the neighboring processes to which local blocks of $x$ must be transferred.
The process of first executing this diffusion-based
rebalancing and then migrating proxy blocks is repeated multiple times,
leading to an iterative load balancing procedure that allows blocks to migrate to distant processes.
Diffusion-based load balancing schemes may not always achieve a perfect global balance,
but we can expect that they will quickly ($\hateq$~few iterations) eliminate processes with high load.
Eliminating peak loads is essential to avoid bottlenecks
and to achieve good parallel efficiency.

Each step of the diffusion-based algorithm only relies on next-neighbor communication.
As a result, if the number of iterations is fixed,
the runtime as well as the memory consumption is independent of the total number of processes.
Whereas the \gls{sfc}-based balancing scheme relies on an octree-like domain partitioning,
the diffusion algorithm operates on the distributed process graph.
This process graph indicates how processes are connected to each other.
We refer to two processes $i$ and $j$ as being connected/neighbors if at least one block on process $i$ is
connected via a face, edge, or corner with a block on process $j$.
This graph is realized in distributed form since every process maintains only links to its neighbor processes.

The implementation of the diffusion-based balancing algorithm consists of two main parts.
In the first part, a diffusion scheme originally proposed by~\cite{Cybenko1989} is used to determine
the flow of process load $f_{ij}$ that is required between all neighboring processes $i$ and $j$ in order to
achieve load balance among all processes.
In the second part, the results from the first part are then used to decide which blocks must be moved to which neighbor process
by matching block weights to process inflow and outflow information.
If load balancing per level is required,
all data is calculated for each level based only on the blocks of the corresponding level.
However, besides calculating
data separately for each level,
program flow is otherwise identical to an application
that does not require per-level balancing.
Consequently, all algorithms in this section can be easily used also 
to perform load balancing per level.

The algorithm is outlined in \cref{algo:dynamic:lb:diffusion}.
The diffusion scheme starts by
calculating the process load $w_i$ of every process by summing up the weights of all process-local blocks.
The flow between all neighboring processes is then determined by an iterative process.
First, every process $i$ calculates the current flow $f_{ij}^{\prime}$ between itself and every neighbor process $j$ with
\begin{equation*}
f_{ij}^{\prime} = \alpha_{ij} \cdot (w_i - w_j)~\text{\cite{Cybenko1989}},
\end{equation*}
where $\alpha_{ij}$ follows the definition of~\cite{boillat1990}
which allows $\alpha_{ij}$ to be determined with next-neighbor communication only.
Hence,
\begin{equation*}
\alpha_{ij} = \frac{1}{\max(d_i,d_j) + 1},
\end{equation*}
with $d_i$ and $d_j$ denoting the number of neighbor processes of $i$ and $j$, respectively.
The current flow $f_{ij}^{\prime}$ is then used to adapt the process loads $w_i$ and $w_j$ of processes $i$ and $j$.
However, no blocks are yet exchanged.
This procedure of calculating flows and adapting process loads accordingly is repeated
for a fixed number of iterations.
We refer to these iterations as ``flow iterations'' as opposed to the number of ``main iterations'' of the
load balancing stage.
Consequently, the diffusion-based load balancing approach represents an
iterative load balancing scheme with nested iterations:
For every main iteration,
first a fixed number of flow iterations is executed
and then proxy blocks are exchanged between neighboring processes.
The flow $f_{ij}$, which is used to determine the blocks that must be exchanged between processes $i$ and $j$,
eventually results from the summation of all $f_{ij}^{\prime}$.
A value for $f_{ij}$ greater than zero indicates outflow,
whereas $f_{ij} < 0$ indicates inflow.

\begin{algorithm2e}[tp]
\begin{spacing}{0.8}
\caption{\footnotesize\itshape
The algorithm outlines the two-part diffusion-based load balancing scheme.
First, the algorithm determines the flow of process load $f_{ij}$
that is required between all neighboring processes $i$ and $j$ in order to
achieve load balance among all processes.
These $f_{ij}$ are then used to decide which blocks must be moved to which neighbor process.
This decision is realized by either a pull or a push scheme outlined in 
\cref{algo:dynamic:lb:diffusion:push,algo:dynamic:lb:diffusion:pull}.
}\label{algo:dynamic:lb:diffusion}
\end{spacing}
\footnotesize
\DontPrintSemicolon
\SetKwComment{Comment}{}{}
\SetKwFunction{PUSH}{DiffusionPush}
\SetKwFunction{PULL}{DiffusionPull}
\SetKwProg{Fn}{Function}{}{end}
\Fn{DiffusionLoadBalancing}
{
calculate process weight $w_i$\tcp*[l]{$\hateq$ sum of all local block weights $\hateq$ process load}
determine number of neighbor processes $d_i$\;
exchange $d_i$ with all neighbor processes\;
\ForAll{neighbor processes $j$}{
$f_{ij} = 0$\tcp*[l]{$\hateq$ flow between current process $i$ and process $j$}
$\alpha_{ij} = \frac{1}{\max(d_i,d_j) + 1}$\;
}
\Repeat(\tcc*[f]{calculate flow $f_{ij}$ ...}){predefined max.\ number of iterations is reached\Comment*[h]{\ // ``flow'' iterations}}{
exchange $w_i$ with all neighbor processes\tcc*[r]{... that is required ...}
$w_i^{\prime} = w_i$\tcc*[r]{... between all ...}
\ForAll(\tcc*[f]{... neighbor processes ...}){neighbor processes $j$}{
$f_{ij}^{\prime} = \alpha_{ij} \cdot (w_i^{\prime} - w_j)$\tcc*[r]{... $i$ and $j$ ...}
$f_{ij} \mathrel{+}= f_{ij}^{\prime}$\tcc*[r]{... in order to ...}
$w_i \mathrel{-}= f_{ij}^{\prime}$\tcc*[r]{... achieve balance}
}
}
call \PUSH{$f_{ij}$} or \PULL{$f_{ij}$} in order to determine which blocks are sent to which neighbor process\;
inform neighbor processes about whether or not blocks are about to be sent\;
}
\end{algorithm2e}

In order to determine which blocks are sent to which neighbor process,
we propose two different approaches that
are referred to as ``push'' and ``pull'' scheme for the rest of this article.
When using the push scheme,
overloaded processes decide which blocks to push to which neighbor,
whereas when using the pull scheme,
underloaded processes choose neighbors from which blocks are requested.
A major challenge arises from the fact that
although every connection to a neighbor process is assigned a certain outflow or inflow value,
these flows are almost always impossible to satisfy because
they rarely match with the weights of the available blocks.
Only entire blocks can be migrated.
It is impossible to only send part of a block to another process.
For \gls{samr} codes,
computation is usually more efficient when choosing larger blocks.
For mesh-based methods like the \gls{lbm},
larger blocks result in less communication overhead.
With larger blocks, the compute kernels
may also take better advantage of \gls{simd} instructions.
Therefore, in practice, large-scale simulations
are often configured to use only few blocks per process.
However, in these cases where only few blocks are assigned to each process,
the weight of any block might exceed every single outflow/inflow value.

Therefore, the central idea of the push scheme is to first calculate the accumulated process outflow
$\text{\it outflow}_i$ of every process $i$ by summing up all $f_{ij} > 0$.
If the weight $w_k$ of a block $k$ is less than or equal to $\text{\it outflow}_i$,
the block $k$ is a viable candidate for being pushed to a neighbor process.
Moreover, all $f_{ij} > 0$ are only used as clues as to which neighbor process $j$
is a viable candidate for receiving blocks.
Similarly, the pull scheme first calculates the accumulated process inflow
$\text{\it inflow}_i$ of every process $i$ by summing up all $f_{ij} < 0$.
Neighbor blocks are viable candidates for being requested if their weights do not exceed the accumulated process inflow.
All $f_{ij} < 0$ are only used as clues as to which neighbor process $j$
is a viable candidate for providing blocks.

The entire push scheme is outlined in \cref{algo:dynamic:lb:diffusion:push}.
Inside a main loop, the algorithm picks the neighbor process~$j$ that is currently associated with the highest outflow value.
If the algorithm is able to identify a block that can be sent to process~$j$,
then the accumulated process outflow $\text{\it outflow}_i$ as well as
the flow $f_{ij}$ towards process~$j$ is decreased by the weight of the block.
If multiple blocks on process~$i$ are viable candidates for being sent to process~$j$,
the algorithm picks the block that is the best fit for migration to process~$j$.
A block is considered a good fit for migration if its connection to process~$i$ is weak and/or its connection to process~$j$ is strong.
If, for example, a block~$m$ is connected to only one other block on process~$i$, but to two blocks on process~$j$,
then~$m$ is considered a better choice for migration to~$j$
than a block~$n$ which is connected to two blocks on process~$i$ and to no block on process~$j$.
Moreover, the type of the connection (face, edge, corner) is also considered while determining the connection strength.

\begin{algorithm2e}[tp]
\begin{spacing}{0.8}
\caption{\footnotesize\itshape
Schematic program flow of the ``push'' approach for determining the blocks that are sent to neighbor processes
on the basis of the inflow/outflow values $f_{ij}$ calculated for every neighbor process
in the diffusion-based balancing algorithm (see \cref{algo:dynamic:lb:diffusion}).
}\label{algo:dynamic:lb:diffusion:push}
\end{spacing}
\footnotesize
\DontPrintSemicolon
\SetKwProg{Fn}{Function}{}{end}
\SetKw{and}{and}
\Fn{DiffusionPush($f_{ij}$)}
{
$\text{\it outflow}_i = 0$\tcp*[l]{accumulated process outflow}
\ForAll{neighbor processes $j$}{
\lIf{$f_{ij} > 0$}{$\text{\it outflow}_i \mathrel{+}= f_{ij}$}
}
\While{$\text{\it outflow}_i > 0$ \and a process $j$ with $f_{ij} > 0$ exists}{
pick process $j$ with largest value for $f_{ij}$\;
construct a list of all the blocks that can be moved to process $j$ and are not yet marked for migration to another process\;
out of the previous list, pick block $k$ that is the best fit for migration to process $j$ and whose weight $w_k \leq \text{\it outflow}_i$\;
\eIf{such a block $k$ exists}{
mark block $k$ for migration to process $j$\;
$f_{ij} \mathrel{-}= w_k$\;
$\text{\it outflow}_i \mathrel{-}= w_k$\;
}{
$f_{ij} = 0$\;
}
}
}
\end{algorithm2e}

The first part of the pull scheme, outlined in \cref{algo:dynamic:lb:diffusion:pull}, 
is conceptually identical to the push algorithm.
Inside a main loop, the algorithm identifies blocks that must be fetched from neighbor processes in order to
satisfy the accumulated process inflow.
In the second part of the pull procedure,
these blocks are then requested from the corresponding neighbor processes.
Every process must comply with these requests.
A process is only allowed to deny such a request if the same block is requested by multiple neighbors.
In that case, only one request can be satisfied.
Ultimately, however, all blocks that are requested are passed on to the appropriate neighbor processes.

\begin{algorithm2e}[tp]
\begin{spacing}{0.8}
\caption{\footnotesize\itshape
Schematic program flow of the ``pull'' approach for determining the blocks that are sent to neighbor processes
on the basis of the inflow/outflow values $f_{ij}$ calculated for every neighbor process
in the diffusion-based balancing algorithm (see \cref{algo:dynamic:lb:diffusion}).
}\label{algo:dynamic:lb:diffusion:pull}
\end{spacing}
\footnotesize
\DontPrintSemicolon
\SetKwProg{Fn}{Function}{}{end}
\SetKw{and}{and}
\Fn{DiffusionPull($f_{ij}$)}
{
$\text{\it inflow}_i = 0$\tcp*[l]{accumulated process inflow}
\ForAll{neighbor processes $j$}{
\lIf{$f_{ij} < 0$}{$\text{\it inflow}_i \mathrel{-}= f_{ij}$}
}
send a list of all local blocks (block \glspl{id} only) and their corresponding weights to all neighbor processes\;
\While{$\text{\it inflow}_i > 0$ \and a process $j$ with $f_{ij} < 0$ exists}{
pick process $j$ with smallest value for $f_{ij}$\tcp*[l]{$\hateq$ largest inflow}
construct a list of all remote blocks that can be fetched from process $j$ and are not yet candidates for migration to the current process $i$\;
out of the previous list, pick remote block $k$ that is the best fit for migration from process $j$ and whose weight $w_k \leq \text{\it inflow}_i$\;
\eIf{such a remote block $k$ exists}{
locally bookmark remote block $k$ as candidate for migration from process $j$ to the current process $i$\;
$f_{ij} \mathrel{+}= w_k$\;
$\text{\it inflow}_i \mathrel{-}= w_k$\;
}{
$f_{ij} = 0$\;
}
}
send a request to every neighbor process containing a list of all the remote blocks that process $i$ wants to fetch\;
\ForAll{local blocks $k$}{
\uIf{block $k$ is requested by one neighbor process $j$}{
mark block $k$ for migration to process $j$\;
}
\ElseIf{block $k$ is requested by multiple neighbor processes}{
out of these processes, mark block $k$ for migration to the neighbor process $j$ with the largest value for $f_{ij}$\;
}
}
}
\end{algorithm2e}

Eventually, the application can choose
the diffusion-based balancing algorithm that uses the push scheme, the pull scheme,
or both the push and the pull scheme in an alternating fashion.
Consequently, the program flow of the entire load balancing procedure
with alternating push/pull schemes and, for example, four main iterations consists of
i)~\cref{algo:dynamic:lb:diffusion,algo:dynamic:lb:diffusion:push} followed by the migration of proxy blocks,
ii)~\cref{algo:dynamic:lb:diffusion,algo:dynamic:lb:diffusion:pull} followed by another migration of proxy blocks,
iii)~\cref{algo:dynamic:lb:diffusion,algo:dynamic:lb:diffusion:push} followed by a third migration of proxy blocks,
and iv)~\cref{algo:dynamic:lb:diffusion,algo:dynamic:lb:diffusion:pull} followed by the fourth and final migration of proxy blocks.

The actual implementation of \cref{algo:dynamic:lb:diffusion} also makes use of a global reduction
for calculating the total simulation load ($\hateq$ sum of all block weights).
This information can be used to adapt the process local inflow/outflow values with respect to the exact globally average process load.
Moreover,
knowledge about the total simulation load enables the algorithm to decide locally if a process
is currently overloaded and whether or not balancing is required.
Another global reduction then allows to synchronize this information among all processes.
As a result,
the entire load balancing procedure can be
terminated early if all processes are already sufficiently in balance.
Ultimately,
this leads to a variable number of main diffusion-based balancing iterations.
Applications only need to define a maximum number of iterations.
But neither of these reductions is mandatory for the algorithm.
Just like during the block-level refinement/coarsening phase (see \cref{sec:dynamic:ref}),
however,
the benefits of aborting the entire load balancing procedure early
can easily amortize the cost of these global reductions.
In \cref{sec:bench} we will demonstrate that this distributed, diffusion-based load balancing pipeline shows
applicability and excellent scalability for extreme-scale parallel simulations that involve trillions of unknowns.

\subsection{Data migration and refinement}\label{sec:dynamic:data}
\begin{figure}[tp]
  \centering
  \includegraphics[width=0.95\textwidth]{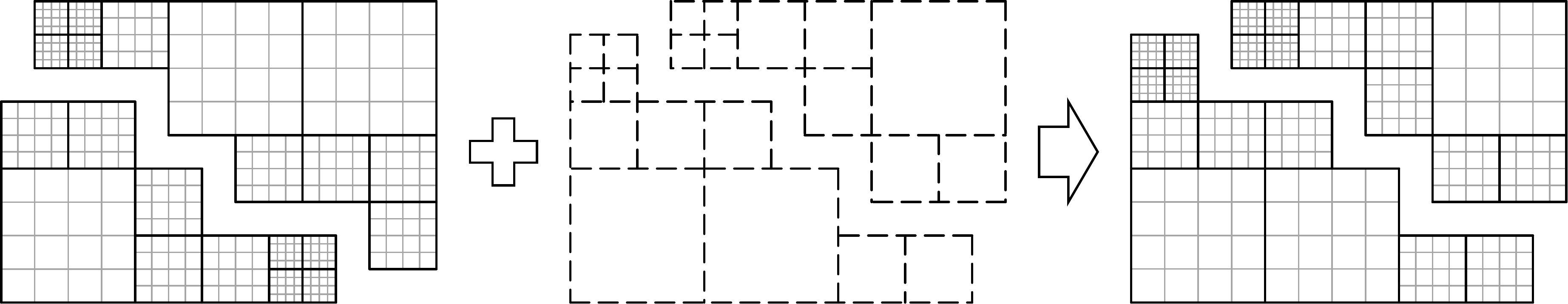}
  \caption{After successfully balancing the proxy data structure (see \cref{fig:dynamic:lb}),
  links between all proxy blocks and their counterparts in the simulation data structure
  enable the refinement/coarsening of simulation data,
  including the migration of simulation data to different processes,
  in one final step.
  Afterwards, the temporarily created proxy data structure is destroyed.
  An important feature also illustrated in this example is that
  in order to merge multiple fine blocks into one coarse block,
  the fine blocks do not have to be located on the same source process.}
  \label{fig:dynamic:data}
\end{figure}

The fourth and final step in the \gls{amr} pipeline is the actual refinement, coarsening, and migration of all simulation data.
This final step is also outlined in \cref{fig:dynamic:data}.
The central idea of this data migration phase is to use the load-balanced state of the proxy data structure
in order to adapt the simulation data structure accordingly.
Because of the bilateral links between proxy blocks and actual blocks,
the refinement, coarsening, and migration of the underlying simulation data is performed in one single step.
The bilateral links also mean that for the migration of the block data only
point-to-point communication via MPI between processes that exchange data is necessary.
Consequently,
the runtime of the final stage of the \gls{amr} pipeline scales linearly with
the amount of block data/the number of blocks that need to be sent,
but its complexity is constant with regard to the total number of processes
in use by the simulation.

Besides the fast and inexpensive migration of proxy blocks during the dynamic load balancing phase,
the refinement, coarsening, and migration of data in one single step proves to be another advantage provided by the proxy data structure.
In mesh-based methods like the \gls{lbm},
octree-based refinement results in eight times the number of grids cells ($\hateq$ eight times the amount of memory)
in refined regions.
As a result,
if dynamic refinement of the simulation data occurs before load balancing is performed,
every process must have eight times the amount of memory available as is currently allocated by the simulation
in case a process is requested to refine all of its data.
Consequently,
in order to avoid running out of memory,
only \sfrac{1}{8} of the available memory can be used for the actual simulation and \sfrac{7}{8} of the available memory must always be kept in reserve
for the refinement procedure.

The existence of the load-balanced proxy data structure, however,
allows an actual block to determine whether or not parts of its data (or even its entire data) end up on a different process after refinement.
In case the refined data ends up on another process (possibly even on multiple other processes),
the implementation of the data migration algorithm enables the sending source processes to only send the data
that is required for the receiving target processes to correctly initialize the refined data.
Allocation of the memory required for the refined data, allocation of the corresponding fine blocks,
as well as the actual refinement of the data only happen on the target processes.
In \cref{fig:dynamic:data}, for example,
the four medium-sized blocks that result from the large block in the upper center being split end up on both processes,
three remain on the original process, while one is assigned to the other process.
Ultimately, our four-step \gls{amr} approach significantly reduces the amount of memory overhead.
Almost the entire available memory can be used for simulation data at all times
without the risk of running out of memory during the refinement phase.

As outlined in the introduction of \cref{sec:dynamic},
the framework does not contain any information on
how block data is serialized or deserialised directly.
However, initially registering block data at the underlying data structure
must include the registration of callback functions
that perform the serialization/deserialization of the corresponding block data.
In total, six such callbacks are required for every block data item:
one pair of serialization/deserialization functions used in case a block is split, multiple blocks are merged,
or a block is migrated without further modification, respectively.
During the data migration phase,
the framework then executes the correct callback functions
in order to perform the actual migration, refinement, and coarsening of the simulation data.
Refinement and coarsening are always performed
via first serializing and then deserializing the corresponding blocks,
even if no migration to another process is necessary.
This software architecture of requiring corresponding serialization functionality
to be also registered when block data is registered
is essential to ensure the extensibility of our
framework to arbitrary simulation data.

\section{Benchmarks}\label{sec:bench}
For the remainder of this article,
we present detailed performance results for a synthetic benchmark application
that executes the entire \gls{amr} pipeline on two current petascale supercomputers.
Finally, we demonstrate the applicability of the \gls{amr} procedure by
outlining performance metrics for a simulation of highly dynamic, turbulent flow.

\subsection{Performance}\label{sec:bench:per}
All benchmarks are run on two petascale supercomputers:
JUQUEEN, an IBM Blue Gene/Q system,
and SuperMUC, a x86-based system build on Intel Xeon \cpu{}s.
JUQUEEN provides 458,752 PowerPC A2 processor cores running at 1.6\,GHz, with each core capable of 4-way multithreading.
Based on our observations in~\cite{Godenschwager2013}, in order to achieve maximal performance,
we make full use of multithreading on JUQUEEN by always placing four threads
(which may either belong to four distinct processes, two different processes, or all to the same process) on one core.
As a result, full-machine jobs consist of 1,835,008 threads.
The SuperMUC system features fewer, but more powerful, cores than JUQUEEN.
It is built out of 18,432 Intel Xeon E5-2680 processors running at 2.7\,GHz, which sums up to a total of 147,456 cores.
During the time the benchmarks were performed, however,
the maximal job size was restricted to 65,536 cores.
The average amount of memory per core on SuperMUC (2\,GiB) is twice the average amount of memory on JUQUEEN (1\,GiB).

In all of the following graphs and tables,
we report the number of processor cores, i.e., the actual amount of hardware resources in use, not the number of processes.
For a fixed number of cores, we can either run the benchmark with $\alpha$ processes (\mpi{} only) or
make use of hybrid execution with $\alpha/\beta$ processes and $\beta$ OpenMP threads per process.
On SuperMUC, we choose $\alpha$ to be equal to the number of cores.
Since we make full use of multithreading on JUQUEEN, we choose $\alpha$ to be equal to four times the number of cores when running the benchmark on JUQUEEN.
For every measurement, we will state the parallelization strategy in use: \mpi{} only or hybrid execution with typically four ($\beta=4$) or eight ($\beta=8$) threads per process.
Since hybrid simulations use $\beta$ times fewer processes, we allocate $\beta$ times more cells for each block (and process).
As a result, for a given benchmark scenario and for a fixed number of processor cores (= fixed amount of hardware resources),
the amount of work remains constant, independent of the parallelization strategy in use.

When using MPI/OpenMP hybrid execution,
computation that involves block data follows a data parallelism model.
First, the block data is uniformly distributed to all available threads.
All threads then perform the same operations on their subset of the data.
When data must be exchanged between neighboring blocks in-between two
iterations/time steps of the simulation,
these messages can be sent and received in parallel by different threads.
In order to send and receive data,
the threads directly call non-blocking, asynchronous send and receive functions of MPI.
Packing data into buffers prior to sending and unpacking data after receiving can
also be done in parallel by different threads if OpenMP is activated.
The OpenMP implementation of the communication module follows a task parallelism model.
First, a unique work package is created for every face, every edge,
and every corner of every block.
These work packages are then processed in parallel by all available threads.
As a result,
all major parts of the simulation are executed thread-parallel.

Since results on JUQUEEN and SuperMUC are often comparable,
we only show a detailed analysis for the performance and scalability of one system: JUQUEEN.
We will, however, discuss the key differences to SuperMUC at the end of each section.

\subsubsection{Setup}\label{sec:bench:per:setup}
\begin{figure}[tp]
  \begin{minipage}{\textwidth}
  \centering
  $\vcenter{\hbox{\includegraphics[width=0.19\textwidth]{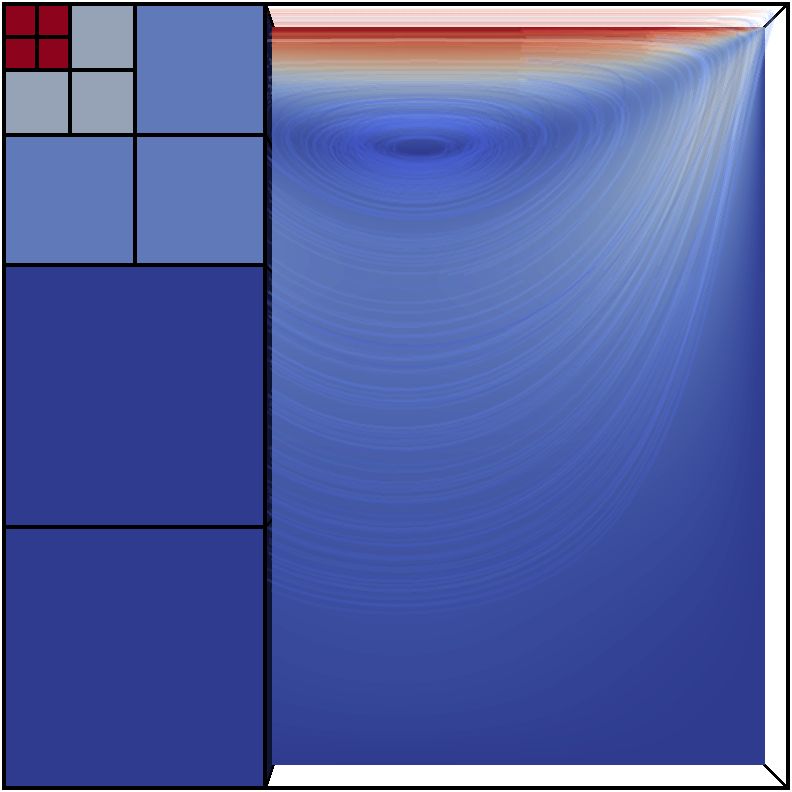}}}$
  \hspace{3mm}
  $\vcenter{\hbox{\includegraphics[height=0.15\textwidth]{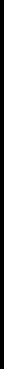}}}$
  \hspace{3mm}  
  $\vcenter{\hbox{\includegraphics[width=0.19\textwidth]{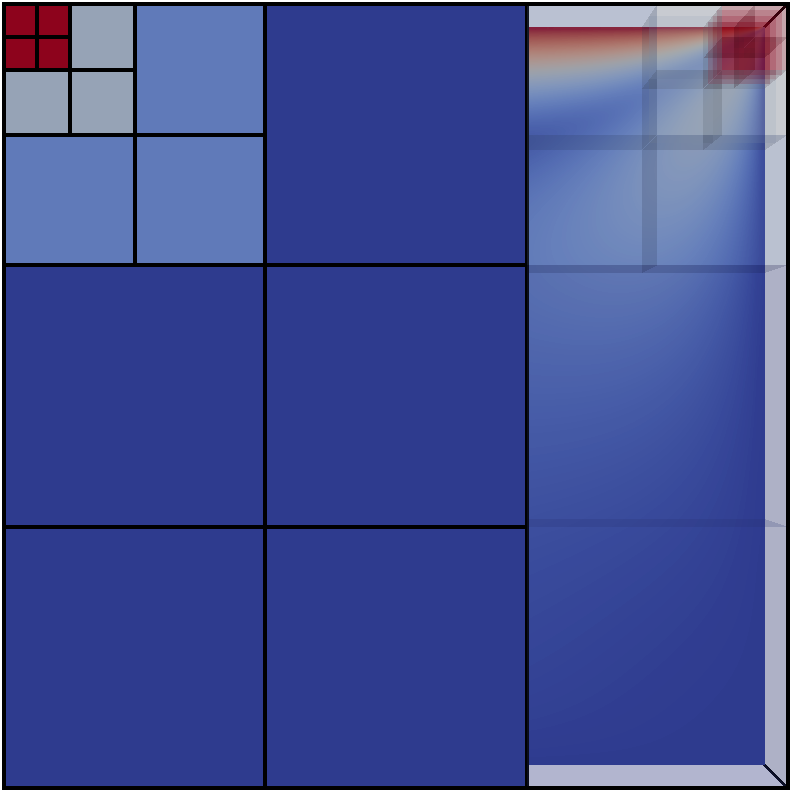}}}$
  \hspace{3mm}
  $\vcenter{\hbox{\includegraphics[height=0.08\textwidth]{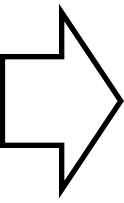}}}$
  \hspace{3mm}
  $\vcenter{\hbox{\includegraphics[width=0.19\textwidth]{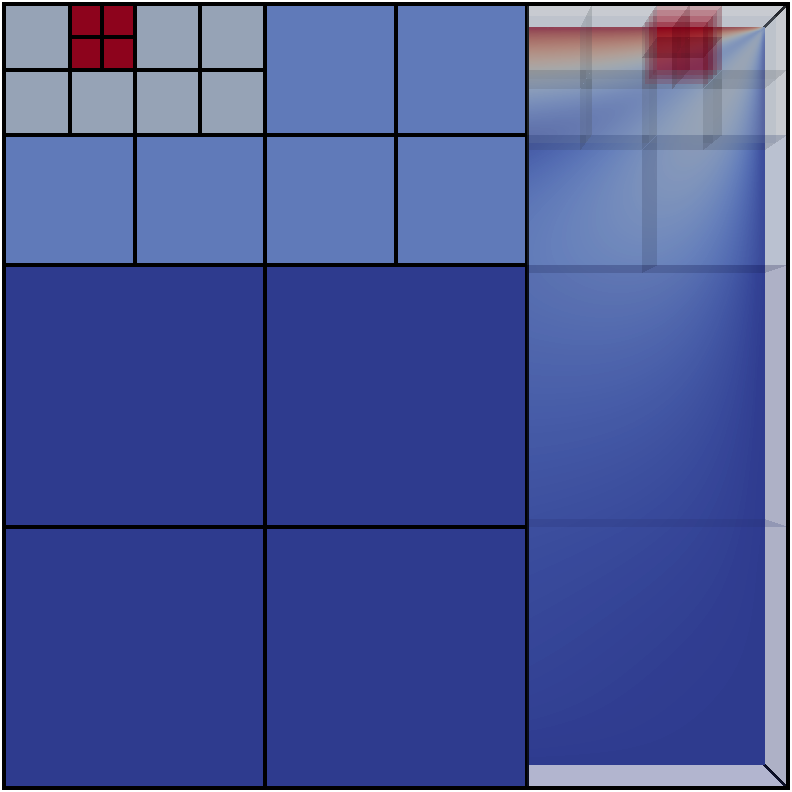}}}$
  \end{minipage}
  \caption{Block partitioning of the benchmark application before (both figures on the left) and after (right) \gls{amr} is triggered.
  \Cref{tab:bench:setup} summarizes corresponding distribution statistics.}
  \label{fig:bench:setup}
\end{figure}

The benchmark application consists of an \gls{lbm}-based simulation of lid-driven cavity flow in 3D.
We employ the \dthqnt{} model that results in 19 unknowns per cell.
Initially, the regions where the moving lid meets the domain boundaries on the side are refined three times.
As a result, the benchmark contains four different levels of refinement,
with level 0 corresponding to the coarsest and level 3 corresponding to the finest level.
At some predefined point in time, \gls{amr} is artificially triggered by
marking all blocks on the finest level for coarsening
and simultaneously generating an equal number of finest cells by marking coarser neighbor blocks for refinement.
Some more blocks are automatically marked for refinement in order to preserve 2:1 balance.
Consequently, the region of finest resolution moves slightly inwards.
Ultimately, 72\,\% of all cells change their size during this \gls{amr} process.
The intent is to put an unusually high amount of stress on the dynamic repartitioning procedure.
The domain partitioning before and after \gls{amr} is triggered is illustrated in \cref{fig:bench:setup}.
During this repartitioning,
the average number of blocks assigned to each process increases by 33\,\%.
As a result,
the amount of data ($\hateq$ number of cells) also increases by the same factor,
since every block, independent of its level, stores a grid of the same size.

When increasing the number of processes by a factor of 2,
the size of the domain is also extended by a factor of 2,
doubling the number of blocks on each level.
As a result,
the distribution characteristics (see \cref{tab:bench:setup}) remain constant,
independent of the number of processes in use.
For a fixed number of grid cells stored at each block,
increasing the number of processes then corresponds to a weak scaling scenario of the simulation
where the global number of blocks and cells increases linearly with the number of processes.
In order to also evaluate the performance of the \gls{amr} pipeline for varying amounts of data per process,
the benchmarks presented in the following subsections are executed multiple times
with different numbers of cells stored at each block.
The stated average number of cells per core values always correspond to the
state of the simulation after \gls{amr} was executed.

\begin{table}[tp]
  \centering
  \caption{Statistics about the distribution of workload and memory among the four available levels
  in the setup illustrated in \cref{fig:bench:setup}.
  Even though the finest cells only cover a small portion of the domain,
  cells on the finest level account for most of the generated workload and memory consumption
  ($\rightarrow$~most blocks are located on the finest level).
  This is true before and after the dynamic repartitioning.}
  \label{tab:bench:setup}
  \footnotesize
  \begin{tabular}{rccccc}
  \toprule
  & $L=0$ & $L=1$ & $L=2$ & $L=3$ &\\
  \midrule
  domain coverage ratio & \tablenum[table-format=2.2]{77.78}\,\% & \tablenum[table-format=2.2]{16.67}\,\% & \tablenum[table-format=2.2]{ 4.17}\,\% & \tablenum[table-format=2.2]{ 1.39}\,\% & \multirow{3}{*}{initially}\\
  workload share        & \tablenum[table-format=2.2]{ 1.10}\,\% & \tablenum[table-format=2.2]{ 3.76}\,\% & \tablenum[table-format=2.2]{15.02}\,\% & \tablenum[table-format=2.2]{80.13}\,\% &\\
  memory/block share    & \tablenum[table-format=2.2]{ 6.54}\,\% & \tablenum[table-format=2.2]{11.22}\,\% & \tablenum[table-format=2.2]{22.43}\,\% & \tablenum[table-format=2.2]{59.81}\,\% &\\
  \midrule
  domain coverage ratio & \tablenum[table-format=2.2]{66.67}\,\% & \tablenum[table-format=2.2]{22.23}\,\% & \tablenum[table-format=2.2]{ 9.72}\,\% & \tablenum[table-format=2.2]{ 1.39}\,\% & \multirow{3}{*}{after \gls{amr}}\\
  workload share        & \tablenum[table-format=2.2]{ 0.78}\,\% & \tablenum[table-format=2.2]{ 4.13}\,\% & \tablenum[table-format=2.2]{28.94}\,\% & \tablenum[table-format=2.2]{66.15}\,\% &\\
  memory/block share    & \tablenum[table-format=2.2]{ 4.23}\,\% & \tablenum[table-format=2.2]{11.27}\,\% & \tablenum[table-format=2.2]{39.44}\,\% & \tablenum[table-format=2.2]{45.07}\,\% &\\
  \bottomrule
  \end{tabular}
\end{table}

\begin{table}[tp]
  \centering
  \caption{Average and maximal number of blocks assigned to each process for the benchmark application illustrated in \cref{fig:bench:setup}.
  These numbers are independent of the total number of processes.
  If more processes are used, the number of blocks on each level is increased accordingly.
  As a consequence, the average as well as the maximal number of blocks per process remain identical for any total number of processes.}
  \label{tab:bench:blocks}
  \footnotesize
  \begin{tabular}{ c C{16mm} C{16mm} C{16mm} }
  \toprule
  & \multicolumn{3}{c}{avg. blocks/proc. (max. blocks/proc.)} \\
  \cmidrule(rl){2-4}
  & & \multicolumn{2}{c}{load balancing} \\
  \cmidrule(rl){3-4}
  level & initially & before & after \\
  \midrule
  0 & 0.383 (1) & 0.328 (\hspace{.5em}1)  & 0.328 (1) \\
  1 & 0.656 (1) & 0.875 (\hspace{.5em}9)  & 0.875 (1) \\
  2 & 1.313 (2) & 3.063 (11) & 3.063 (4) \\
  3 & 3.500 (4) & 3.500 (16) & 3.500 (4) \\
  \bottomrule
  \end{tabular}
\end{table}

The average and maximal number of blocks assigned to each process are
listed in \cref{tab:bench:blocks}.
Before load balancing is executed during the \gls{amr} procedure,
the distribution of blocks (more precisely:\ proxy blocks)
is highly irregular
with some processes containing far more blocks on certain levels than the average number of blocks per process would suggest.
Only after
load balancing,
a perfect distribution is achieved with no single process containing more blocks
than expected.

\subsubsection{Space filling curves}\label{sec:bench:per:sfc}
First, we evaluate the performance of the entire \gls{amr} pipeline
when using our \gls{sfc}-based load balancing scheme during the dynamic balancing stage.
Since \gls{lbm}-based simulations require per-level balancing,
a global synchronization of all block \glspl{id} using an \mpi{} \code{allgather} operation is necessary (cf.\ \cref{sec:dynamic:lb:sfc}).
Moreover, we also make use of hybrid parallel execution.
On JUQUEEN, we use eight threads per \mpi{} process in order to reduce the total number of processes.
Fewer processes result in a smaller number of globally available blocks,
since the more threads are assigned to one process, the more cells are stored at each block
(cf.\ second paragraph of \cref{sec:bench:per}).
Using fewer processes for the same number of allocated cores also
results in more memory available for one process.
Dealing with fewer processes, fewer blocks, and more memory per process
is crucial for the \gls{sfc}-based balancing performance
(see \cref{sec:bench:per:comp}).
Moreover, for hybrid parallel execution with eight threads per process,
the implementation of the \gls{lbm} still achieves close to peak performance~\cite{schornbaum16}.
On JUQUEEN, Morton order-based balancing is approximately twice as fast as Hilbert order-based balancing.
The irregular, indirect memory access caused by an additional access order lookup
required for Hilbert order-based balancing
results in a notable penalty on the Blue Gene/Q architecture.

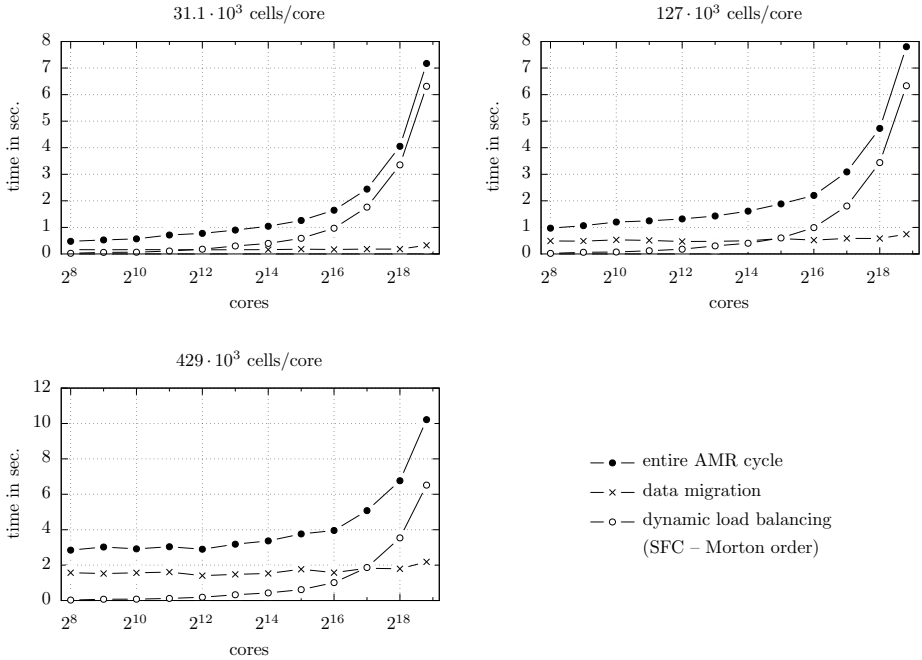
\begin{figure}[tp]
   \centering
   \resizebox{\textwidth}{!}{\input{jsfc_.tex}}
   \caption{Detailed runtime for the entire \gls{amr} cycle on JUQUEEN
   when using \gls{sfc}-based dynamic load balancing for three different
   benchmarks that only vary in the amount of data.}
   \label{fig:bench:per:sfc:juqueen}
\end{figure}

\Cref{fig:bench:per:sfc:juqueen} illustrates the performance of the \gls{amr} procedure
when employing the \gls{sfc}-based balancing scheme using Morton order.
As expected,
the runtime of the balancing algorithm is independent of the amount of data stored on each block,
but it increases significantly the more processes are used.
With the large number of cores on JUQUEEN, the disadvantages of
a global synchronization based on an \mpi{} \code{allgather} operation
becomes clearly visible in the timings.
This approach does not scale to extreme numbers of processes and, as a consequence,
will not be feasible if the number of cores continues to increase.
The time required for the migration procedure is,
as expected, proportional to the amount of data stored on each block.
The communication network on JUQUEEN shows homogeneous performance across almost the entire system.
As a result, for the same amount of data per process,
the runtime of the migration procedure is almost completely independent of the number of processes.
Ultimately, in large simulations,
the runtime of the entire \gls{amr} algorithm is dominated by the \gls{sfc}-based dynamic balancing step.

Results on SuperMUC are similar.
The O(N\,log\,N) scaling properties of the \gls{sfc}-based balancing can be seen in the measurements.
However, since we only use up to $2^{16}$ cores on SuperMUC,
the impact of the global synchronization is much less severe than with larger numbers of cores on JUQUEEN.
Moreover,
due to the different architecture of the processor cores,
there is barely any difference between using Morton or Hilbert order for the \gls{sfc}-based balancing algorithm.
For 13.8 billion cells ($\hateq$ 261 billion unknowns) distributed to
65,536 cores ($\hateq 210 \cdot 10^3$ cells/core),
the entire \gls{amr} procedure is finished in less than one second on SuperMUC.

\subsubsection{Diffusion-based load balancing}\label{sec:bench:per:diffusion}
In this subsection,
we evaluate the runtime of the \gls{amr} procedure
when using the diffusion-based load balancing approach from \cref{sec:dynamic:lb:diffusion}.
We compare two different configurations.
The first configuration exclusively uses the push scheme,
with 15 flow iterations during each execution of the push algorithm.
The other configuration alternates between calling the push and the pull algorithm for every execution of
one main iteration of the diffusion procedure.
Here, each call to the push or pull algorithm only executes 5 flow iterations.
For the rest of this article,
we refer to these two configurations as ``push'' and ``push/pull'' configuration, respectively.
Both variants always converge towards perfect balance as shown in \cref{tab:bench:blocks},
for every number of cores and on both systems, SuperMUC and JUQUEEN.
When using the ``push'' configuration,
executing fewer flow iterations (5, 8, 10, 12, etc.) does not always result in perfect balance,
hence the 15 flow iterations that are used for this configuration.

\begin{figure}[tp]
   \centering
   \scalebox{.7}{\input{jdifit_.tex}}
   \caption{Number of main iterations that are required for the diffusion procedure
   until perfect balance is achieved on JUQUEEN.}
   \label{fig:bench:per:diff:juqueen_iter}
\end{figure}
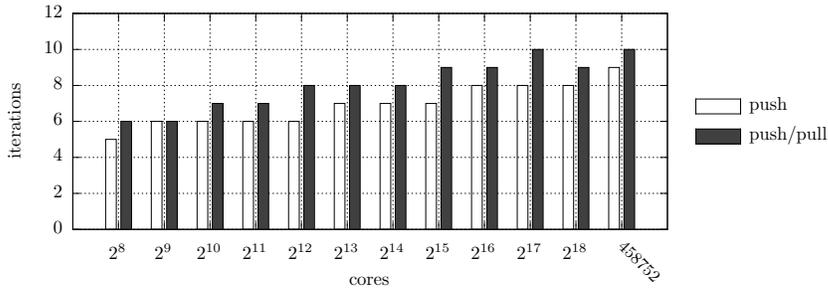

We now use four threads per process on JUQUEEN,
different from the eight threads that were employed for \gls{sfc}-based balancing.
For diffusion-based balancing,
we do not need to restrict the total number of processes
and the version utilizing four instead of eight threads per process results in the best performance
for the entire simulation, including the algorithm for the \gls{lbm} on nonuniform grids~\cite{schornbaum16}.
\Cref{fig:bench:per:diff:juqueen_iter} lists the number of main iterations that are required
for the diffusion procedure until perfect balance is achieved\footnote{number of main iterations $\hateq$ number of times \cref{algo:dynamic:lb:diffusion} is executed during the dynamic load balancing stage}.
The number of iterations slightly increases as the number of processes/utilized cores increases exponentially.
The push/pull version, on average, requires one more iteration than the push only version.
Moreover, executing the pull algorithm is more expensive than executing the push algorithm
(see \cref{algo:dynamic:lb:diffusion:push,algo:dynamic:lb:diffusion:pull}).
However, the push/pull version performs considerably fewer flow iterations (5 instead of 15).
Ultimately,
both versions result in almost identical times for the entire \gls{amr} procedure.

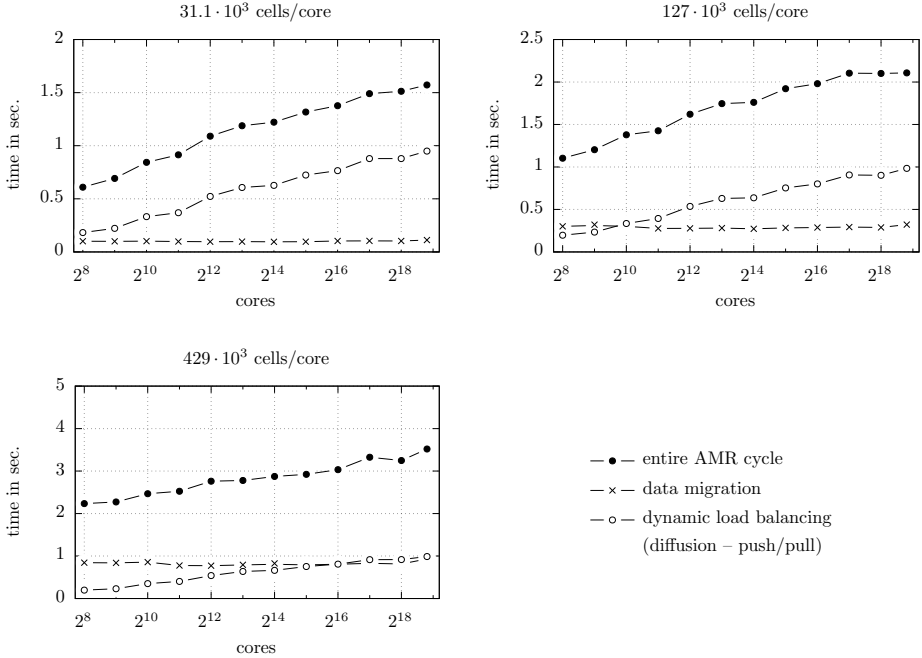
\begin{figure}[tp]
   \centering
   \resizebox{\textwidth}{!}{\input{jdiff_.tex}}
   \caption{Detailed runtime for the entire \gls{amr} cycle on JUQUEEN
   when using diffusion-based dynamic load balancing for three different
   benchmarks that only vary in the amount of data.}
   \label{fig:bench:per:diff:juqueen}
\end{figure}

For a more detailed analysis of the performance results,
we therefore only focus on the push/pull version.
These detailed results are presented in \cref{fig:bench:per:diff:juqueen}.
The time required for the balancing algorithm is, again, independent of
the amount of simulation data stored on each actual block.
Contrary to \gls{sfc}-based balancing, however,
the time required for the diffusion-based balancing increases much slower
and mainly depends on the number of main iterations required for the diffusion approach.
If the number of main iterations is identical (as is the case for $2^{15}$ and $2^{16}$ cores, for example),
the time required for the dynamic load balancing stage also remains almost identical.
Furthermore,
since diffusion-based balancing
requires only communication between neighboring processes,
the time required for the data migration stage proves to be
virtually independent of the total number of processes and
it remains nearly unchanged from 256 up to 458,752 cores.
Consequently, for massively parallel simulations,
the diffusion-based dynamic load balancing approach promises superior scaling characteristics
as compared to an \gls{sfc}-based, global load balancing scheme.
When using the diffusion-based approach,
executing one entire \gls{amr} cycle for a simulation that consists of
197 billion cells ($\hateq$~3.7 trillion unknowns) and runs on all 458,752 cores of JUQUEEN
only takes 3.5 seconds to complete,
as opposed to the 10 seconds when using the \gls{sfc}-based balancing scheme.

Results on SuperMUC are again similar.
Just as on JUQUEEN,
the number of main iterations required to achieve perfect balance
slightly increases while the number of processes increases exponentially.
Also, the runtime of the entire \gls{amr} procedure shows hardly any difference
between the push/pull and the push only version.
Ultimately,
the time required for the dynamic load balancing stage always remains below 90\,ms.
For 13.8 billion cells ($\hateq$ 261 billion unknowns) on 65,536 cores
($\hateq 210 \cdot 10^3$ cells/core) of SuperMUC,
the entire \gls{amr} procedure completes in half a second.
This is almost twice as fast when compared to the \gls{sfc}-based balancing scheme.

\subsubsection{Comparison between different strategies}\label{sec:bench:per:comp}
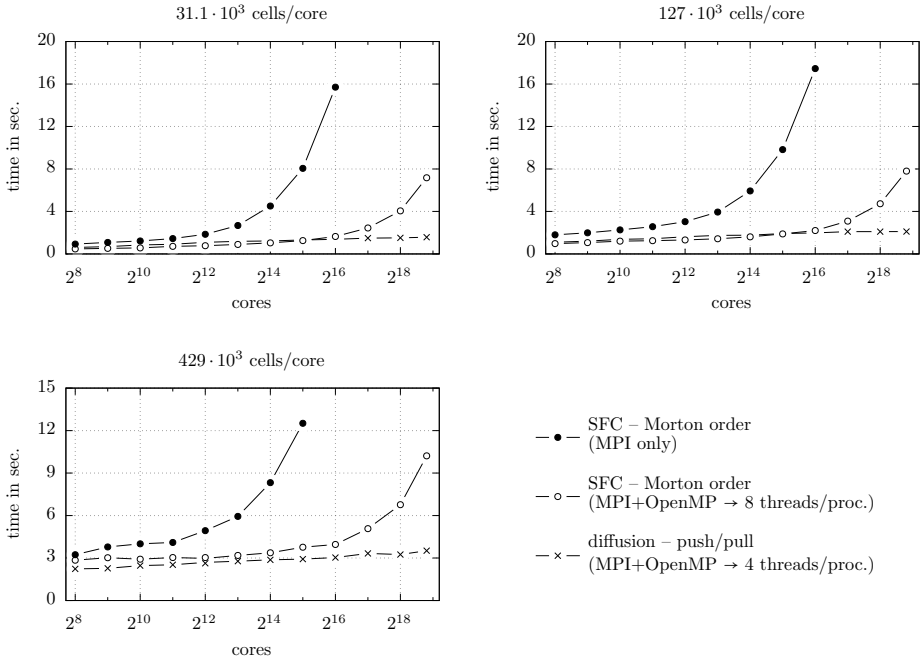
\begin{figure}[tp]
   \centering
   \resizebox{\textwidth}{!}{\input{jamr_.tex}}
   \caption{Runtime of one entire \gls{amr} cycle, including dynamic load balancing and data migration, on JUQUEEN.
   Results are compared for different parallelization and balancing strategies.}
   \label{fig:bench:per:comp:juqueen}
\end{figure}

Finally, \cref{fig:bench:per:comp:juqueen} provide a direct
comparison between different parallelization and dynamic load balancing strategies on JUQUEEN.
The advantages of the diffusion-based balancing approach are obvious.
\gls{amr} that relies on the \gls{sfc}-based balancing scheme presented in \cref{sec:dynamic:lb:sfc}
suffers from the scheme's O(N\,log\,N) scaling properties,
whereas, in direct comparison, \gls{amr} that relies on diffusion-based balancing shows nearly constant runtime.
For the \gls{sfc}-based balancing and \mpi{} only parallelization,
some results are missing in the graphs of \cref{fig:bench:per:comp:juqueen}.
The corresponding simulations cannot complete the \gls{amr} procedure successfully
since they run out of memory during the global \code{allgather} synchronization of the balancing stage.
In order to successfully use the \gls{sfc}-based balancing algorithm on all 458,752 cores of JUQUEEN,
we must use an OpenMP/\mpi{} hybrid parallelization scheme
in order to reduce the number of processes.
As a consequence, more memory is available for each individual process
and the \code{allgather} operation can be executed successfully.
Ultimately,
the diagrams in \cref{fig:bench:per:comp:juqueen} show the superior performance and scaling characteristics
of a fully distributed \gls{amr} pipeline that relies on our diffusion-based instead of
our \gls{sfc}-based dynamic load balancing algorithm.
On SuperMUC, we observed the same behavior.
Since SuperMUC consists of considerably fewer cores with more memory per core,
the differences between the two load balancing strategies are, however, not as large.
Still, the more cores are used for a single simulation,
the greater the benefits of the distributed \gls{amr} pipeline that relies on diffusion-based balancing. 

However, since the diffusion-based balancing approach represents an iterative, local balancing scheme,
perfect balance cannot be guaranteed, as opposed to a global, \gls{sfc}-based balancing scheme.
Consequently, although the chosen benchmark puts a lot of pressure on the repartitioning procedure by
triggering refinement/coarsening for more than two thirds of the global mesh,
other scenarios might not result in perfect balance for the two configurations used in \cref{sec:bench:per:diffusion}.
The advantage of the iterative, fully distributed balancing approach is, however,
that it makes the memory requirement of the entire \gls{amr} pipeline completely
independent of the total number of processes.
The memory required by a process then only depends on the amount of simulation data
that is assigned to this process,
but it is independent of the global amount of data and the total number of processes.
In consequence, as our results indicate, this approach will scale to millions of processes and beyond
and is thus well prepared for the upcoming generation of exascale supercomputers.
Furthermore,
our results indicate that even if perfect balance is not achieved,
the peak workloads
are greatly reduced already after the first few main iterations of the diffusion algorithm.

For the \gls{lbm}-based simulations that we study in our work,
the sizes of the blocks are always chosen such that each process
ends up with only few blocks per level.
Typically, each process will contain no more than around four blocks of each level.
As a result,
most of the time required for one iteration of the \gls{lbm} is spent executing
the compute kernel that updates the 19 (for the \dthqnt{} model) or 27 (\dthqts{})
values stored in each cell of the grid contained within each block.
Less time is spent for communication and the synchronization of data between neighboring blocks.
These kinds of simulations that only contain very few blocks per process
(with hundreds to thousands of cells per block)
do not face the same partitioning quality challenges that
unstructured, cell-based codes are facing where for each process
thousands of individual cells must be kept 
as compact agglomerations with low surface to volume ratios.
For the \gls{lbm}-based simulations with few blocks per process,
the partitioning quality is mainly determined by the balance/imbalance of
the number of blocks per process.
As shown in the previous benchmark,
few iterations ($\leq 10$) of the iterative scheme have been enough to eliminate all imbalances.
Future work that builds on the \gls{amr} pipeline presented in this article
will further study and analyze the current partitioning quality and its influence on
different simulations.
Furthermore, as noted in \cref{sec:dynamic:lb},
future work will include the integration of and comparison with
other specialized dynamic load balancing libraries.

\subsection{Example application}\label{sec:bench:vf}
\begin{figure}[tp]
  \centering
  \includegraphics[width=0.95\textwidth]{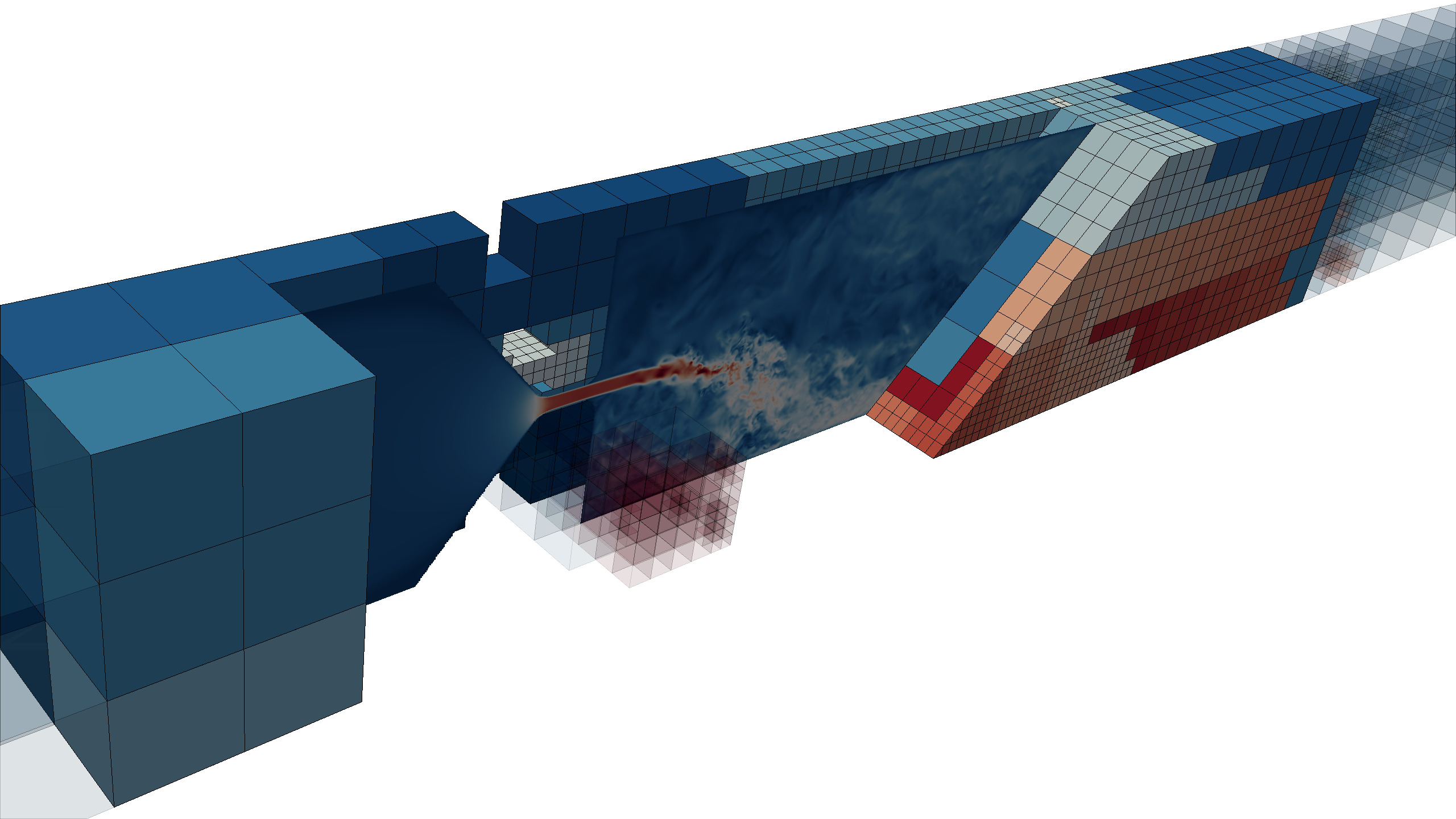}
  \caption{3D simulation using a phantom geometry of the human vocal fold.
  The figure corresponds to the final time step 180,000 ($\hateq$ time step 2,880,000 on the finest level)
  and only shows the domain partitioning into blocks.
  Each block consists of $34 \times 34 \times 34$ cells.
  The different colors of the blocks depict their process association.}
  \label{fig:bench:vf}
\end{figure}

In order to demonstrate the capability of the presented algorithms,
we finally turn to an application-oriented example.
\Cref{fig:bench:vf} illustrates a phantom geometry of the vocal fold
as it is used to study the voice generation within the human throat~\cite{becker2009}.
For this direct numerical simulation with a Reynolds number of 2,500,
we use the \gls{lbm} with the \dthqts{} lattice and the \gls{trt} collision model.
The simulation runs on 3,584 cores of SuperMUC and starts with a
completely uniform partitioning of the entire domain into a total of 23.8 million fluid cells.
\gls{amr} with a refinement criterion based on velocity gradients causes
the simulation to end up with 308 million fluid cells distributed to 5 different levels.
The time spent executing the \gls{amr} algorithm (see \cref{algo:dynamic:proc})
accounts for 17\,\% of the total runtime.
95\,\% of that time is spent on the first \gls{amr} pipeline stage and evaluating the refinement criterion,
i.e., deciding whether blocks require refinement.
Consequently, only 5\,\% of the time spent for \gls{amr} ($\hateq$ 1\,\% of the total runtime)
is consumed by dynamic load balancing and data migration.
During the final phase of the simulation depicted in \cref{fig:bench:vf},
311 times less memory is required and 701 times less workload is generated
as compared to the same simulation with the entire domain refined to the finest level.

\section{Conclusion}\label{sec:conclusion}
In this article,
we have presented an approach for \gls{samr} that exploits the hierarchical nature of a block-structured domain partitioning
by using a lightweight, temporary copy of the core data structure during the \gls{amr} process.
The temporarily created data structure
does not contain any of the simulation data and only
acts as a proxy for the actual data structure.
This proxy data structure enables inexpensive, iterative, diffusion-based dynamic load balancing schemes
that do not require to communicate the actual simulation data
during the entire load balancing phase.
All data structures are stored in a perfectly distributed manner, i.e.,
meta data memory consumption is completely independent of the total number of processes.
Ultimately,
the \gls{samr} approach presented in this article allows simulations that make use of dynamic \gls{amr}
to efficiently scale to extreme-scale parallel machines.

We have demonstrated that an entire \gls{amr} cycle can be executed in half a second
for a mesh that consists of 13.8 billion cells when using 65,536 processor cores.
We have also confirmed the applicability of our \gls{amr} approach for meshes with up to
197 billion cells, distributed to almost half a million cores.
As such,
the approach demonstrates state-of-the-art scalability.
To the best knowledge of the authors,
the scale as well as the performance of the benchmarks presented in this article
significantly exceed the data previously published for \gls{lbm}-based simulations capable of \gls{amr}~\cite{Freudiger08,Hasert2014784,Lahnert2016,FLD:FLD2469,Neumann2013,Schoenherr20113730,Yu20096456}.

For future work,
we will use the underlying distributed data structures combined with the presented \gls{amr} algorithm
for meshfree simulation methods that work fundamentally different to the \gls{lbm}.
Future work will also look into further improving the current implementation of the
load balancing schemes
and see the integration of additional dynamic load balancing algorithms
based on specialized libraries like ParMETIS~\cite{parmetisweb,parmetis},
Zoltan~\cite{zoltan,zoltanweb}, or PT-Scotch~\cite{ptscotch,ptscotchweb}.

\section*{Reproducibility}
All concepts and algorithms presented in this article are implemented in the \Walberla{} software framework.
The \gls{lbm}-based benchmark application presented in \cref{sec:bench:per} was also added to the framework and is available as part of the software.
The \Walberla{} software framework is available under an open source license and
can be freely downloaded at \url{http://walberla.net}.

\section*{Acknowledgments}
The authors would like to thank Prof.\ Henning Meyerhenke, Christian Godenschwager, and Martin Bauer for valuable discussions.
We are also grateful to the J\"ulich Supercomputing Center and the Leibniz Rechenzentrum in Munich for providing
access to the supercomputers JUQUEEN and SuperMUC.

\clearpage

\appendix

\setcounter{table}{0}
\renewcommand{\thetable}{A\arabic{table}}
\setcounter{figure}{0}
\renewcommand{\thefigure}{A\arabic{figure}}

\renewcommand{\bottomfraction}{0.8}
\setcounter{bottomnumber}{3}

\section[JUQUEEN timing details]{Additional timing details for benchmarks on JUQUEEN in \cref{sec:bench:per:sfc}}\label{app:sfc:juqueen}
\Cref{tab:bench:per:sfc:juqueen} lists the exact runtime for one entire \gls{amr} cycle on JUQUEEN
when using \gls{sfc}-based dynamic load balancing.
These timings show that on JUQUEEN Morton order-based balancing is approximately twice as fast as
Hilbert order-based balancing.
\Cref{tab:tables:sfc:juqueen} lists a breakdown of all times that were used to generate the graphs
in \cref{fig:bench:per:sfc:juqueen}.

\begin{table}[hb]
  \centering
  \caption{Time for one entire \gls{amr} cycle, including dynamic load balancing and data migration on JUQUEEN (in seconds).
  The table compares \gls{sfc}-based load balancing using Hilbert order with \gls{sfc}-based load balancing using Morton order.
  Times are listed for three benchmarks that only vary in the amount of data
  assigned to each block (and therefore each core).}
  \label{tab:bench:per:sfc:juqueen}
  \scriptsize
  \begin{tabular}{rccccccccc}
  \toprule
  & & \multicolumn{2}{c}{$31.1 \cdot 10^3$ cells/core} & & \multicolumn{2}{c}{$127 \cdot 10^3$ cells/core} & & \multicolumn{2}{c}{$429 \cdot 10^3$ cells/core} \\
  \cmidrule(rl){3-4}
  \cmidrule(rl){6-7}
  \cmidrule(rl){9-10}
  cores & & Hilbert & Morton & & Hilbert & Morton & & Hilbert & Morton \\
  \midrule
  256      & & \tablenum[table-format=2.2]{ 0.48} & 0.47 & & \tablenum[table-format=2.2]{ 1.02} & 0.97 & & \tablenum[table-format=2.2]{ 2.83} & \tablenum[table-format=2.2]{ 2.85} \\
  512      & & \tablenum[table-format=2.2]{ 0.57} & 0.53 & & \tablenum[table-format=2.2]{ 1.18} & 1.06 & & \tablenum[table-format=2.2]{ 2.64} & \tablenum[table-format=2.2]{ 3.02} \\
  1\,024   & & \tablenum[table-format=2.2]{ 0.60} & 0.57 & & \tablenum[table-format=2.2]{ 1.23} & 1.20 & & \tablenum[table-format=2.2]{ 2.93} & \tablenum[table-format=2.2]{ 2.92} \\
  2\,048   & & \tablenum[table-format=2.2]{ 0.75} & 0.71 & & \tablenum[table-format=2.2]{ 1.39} & 1.25 & & \tablenum[table-format=2.2]{ 3.12} & \tablenum[table-format=2.2]{ 3.04} \\
  4\,096   & & \tablenum[table-format=2.2]{ 0.88} & 0.78 & & \tablenum[table-format=2.2]{ 1.52} & 1.32 & & \tablenum[table-format=2.2]{ 3.14} & \tablenum[table-format=2.2]{ 2.90} \\
  8\,192   & & \tablenum[table-format=2.2]{ 1.16} & 0.90 & & \tablenum[table-format=2.2]{ 1.68} & 1.43 & & \tablenum[table-format=2.2]{ 3.25} & \tablenum[table-format=2.2]{ 3.18} \\
  16\,384  & & \tablenum[table-format=2.2]{ 1.51} & 1.04 & & \tablenum[table-format=2.2]{ 2.23} & 1.61 & & \tablenum[table-format=2.2]{ 3.85} & \tablenum[table-format=2.2]{ 3.37} \\
  32\,768  & & \tablenum[table-format=2.2]{ 2.12} & 1.26 & & \tablenum[table-format=2.2]{ 2.78} & 1.88 & & \tablenum[table-format=2.2]{ 4.73} & \tablenum[table-format=2.2]{ 3.76} \\
  65\,536  & & \tablenum[table-format=2.2]{ 3.35} & 1.65 & & \tablenum[table-format=2.2]{ 3.93} & 2.20 & & \tablenum[table-format=2.2]{ 5.82} & \tablenum[table-format=2.2]{ 3.97} \\
  131\,072 & & \tablenum[table-format=2.2]{ 5.82} & 2.44 & & \tablenum[table-format=2.2]{ 6.50} & 3.09 & & \tablenum[table-format=2.2]{ 8.63} & \tablenum[table-format=2.2]{ 5.08} \\
  262\,144 & & \tablenum[table-format=2.2]{10.63} & 4.05 & & \tablenum[table-format=2.2]{11.41} & 4.73 & & \tablenum[table-format=2.2]{13.57} & \tablenum[table-format=2.2]{ 6.77} \\
  458\,752 & & \tablenum[table-format=2.2]{18.54} & 7.17 & & \tablenum[table-format=2.2]{19.22} & 7.80 & & \tablenum[table-format=2.2]{21.62} & \tablenum[table-format=2.2]{10.22} \\
  \bottomrule
  \end{tabular}
\end{table}

\begin{table}[hb]
  \centering
  \caption{Breakdown of all times (in sec.) from the benchmark outlined in \cref{fig:bench:per:sfc:juqueen}:
  One complete \gls{amr} cycle with \gls{sfc}-based (Morton order) load balancing on JUQUEEN.}
  \label{tab:tables:sfc:juqueen}
  \scriptsize
  \begin{tabular}{rcccccccccccc}
  \toprule
  & & \multicolumn{3}{c}{entire \gls{amr}} & & \multicolumn{3}{c}{load balancing} & & \multicolumn{3}{c}{data migration} \\
  \cmidrule(rl){3-5}
  \cmidrule(rl){7-9}
  \cmidrule(rl){11-13}
  & & \multicolumn{3}{c}{cells/core (in $10^3$)} & & \multicolumn{3}{c}{cells/core (in $10^3$)} & & \multicolumn{3}{c}{cells/core (in $10^3$)} \\[0.75mm]
  cores & & 31.1 & 127 & 429 & & 31.1 & 127 & 429 & & 31.1 & 127 & 429 \\
  \midrule
  256      & & 0.47 & 0.97 & \tablenum[table-format=2.2]{ 2.85} & & 0.02 & 0.02 & 0.02 & & 0.16 & 0.49 & 1.57 \\
  512      & & 0.53 & 1.06 & \tablenum[table-format=2.2]{ 3.02} & & 0.06 & 0.06 & 0.07 & & 0.16 & 0.48 & 1.53 \\
  1\,024   & & 0.57 & 1.20 & \tablenum[table-format=2.2]{ 2.92} & & 0.06 & 0.07 & 0.08 & & 0.16 & 0.53 & 1.56 \\
  2\,048   & & 0.71 & 1.25 & \tablenum[table-format=2.2]{ 3.04} & & 0.11 & 0.12 & 0.12 & & 0.16 & 0.51 & 1.61 \\
  4\,096   & & 0.78 & 1.32 & \tablenum[table-format=2.2]{ 2.90} & & 0.18 & 0.18 & 0.19 & & 0.15 & 0.47 & 1.40 \\
  8\,192   & & 0.90 & 1.43 & \tablenum[table-format=2.2]{ 3.18} & & 0.30 & 0.30 & 0.33 & & 0.16 & 0.48 & 1.48 \\
  16\,384  & & 1.04 & 1.61 & \tablenum[table-format=2.2]{ 3.37} & & 0.39 & 0.40 & 0.43 & & 0.16 & 0.50 & 1.53 \\
  32\,768  & & 1.26 & 1.88 & \tablenum[table-format=2.2]{ 3.76} & & 0.59 & 0.60 & 0.62 & & 0.18 & 0.58 & 1.77 \\
  65\,536  & & 1.65 & 2.20 & \tablenum[table-format=2.2]{ 3.96} & & 0.97 & 0.99 & 1.01 & & 0.17 & 0.52 & 1.58 \\
  131\,072 & & 2.44 & 3.09 & \tablenum[table-format=2.2]{ 5.08} & & 1.76 & 1.80 & 1.86 & & 0.18 & 0.59 & 1.82 \\
  262\,144 & & 4.05 & 4.73 & \tablenum[table-format=2.2]{ 6.77} & & 3.35 & 3.44 & 3.54 & & 0.18 & 0.58 & 1.79 \\
  458\,752 & & 7.17 & 7.80 & \tablenum[table-format=2.2]{10.22} & & 6.31 & 6.33 & 6.52 & & 0.33 & 0.74 & 2.19 \\
  \bottomrule
  \end{tabular}
\end{table}

\clearpage
\section[SuperMUC benchmark results]{Results of benchmarks on SuperMUC of \cref{sec:bench:per:sfc}}\label{app:sfc:supermuc}
\Cref{tab:bench:per:sfc:supermuc} lists the exact runtime for one entire \gls{amr} cycle on SuperMUC
when using \gls{sfc}-based dynamic load balancing.
These timings show that on SuperMUC
there is barely any difference between using Morton or Hilbert order
for the \gls{sfc}-based balancing algorithm.
If a small difference in execution time can be measured,
the \gls{amr} procedure that uses the Hilbert order-based balancing is a bit slower
due to the additional effort that is required for accessing a lookup table (cf.\ \cref{sec:dynamic:lb:sfc}).
On SuperMUC, the benchmark always makes use of
hybrid parallel execution with four OpenMP threads per \mpi{} process.

\Cref{fig:bench:per:sfc:supermuc} shows detailed results for \gls{sfc}-based balancing using Morton order.
The time required for the load balancing stage only depends on the total number of globally available proxy blocks, but is
independent of the amount of simulation data stored in each actual block.
Consequently, the time required for the load balancing stage is identical in all three scenarios.
As expected, the runtime of the \gls{sfc}-based balancing algorithm
increases with the number of processes due to
the \code{allgather} operation and the subsequent sorting of all block \glspl{id} (cf.\ \cref{sec:dynamic:lb:sfc}).
Just as on JUQUEEN, if the amount of data per block increases,
the time required for the migration stage increases proportionally.
Furthermore, migration time also increases with the number of processes
since \gls{sfc}-based balancing results in a global reassignment of all blocks regardless of
the blocks' previous process associations.
As a consequence,
some of the data must be migrated between distant processes,
and this distance increases the more processes are utilized\footnote{On SuperMUC,
more bandwidth is available for communication within the same compute island (1 island $\hateq$ 8,192 cores)
than for inter-island communication.}.
Ultimately,
\gls{sfc}-based dynamic load balancing shows good performance (140\,ms on 65,536 cores),
with the runtime of the entire \gls{amr} cycle being dominated by the migration, refinement, and coarsening of the cell data.
A breakdown of all times that were used to generate the graphs in \cref{fig:bench:per:sfc:juqueen}
is also presented in \cref{tab:tables:sfc:supermuc}.
For 13.8 billion cells ($\hateq$ 261 billion unknowns) on 65,536 cores ($\hateq 210 \cdot 10^3$ cells/core),
the entire \gls{amr} procedure is finished in less than one second.

\begin{table}[hb]
  \centering
  \caption{Time for one entire \gls{amr} cycle, including dynamic load balancing and data migration on SuperMUC (in seconds).
  The table compares \gls{sfc}-based load balancing using Hilbert order with \gls{sfc}-based load balancing using Morton order.
  Times are listed for three benchmarks that only vary in the amount of data assigned to each block (and therefore each core).}
  \label{tab:bench:per:sfc:supermuc}
  \scriptsize
  \begin{tabular}{rccccccccc}
  \toprule
  & & \multicolumn{2}{c}{$62.1 \cdot 10^3$ cells/core} & & \multicolumn{2}{c}{$210 \cdot 10^3$ cells/core} & & \multicolumn{2}{c}{$971 \cdot 10^3$ cells/core} \\
  \cmidrule(rl){3-4}
  \cmidrule(rl){6-7}
  \cmidrule(rl){9-10}
  cores & & Hilbert & Morton & & Hilbert & Morton & & Hilbert & Morton \\
  \midrule
  512     & & 0.15 & 0.15 & & 0.40 & 0.40 & & 1.77 & 1.75 \\
  1\,024  & & 0.17 & 0.17 & & 0.43 & 0.43 & & 1.86 & 1.86 \\
  2\,048  & & 0.18 & 0.18 & & 0.45 & 0.45 & & 1.96 & 1.95 \\
  4\,096  & & 0.20 & 0.20 & & 0.48 & 0.47 & & 2.01 & 2.02 \\
  8\,192  & & 0.22 & 0.20 & & 0.50 & 0.48 & & 2.06 & 2.04 \\
  16\,384 & & 0.28 & 0.26 & & 0.63 & 0.61 & & 2.56 & 2.50 \\
  32\,768 & & 0.33 & 0.29 & & 0.75 & 0.71 & & 3.01 & 2.93 \\
  65\,536 & & 0.57 & 0.42 & & 1.04 & 0.93 & & 3.50 & 3.38 \\
  \bottomrule
  \end{tabular}
\end{table}

\begin{figure}[tbp]
   \centering
   \resizebox{\textwidth}{!}{\input{ssfc_.tex}}
   \caption{Detailed runtime for the entire \gls{amr} cycle on SuperMUC
   when using \gls{sfc}-based dynamic load balancing for three different
   benchmarks that only vary in the amount of data.
   A breakdown of all times used to generate these graphs
   is available in \cref{tab:tables:sfc:supermuc}.}
   \label{fig:bench:per:sfc:supermuc}
\end{figure}

\begin{table}[tbp]
  \centering
  \caption{Breakdown of all times (in sec.) from the benchmark outlined in \cref{fig:bench:per:sfc:supermuc}:
  One complete \gls{amr} cycle with \gls{sfc}-based (Morton order) load balancing on SuperMUC.}
  \label{tab:tables:sfc:supermuc}
  \scriptsize
  \begin{tabular}{rcccccccccccc}
  \toprule
  & & \multicolumn{3}{c}{entire \gls{amr}} & & \multicolumn{3}{c}{load balancing} & & \multicolumn{3}{c}{data migration} \\
  \cmidrule(rl){3-5}
  \cmidrule(rl){7-9}
  \cmidrule(rl){11-13}
  & & \multicolumn{3}{c}{cells/core (in $10^3$)} & & \multicolumn{3}{c}{cells/core (in $10^3$)} & & \multicolumn{3}{c}{cells/core (in $10^3$)} \\[0.75mm]
  cores & & 62.1 & 210  & 971 & & 62.1 & 210  & 971 & & 62.1 & 210  & 971 \\
  \midrule
    512   & & 0.15 & 0.40 & 1.75 & & 0.003 & 0.003 & 0.003 & & 0.08 & 0.24 & 1.09 \\
  1\,024  & & 0.17 & 0.43 & 1.86 & & 0.007 & 0.007 & 0.006 & & 0.09 & 0.25 & 1.15 \\
  2\,048  & & 0.18 & 0.45 & 1.95 & & 0.009 & 0.008 & 0.009 & & 0.09 & 0.27 & 1.20 \\
  4\,096  & & 0.20 & 0.47 & 2.02 & & 0.016 & 0.017 & 0.017 & & 0.09 & 0.28 & 1.26 \\
  8\,192  & & 0.20 & 0.48 & 2.04 & & 0.020 & 0.020 & 0.020 & & 0.09 & 0.28 & 1.25 \\
  16\,384 & & 0.26 & 0.61 & 2.50 & & 0.043 & 0.045 & 0.043 & & 0.12 & 0.36 & 1.66 \\
  32\,768 & & 0.29 & 0.71 & 2.93 & & 0.052 & 0.053 & 0.052 & & 0.14 & 0.46 & 2.08 \\
  65\,536 & & 0.42 & 0.93 & 3.38 & & 0.140 & 0.133 & 0.138 & & 0.19 & 0.53 & 2.41 \\
  \bottomrule
  \end{tabular}
\end{table}

\clearpage
\section[JUQUEEN timing details]{Additional timing details for benchmarks on JUQUEEN in \cref{sec:bench:per:diffusion}}\label{app:diff:juqueen}
\Cref{tab:bench:per:diff:juqueen} lists the exact runtime for one entire \gls{amr} cycle on JUQUEEN
when using diffusion-based dynamic load balancing.
These timings show that the push scheme as well as the push/pull scheme both result
in almost identical runtimes.
\Cref{tab:tables:diff:juqueen} lists a breakdown of all times that were used to generate the graphs
in \cref{fig:bench:per:diff:juqueen}.

\begin{table}[hbp]
  \centering
  \caption{Time for one entire \gls{amr} cycle, including dynamic load balancing and data migration on JUQUEEN (in seconds).
  The table compares two different versions for the diffusion-based load balancing.
  Times are, again, listed for three benchmarks that only vary in the amount of data assigned to each block.}
  \label{tab:bench:per:diff:juqueen}
  \scriptsize
  \begin{tabular}{rccccccccc}
  \toprule
  & & \multicolumn{2}{c}{$31.1 \cdot 10^3$ cells/core} & & \multicolumn{2}{c}{$127 \cdot 10^3$ cells/core} & & \multicolumn{2}{c}{$429 \cdot 10^3$ cells/core} \\
  \cmidrule(rl){3-4}
  \cmidrule(rl){6-7}
  \cmidrule(rl){9-10}
  cores & & push & push/pull & & push & push/pull & & push & push/pull \\
  \midrule
  256      & & 0.60 & 0.61 & & 1.05 & 1.10 & & 2.17 & 2.23 \\
  512      & & 0.71 & 0.69 & & 1.22 & 1.20 & & 2.40 & 2.27 \\
  1\,024   & & 0.80 & 0.84 & & 1.34 & 1.38 & & 2.48 & 2.47 \\
  2\,048   & & 0.91 & 0.91 & & 1.44 & 1.43 & & 2.52 & 2.52 \\
  4\,096   & & 1.03 & 1.09 & & 1.56 & 1.62 & & 2.67 & 2.76 \\
  8\,192   & & 1.09 & 1.19 & & 1.65 & 1.75 & & 2.67 & 2.78 \\
  16\,384  & & 1.18 & 1.22 & & 1.74 & 1.76 & & 2.80 & 2.87 \\
  32\,768  & & 1.23 & 1.32 & & 1.85 & 1.92 & & 2.87 & 2.92 \\
  65\,536  & & 1.35 & 1.38 & & 1.86 & 1.98 & & 3.02 & 3.03 \\
  131\,072 & & 1.36 & 1.49 & & 1.79 & 2.10 & & 3.04 & 3.33 \\
  262\,144 & & 1.48 & 1.51 & & 1.94 & 2.10 & & 3.15 & 3.25 \\
  458\,752 & & 1.55 & 1.57 & & 2.06 & 2.11 & & 3.26 & 3.52 \\
  \bottomrule
  \end{tabular}
\end{table}

\begin{table}[hbp]
  \centering
  \caption{Breakdown of all times (in sec.) from the benchmark outlined in \cref{fig:bench:per:diff:juqueen}:
  One complete \gls{amr} cycle with diffusion-based (push/pull scheme with 5 flow iterations) load balancing on JUQUEEN.}
  \label{tab:tables:diff:juqueen}
  \scriptsize
  \begin{tabular}{rcccccccccccc}
  \toprule
  & & \multicolumn{3}{c}{entire \gls{amr}} & & \multicolumn{3}{c}{load balancing} & & \multicolumn{3}{c}{data migration} \\
  \cmidrule(rl){3-5}
  \cmidrule(rl){7-9}
  \cmidrule(rl){11-13}
  & & \multicolumn{3}{c}{cells/core (in $10^3$)} & & \multicolumn{3}{c}{cells/core (in $10^3$)} & & \multicolumn{3}{c}{cells/core (in $10^3$)} \\[0.75mm]
  cores & & 31.1 & 127 & 429 & & 31.1 & 127 & 429 & & 31.1 & 127 & 429 \\
  \midrule  
  256      & & 0.61 & 1.10 & 2.23 & & 0.18 & 0.20 & 0.20 & & 0.10 & 0.30 & 0.84 \\
  512      & & 0.69 & 1.20 & 2.27 & & 0.22 & 0.23 & 0.23 & & 0.10 & 0.30 & 0.84 \\
  1\,024   & & 0.84 & 1.38 & 2.47 & & 0.33 & 0.33 & 0.35 & & 0.10 & 0.30 & 0.86 \\
  2\,048   & & 0.91 & 1.43 & 2.53 & & 0.37 & 0.39 & 0.40 & & 0.10 & 0.28 & 0.78 \\
  4\,096   & & 1.09 & 1.62 & 2.76 & & 0.52 & 0.54 & 0.54 & & 0.10 & 0.28 & 0.77 \\
  8\,192   & & 1.19 & 1.75 & 2.78 & & 0.61 & 0.63 & 0.64 & & 0.10 & 0.28 & 0.78 \\
  16\,384  & & 1.22 & 1.76 & 2.87 & & 0.63 & 0.64 & 0.66 & & 0.10 & 0.27 & 0.77 \\
  32\,768  & & 1.32 & 1.92 & 2.92 & & 0.72 & 0.75 & 0.75 & & 0.10 & 0.28 & 0.79 \\
  65\,536  & & 1.38 & 1.98 & 3.03 & & 0.77 & 0.80 & 0.81 & & 0.10 & 0.29 & 0.81 \\
  131\,072 & & 1.49 & 2.10 & 3.33 & & 0.88 & 0.91 & 0.92 & & 0.10 & 0.29 & 0.83 \\
  262\,144 & & 1.51 & 2.10 & 3.25 & & 0.88 & 0.90 & 0.92 & & 0.10 & 0.29 & 0.81 \\
  458\,752 & & 1.57 & 2.11 & 3.52 & & 0.95 & 0.98 & 0.99 & & 0.11 & 0.32 & 0.93 \\
  \bottomrule
  \end{tabular}
\end{table}

\clearpage
\section[SuperMUC benchmark results]{Results of benchmarks on SuperMUC of \cref{sec:bench:per:diffusion}}\label{app:diff:supermuc}
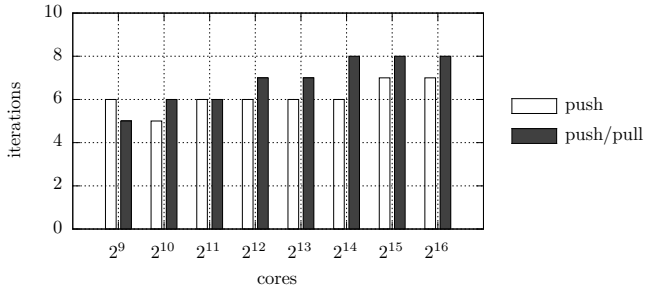
\begin{figure}[!hb]
   \centering
   \scalebox{.7}{\input{sdifit_.tex}}
   \caption{Number of main iterations that are required for the diffusion procedure
   until perfect balance is achieved on SuperMUC.}
   \label{fig:bench:per:diff:supermuc_iter}
\end{figure}

\Cref{fig:bench:per:diff:supermuc_iter} lists the number of main iterations that are required
for the diffusion procedure until perfect balance is achieved on SuperMUC\footnote{number of main iterations $\hateq$ number of times \cref{algo:dynamic:lb:diffusion} is executed during the dynamic load balancing stage}.
The number of iterations increases very little as the number of processes/utilized cores increases exponentially.
Just as on JUQUEEN,
the push/pull version, on average, requires one more iteration than the push only version.
Ultimately,
both versions result in almost identical times for the entire \gls{amr} procedure
as shown in \cref{tab:bench:per:diff:supermuc}.

Detailed results when using the push/pull scheme
are presented in \cref{fig:bench:per:diff:supermuc}.
Here, we again use hybrid parallelization with four threads per process.
Contrary to \gls{sfc}-based balancing,
the time required for the diffusion-based balancing increases much slower
and mainly depends on the number of main iterations required for the diffusion approach.
Just as on JUQUEEN,
if the number of main iterations is identical,
the time required for the dynamic load balancing stage also remains almost identical.
Contrary to \gls{sfc}-based balancing on SuperMUC (see \cref{fig:bench:per:sfc:supermuc}),
the time required for the data migration stage proves to be
virtually independent of the total number of processes
since diffusion-based balancing requires only communication between neighboring processes.
For simulations with large amounts of data that must be communicated during the \gls{amr} procedure,
the runtime of the entire \gls{amr} algorithm
remains almost constant for any number of processes.

\begin{table}[!hb]
  \centering
  \caption{Time for one entire \gls{amr} cycle, including dynamic load balancing and data migration on SuperMUC (in seconds).
  The table compares two different versions for the diffusion-based load balancing.
  Times are listed for three benchmarks that only vary in the amount of data assigned to each block (and therefore each core).}  
  \label{tab:bench:per:diff:supermuc}
  \scriptsize
  \begin{tabular}{rcccccccccccc}
  \toprule
  & & \multicolumn{2}{c}{$62.1 \cdot 10^3$ cells/core} & & \multicolumn{2}{c}{$210 \cdot 10^3$ cells/core} & & \multicolumn{2}{c}{$971 \cdot 10^3$ cells/core} \\
  \cmidrule(rl){3-4}
  \cmidrule(rl){6-7}
  \cmidrule(rl){9-10}
  cores & & push & push/pull & & push & push/pull & & push & push/pull \\
  \midrule
  512     & & 0.14 & 0.15 & & 0.36 & 0.36 & & 1.53 & 1.58 \\
  1\,024  & & 0.16 & 0.17 & & 0.38 & 0.39 & & 1.60 & 1.62 \\
  2\,048  & & 0.18 & 0.18 & & 0.41 & 0.40 & & 1.63 & 1.65 \\
  4\,096  & & 0.20 & 0.20 & & 0.42 & 0.43 & & 1.65 & 1.69 \\
  8\,192  & & 0.21 & 0.21 & & 0.43 & 0.44 & & 1.65 & 1.69 \\
  16\,384 & & 0.23 & 0.23 & & 0.46 & 0.46 & & 1.69 & 1.80 \\
  32\,768 & & 0.24 & 0.26 & & 0.48 & 0.49 & & 1.75 & 1.86 \\
  65\,536 & & 0.27 & 0.26 & & 0.53 & 0.53 & & 1.80 & 1.90 \\
  \bottomrule
  \end{tabular}
\end{table}

\clearpage
\begin{figure}[tbp]
   \centering
   \resizebox{\textwidth}{!}{\input{sdiff_.tex}}
   \caption{Detailed runtime for the entire \gls{amr} cycle on SuperMUC
   when using diffusion-based dynamic load balancing for three different
   benchmarks that only vary in the amount of data.
   A breakdown of all times used to generate these graphs
   is available in \cref{tab:tables:diff:supermuc}.}
   \label{fig:bench:per:diff:supermuc}
\end{figure}

\begin{table}[tbp]
  \centering
  \caption{Breakdown of all times (in sec.) from the benchmark outlined in \cref{fig:bench:per:diff:supermuc}:
  One complete \gls{amr} cycle with diffusion-based (push/pull scheme with 5 flow iterations) load balancing on SuperMUC.}
  \label{tab:tables:diff:supermuc}
  \scriptsize
  \begin{tabular}{rccccccccccccccc}
  \toprule
  & & \multicolumn{3}{c}{entire \gls{amr}} & & \multicolumn{3}{c}{load balancing} & & \multicolumn{3}{c}{data migration} \\
  \cmidrule(rl){3-5}
  \cmidrule(rl){7-9}
  \cmidrule(rl){11-13}
  & & \multicolumn{3}{c}{cells/core (in $10^3$)} & & \multicolumn{3}{c}{cells/core (in $10^3$)} & & \multicolumn{3}{c}{cells/core (in $10^3$)} \\[0.75mm]
  cores & & 62.1 & 210 & 971 & & 62.1 & 210 & 971 & & 62.1 & 210 & 971 \\
  \midrule
  512     & & 0.15 & 0.36 & 1.58 & & 0.014 & 0.013 & 0.013 & & 0.07 & 0.20 & 0.88 \\
  1\,024  & & 0.17 & 0.39 & 1.62 & & 0.024 & 0.023 & 0.023 & & 0.07 & 0.20 & 0.87 \\
  2\,048  & & 0.18 & 0.40 & 1.65 & & 0.031 & 0.030 & 0.031 & & 0.07 & 0.20 & 0.88 \\
  4\,096  & & 0.20 & 0.43 & 1.69 & & 0.046 & 0.045 & 0.045 & & 0.07 & 0.20 & 0.88 \\
  8\,192  & & 0.21 & 0.44 & 1.69 & & 0.052 & 0.051 & 0.052 & & 0.07 & 0.20 & 0.87 \\
  16\,384 & & 0.23 & 0.46 & 1.80 & & 0.069 & 0.071 & 0.072 & & 0.07 & 0.20 & 0.89 \\
  32\,768 & & 0.26 & 0.49 & 1.86 & & 0.085 & 0.086 & 0.086 & & 0.07 & 0.23 & 0.97 \\
  65\,536 & & 0.26 & 0.53 & 1.90 & & 0.087 & 0.088 & 0.087 & & 0.07 & 0.23 & 1.01 \\
  \bottomrule
  \end{tabular}
\end{table}

\clearpage
\section[Example application]{Time evolution of the application in \cref{sec:bench:vf}}\label{app:pics}
\begin{figure}[!hbp]
  \centering
  \includegraphics[width=0.9\textwidth]{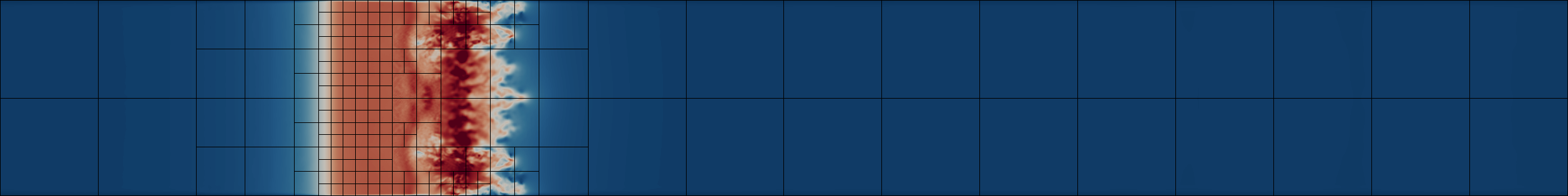}\\
  \vspace{1mm}
  \includegraphics[width=0.9\textwidth]{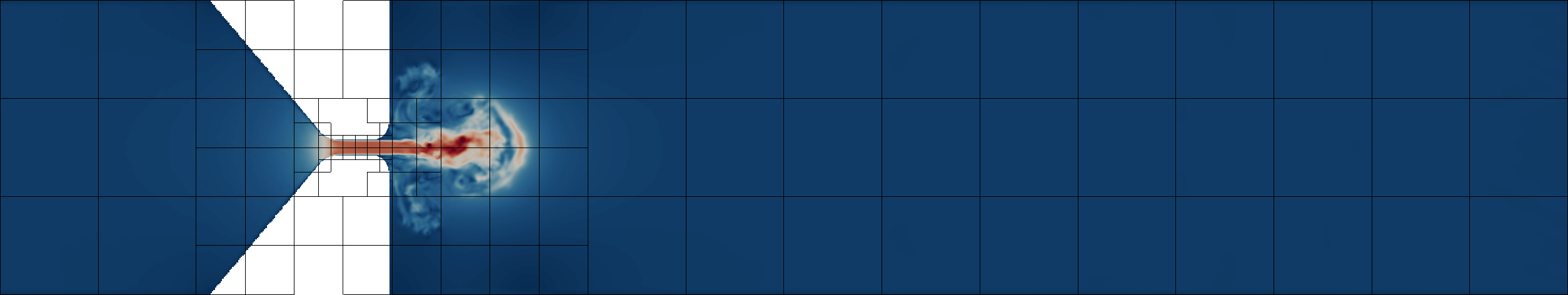}\\
  \vspace{5mm}
  \includegraphics[width=0.9\textwidth]{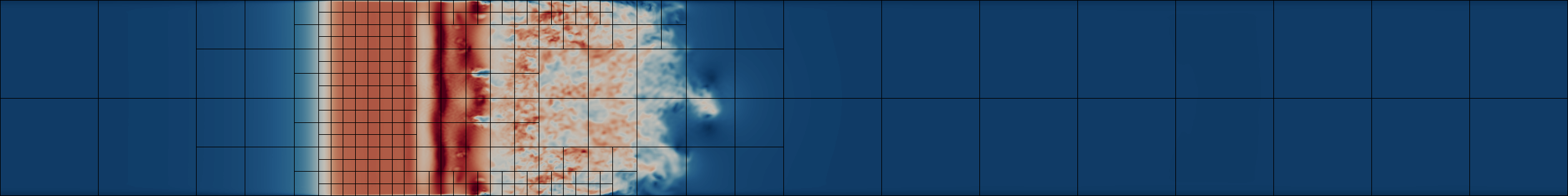}\\
  \vspace{1mm}
  \includegraphics[width=0.9\textwidth]{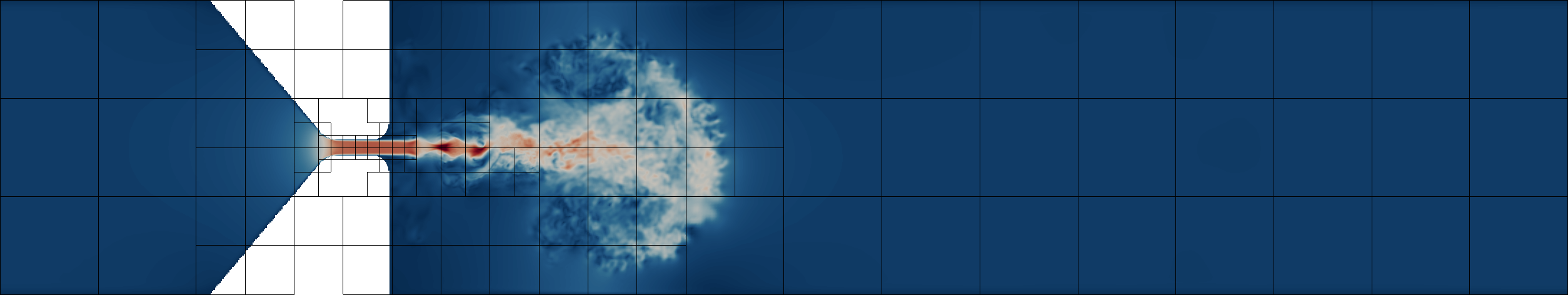}\\
  \vspace{5mm}
  \includegraphics[width=0.9\textwidth]{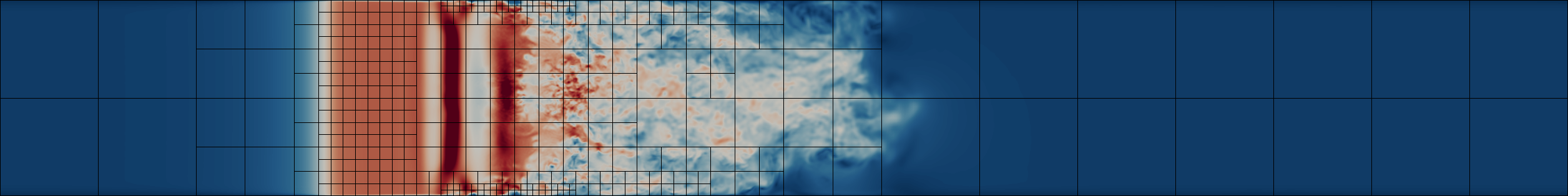}\\
  \vspace{1mm}
  \includegraphics[width=0.9\textwidth]{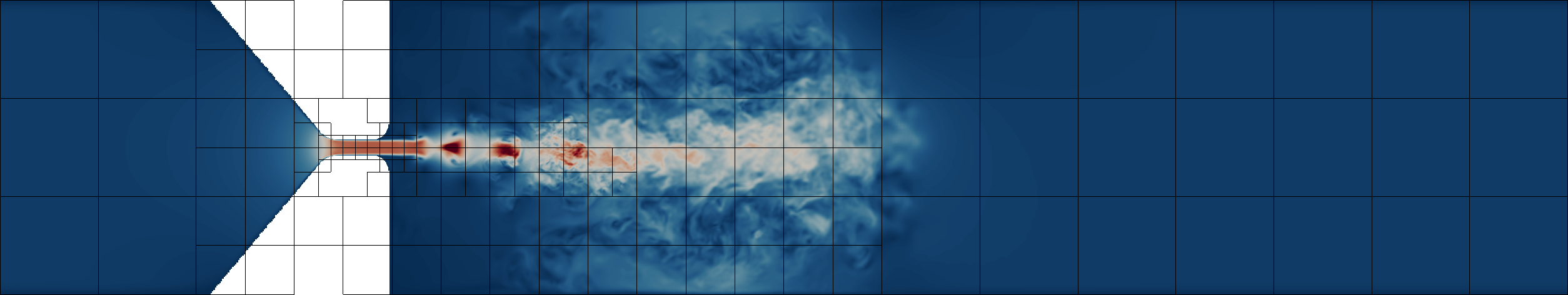}\\
  \vspace{5mm}
  \includegraphics[width=0.9\textwidth]{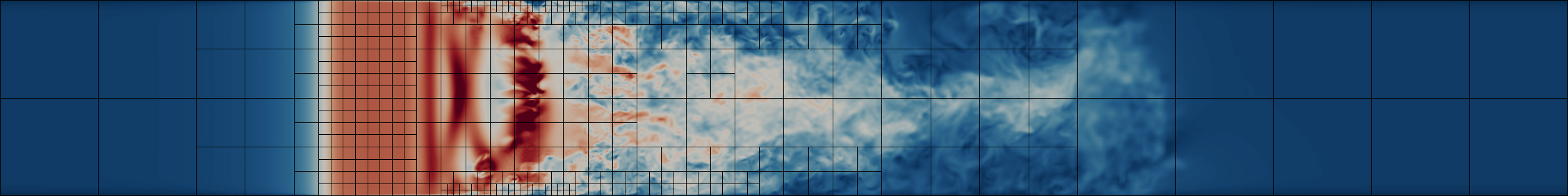}\\
  \vspace{1mm}
  \includegraphics[width=0.9\textwidth]{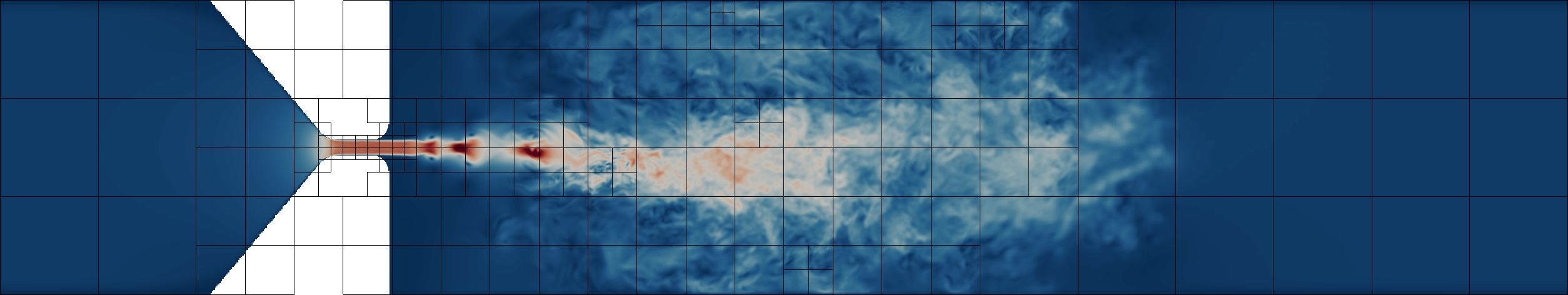}\\
  \caption{Evolution of the simulation introduced in \cref{sec:bench:vf} over time.
  The figures represent 2D slices through the simulation as viewed from the top and from the side.
  They show the state of the simulation during the build-up phase at time step 6,000, 11,000, 20,000, and 36,000.
  These figures, however, do not show the entire domain.
  The actual domain extends slightly further to the left and a lot further to the right.}
  \label{fig:app:vf}
\end{figure}

\Cref{fig:app:vf} provides an illustration of the evolution of the simulation
introduced in \cref{sec:bench:vf} over time.
The simulation runs for approx.\ 24 hours on 3,584 cores of SuperMUC and
spans 180,000 time steps on the coarsest and 2,880,000 time steps on the finest grid.

\noindent
Of the 308 million fluid cells at the end of the simulation,
almost 50\,\% are located on the second finest level.
Over the course of the simulation,
the total number of blocks increases from 612 (all on the coarsest level) to 8030 (distributed among all 5 levels).
Even though the \gls{amr} algorithm is executed in every time step of the 180,000 coarse time steps,
actual dynamic refinement/coarsening of the grid data is only triggered 537 times.
Ultimately,
executing the entire \gls{amr} pipeline,
including dynamic load balancing and the migration of the data,
only happens every 335 time steps (on average).

\end{document}

%% file: dyn1_v2_.tex
\begingroup%
  \makeatletter%
  \providecommand\color[2][]{%
    \errmessage{(Inkscape) Color is used for the text in Inkscape, but the package 'color.sty' is not loaded}%
    \renewcommand\color[2][]{}%
  }%
  \providecommand\transparent[1]{%
    \errmessage{(Inkscape) Transparency is used (non-zero) for the text in Inkscape, but the package 'transparent.sty' is not loaded}%
    \renewcommand\transparent[1]{}%
  }%
  \providecommand\rotatebox[2]{#2}%
  \ifx\svgwidth\undefined%
    \setlength{\unitlength}{1050.54003906bp}%
    \ifx\svgscale\undefined%
      \relax%
    \else%
      \setlength{\unitlength}{\unitlength * \real{\svgscale}}%
    \fi%
  \else%
    \setlength{\unitlength}{\svgwidth}%
  \fi%
  \global\let\svgwidth\undefined%
  \global\let\svgscale\undefined%
  \makeatother%
  \begin{picture}(1,0.48399735)%
  \huge
    \put(0,0){\includegraphics[width=\unitlength]{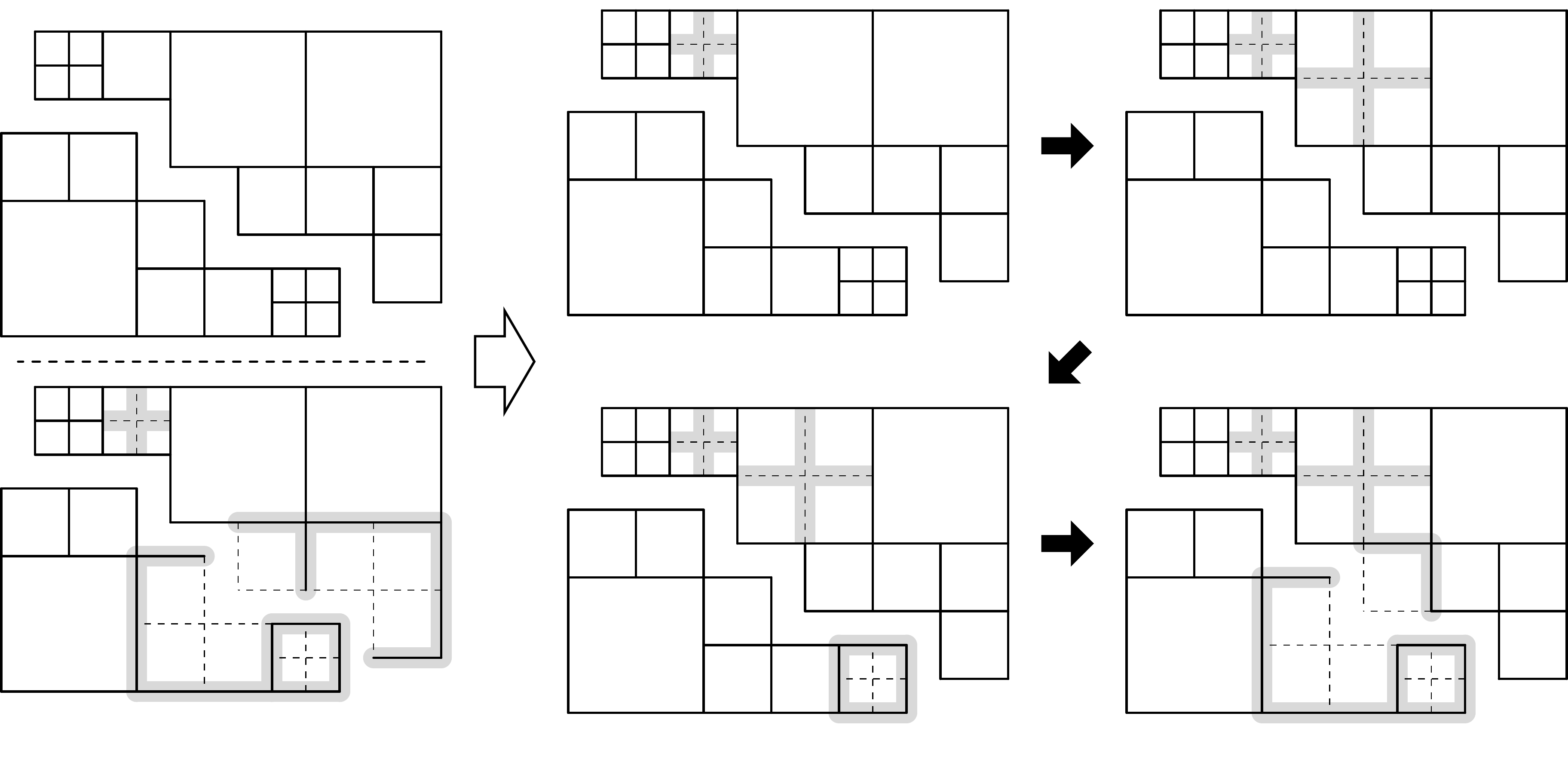}}%
    \put(0.49427911,0.25625734){\makebox(0,0)[lb]{\smash{(1)}}}%
    \put(0.85051494,0.25625734){\makebox(0,0)[lb]{\smash{(2)}}}%
    \put(0.49427911,0.00255925){\makebox(0,0)[lb]{\smash{(3)}}}%
    \put(0.85051494,0.00255925){\makebox(0,0)[lb]{\smash{(4)}}}%
  \end{picture}%
\endgroup%

%% file: sfc_.tex
\begingroup%
  \makeatletter%
  \providecommand\color[2][]{%
    \errmessage{(Inkscape) Color is used for the text in Inkscape, but the package 'color.sty' is not loaded}%
    \renewcommand\color[2][]{}%
  }%
  \providecommand\transparent[1]{%
    \errmessage{(Inkscape) Transparency is used (non-zero) for the text in Inkscape, but the package 'transparent.sty' is not loaded}%
    \renewcommand\transparent[1]{}%
  }%
  \providecommand\rotatebox[2]{#2}%
  \ifx\svgwidth\undefined%
    \setlength{\unitlength}{1043.75996094bp}%
    \ifx\svgscale\undefined%
      \relax%
    \else%
      \setlength{\unitlength}{\unitlength * \real{\svgscale}}%
    \fi%
  \else%
    \setlength{\unitlength}{\svgwidth}%
  \fi%
  \global\let\svgwidth\undefined%
  \global\let\svgscale\undefined%
  \makeatother%
  \begin{picture}(1,0.43082308)%
  \huge
    \put(0,0){\includegraphics[width=\unitlength]{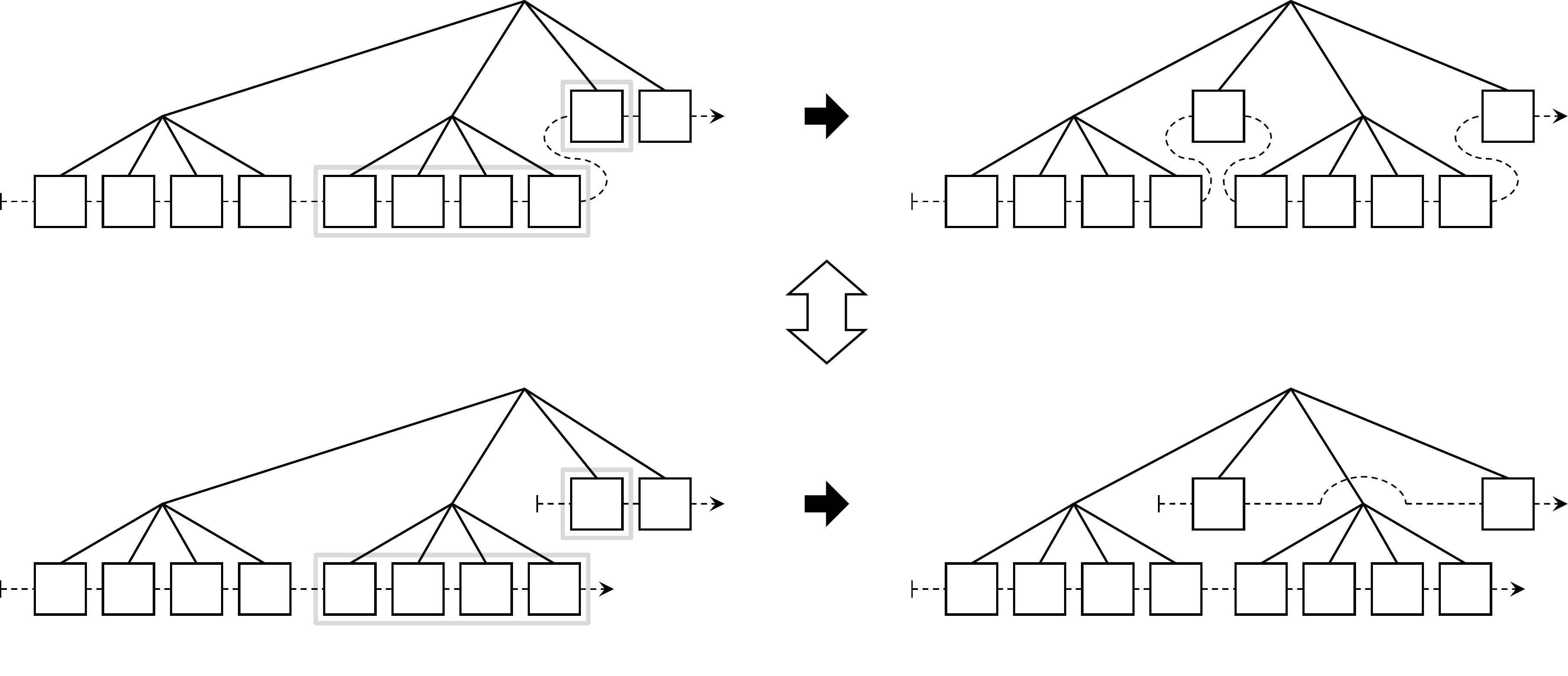}}%
    \put(0.03367633,0.29597407){\makebox(0,0)[lb]{\smash{1}}}%
    \put(0.07713459,0.29597407){\makebox(0,0)[lb]{\smash{1}}}%
    \put(0.12059286,0.29597407){\makebox(0,0)[lb]{\smash{2}}}%
    \put(0.16405113,0.29597407){\makebox(0,0)[lb]{\smash{2}}}%
    \put(0.21838354,0.29597407){\makebox(0,0)[lb]{\smash{3}}}%
    \put(0.26184181,0.29597407){\makebox(0,0)[lb]{\smash{3}}}%
    \put(0.30530008,0.29597407){\makebox(0,0)[lb]{\smash{3}}}%
    \put(0.34875835,0.29597407){\makebox(0,0)[lb]{\smash{4}}}%
    \put(0.37591977,0.35030648){\makebox(0,0)[lb]{\smash{4}}}%
    \put(0.41937803,0.35030648){\makebox(0,0)[lb]{\smash{4}}}%
    \put(0.6149594,0.29597407){\makebox(0,0)[lb]{\smash{1}}}%
    \put(0.65841767,0.29597407){\makebox(0,0)[lb]{\smash{1}}}%
    \put(0.70187594,0.29597407){\makebox(0,0)[lb]{\smash{2}}}%
    \put(0.7453342,0.29597407){\makebox(0,0)[lb]{\smash{2}}}%
    \put(0.79966662,0.29597407){\makebox(0,0)[lb]{\smash{4}}}%
    \put(0.84312489,0.29597407){\makebox(0,0)[lb]{\smash{4}}}%
    \put(0.88658316,0.29597407){\makebox(0,0)[lb]{\smash{4}}}%
    \put(0.93004142,0.29597407){\makebox(0,0)[lb]{\smash{4}}}%
    \put(0.95720284,0.35027774){\makebox(0,0)[lb]{\smash{4}}}%
    \put(0.77249562,0.35027774){\makebox(0,0)[lb]{\smash{3}}}%
    \put(0.03365716,0.04876205){\makebox(0,0)[lb]{\smash{1}}}%
    \put(0.07713459,0.04876205){\makebox(0,0)[lb]{\smash{1}}}%
    \put(0.12059286,0.04876205){\makebox(0,0)[lb]{\smash{2}}}%
    \put(0.16405113,0.04876205){\makebox(0,0)[lb]{\smash{2}}}%
    \put(0.21838354,0.04876205){\makebox(0,0)[lb]{\smash{3}}}%
    \put(0.26184181,0.04876205){\makebox(0,0)[lb]{\smash{3}}}%
    \put(0.30530008,0.04876205){\makebox(0,0)[lb]{\smash{4}}}%
    \put(0.34875835,0.04876205){\makebox(0,0)[lb]{\smash{4}}}%
    \put(0.37591977,0.10309447){\makebox(0,0)[lb]{\smash{1}}}%
    \put(0.41937803,0.10309447){\makebox(0,0)[lb]{\smash{2}}}%
    \put(0.6149594,0.04876205){\makebox(0,0)[lb]{\smash{1}}}%
    \put(0.65841767,0.04876205){\makebox(0,0)[lb]{\smash{1}}}%
    \put(0.70187594,0.04876205){\makebox(0,0)[lb]{\smash{2}}}%
    \put(0.7453342,0.04876205){\makebox(0,0)[lb]{\smash{2}}}%
    \put(0.79966662,0.04876205){\makebox(0,0)[lb]{\smash{1}}}%
    \put(0.84312489,0.04876205){\makebox(0,0)[lb]{\smash{1}}}%
    \put(0.88658316,0.04876205){\makebox(0,0)[lb]{\smash{1}}}%
    \put(0.93004142,0.04876205){\makebox(0,0)[lb]{\smash{1}}}%
    \put(0.95720284,0.10309447){\makebox(0,0)[lb]{\smash{2}}}%
    \put(0.77249562,0.10309447){\makebox(0,0)[lb]{\smash{3}}}%
    \put(0.21335365,0.25041761){\makebox(0,0)[lb]{\smash{(1.1)}}}%
    \put(0.21335365,0.00323434){\makebox(0,0)[lb]{\smash{(2.1}}}%
    \put(0.24359049,0.00323434){\makebox(0,0)[lb]{\smash{)}}}%
    \put(0.77290759,0.25041761){\makebox(0,0)[lb]{\smash{(}}}%
    \put(0.77877101,0.25041761){\makebox(0,0)[lb]{\smash{1.2)}}}%
    \put(0.77290759,0.00323434){\makebox(0,0)[lb]{\smash{(}}}%
    \put(0.77877101,0.00323434){\makebox(0,0)[lb]{\smash{2}}}%
    \put(0.78854337,0.00323434){\makebox(0,0)[lb]{\smash{.2)}}}%
  \end{picture}%
\endgroup%

%% file: jsfc_.tex
\begingroup
  \makeatletter
  \providecommand\color[2][]{%
    \GenericError{(gnuplot) \space\space\space\@spaces}{%
      Package color not loaded in conjunction with
      terminal option `colourtext'%
    }{See the gnuplot documentation for explanation.%
    }{Either use 'blacktext' in gnuplot or load the package
      color.sty in LaTeX.}%
    \renewcommand\color[2][]{}%
  }%
  \providecommand\includegraphics[2][]{%
    \GenericError{(gnuplot) \space\space\space\@spaces}{%
      Package graphicx or graphics not loaded%
    }{See the gnuplot documentation for explanation.%
    }{The gnuplot epslatex terminal needs graphicx.sty or graphics.sty.}%
    \renewcommand\includegraphics[2][]{}%
  }%
  \providecommand\rotatebox[2]{#2}%
  \@ifundefined{ifGPcolor}{%
    \newif\ifGPcolor
    \GPcolortrue
  }{}%
  \@ifundefined{ifGPblacktext}{%
    \newif\ifGPblacktext
    \GPblacktexttrue
  }{}%
  \let\gplgaddtomacro\g@addto@macro
  \gdef\gplbacktext{}%
  \gdef\gplfronttext{}%
  \makeatother
  \ifGPblacktext
    \def\colorrgb#1{}%
    \def\colorgray#1{}%
  \else
    \ifGPcolor
      \def\colorrgb#1{\color[rgb]{#1}}%
      \def\colorgray#1{\color[gray]{#1}}%
      \expandafter\def\csname LTw\endcsname{\color{white}}%
      \expandafter\def\csname LTb\endcsname{\color{black}}%
      \expandafter\def\csname LTa\endcsname{\color{black}}%
      \expandafter\def\csname LT0\endcsname{\color[rgb]{1,0,0}}%
      \expandafter\def\csname LT1\endcsname{\color[rgb]{0,1,0}}%
      \expandafter\def\csname LT2\endcsname{\color[rgb]{0,0,1}}%
      \expandafter\def\csname LT3\endcsname{\color[rgb]{1,0,1}}%
      \expandafter\def\csname LT4\endcsname{\color[rgb]{0,1,1}}%
      \expandafter\def\csname LT5\endcsname{\color[rgb]{1,1,0}}%
      \expandafter\def\csname LT6\endcsname{\color[rgb]{0,0,0}}%
      \expandafter\def\csname LT7\endcsname{\color[rgb]{1,0.3,0}}%
      \expandafter\def\csname LT8\endcsname{\color[rgb]{0.5,0.5,0.5}}%
    \else
      \def\colorrgb#1{\color{black}}%
      \def\colorgray#1{\color[gray]{#1}}%
      \expandafter\def\csname LTw\endcsname{\color{white}}%
      \expandafter\def\csname LTb\endcsname{\color{black}}%
      \expandafter\def\csname LTa\endcsname{\color{black}}%
      \expandafter\def\csname LT0\endcsname{\color{black}}%
      \expandafter\def\csname LT1\endcsname{\color{black}}%
      \expandafter\def\csname LT2\endcsname{\color{black}}%
      \expandafter\def\csname LT3\endcsname{\color{black}}%
      \expandafter\def\csname LT4\endcsname{\color{black}}%
      \expandafter\def\csname LT5\endcsname{\color{black}}%
      \expandafter\def\csname LT6\endcsname{\color{black}}%
      \expandafter\def\csname LT7\endcsname{\color{black}}%
      \expandafter\def\csname LT8\endcsname{\color{black}}%
    \fi
  \fi
    \setlength{\unitlength}{0.0500bp}%
    \ifx\gptboxheight\undefined%
      \newlength{\gptboxheight}%
      \newlength{\gptboxwidth}%
      \newsavebox{\gptboxtext}%
    \fi%
    \setlength{\fboxrule}{0.5pt}%
    \setlength{\fboxsep}{1pt}%
\begin{picture}(10700.00,7360.00)%
    \gplgaddtomacro\gplbacktext{%
      \csname LTb\endcsname%
      \put(917,4478){\makebox(0,0)[r]{\strut{}$0$}}%
      \csname LTb\endcsname%
      \put(917,4768){\makebox(0,0)[r]{\strut{}$1$}}%
      \csname LTb\endcsname%
      \put(917,5059){\makebox(0,0)[r]{\strut{}$2$}}%
      \csname LTb\endcsname%
      \put(917,5349){\makebox(0,0)[r]{\strut{}$3$}}%
      \csname LTb\endcsname%
      \put(917,5640){\makebox(0,0)[r]{\strut{}$4$}}%
      \csname LTb\endcsname%
      \put(917,5930){\makebox(0,0)[r]{\strut{}$5$}}%
      \csname LTb\endcsname%
      \put(917,6220){\makebox(0,0)[r]{\strut{}$6$}}%
      \csname LTb\endcsname%
      \put(917,6511){\makebox(0,0)[r]{\strut{}$7$}}%
      \csname LTb\endcsname%
      \put(917,6801){\makebox(0,0)[r]{\strut{}$8$}}%
      \csname LTb\endcsname%
      \put(1122,4199){\makebox(0,0){\strut{}$2^{8}$}}%
      \csname LTb\endcsname%
      \put(1841,4199){\makebox(0,0){\strut{}$2^{10}$}}%
      \csname LTb\endcsname%
      \put(2561,4199){\makebox(0,0){\strut{}$2^{12}$}}%
      \csname LTb\endcsname%
      \put(3281,4199){\makebox(0,0){\strut{}$2^{14}$}}%
      \csname LTb\endcsname%
      \put(4000,4199){\makebox(0,0){\strut{}$2^{16}$}}%
      \csname LTb\endcsname%
      \put(4720,4199){\makebox(0,0){\strut{}$2^{18}$}}%
    }%
    \gplgaddtomacro\gplfronttext{%
      \csname LTb\endcsname%
      \put(518,5639){\rotatebox{-270}{\makebox(0,0){\strut{}time in sec.}}}%
      \csname LTb\endcsname%
      \put(3084,3920){\makebox(0,0){\strut{}cores}}%
      \csname LTb\endcsname%
      \put(3084,7080){\makebox(0,0){\strut{}$31.1 \cdot 10^3$ cells/core}}%
    }%
    \gplgaddtomacro\gplbacktext{%
      \csname LTb\endcsname%
      \put(6160,4478){\makebox(0,0)[r]{\strut{}$0$}}%
      \csname LTb\endcsname%
      \put(6160,4768){\makebox(0,0)[r]{\strut{}$1$}}%
      \csname LTb\endcsname%
      \put(6160,5059){\makebox(0,0)[r]{\strut{}$2$}}%
      \csname LTb\endcsname%
      \put(6160,5349){\makebox(0,0)[r]{\strut{}$3$}}%
      \csname LTb\endcsname%
      \put(6160,5640){\makebox(0,0)[r]{\strut{}$4$}}%
      \csname LTb\endcsname%
      \put(6160,5930){\makebox(0,0)[r]{\strut{}$5$}}%
      \csname LTb\endcsname%
      \put(6160,6220){\makebox(0,0)[r]{\strut{}$6$}}%
      \csname LTb\endcsname%
      \put(6160,6511){\makebox(0,0)[r]{\strut{}$7$}}%
      \csname LTb\endcsname%
      \put(6160,6801){\makebox(0,0)[r]{\strut{}$8$}}%
      \csname LTb\endcsname%
      \put(6365,4199){\makebox(0,0){\strut{}$2^{8}$}}%
      \csname LTb\endcsname%
      \put(7084,4199){\makebox(0,0){\strut{}$2^{10}$}}%
      \csname LTb\endcsname%
      \put(7804,4199){\makebox(0,0){\strut{}$2^{12}$}}%
      \csname LTb\endcsname%
      \put(8524,4199){\makebox(0,0){\strut{}$2^{14}$}}%
      \csname LTb\endcsname%
      \put(9243,4199){\makebox(0,0){\strut{}$2^{16}$}}%
      \csname LTb\endcsname%
      \put(9963,4199){\makebox(0,0){\strut{}$2^{18}$}}%
    }%
    \gplgaddtomacro\gplfronttext{%
      \csname LTb\endcsname%
      \put(5761,5639){\rotatebox{-270}{\makebox(0,0){\strut{}time in sec.}}}%
      \csname LTb\endcsname%
      \put(8327,3920){\makebox(0,0){\strut{}cores}}%
      \csname LTb\endcsname%
      \put(8327,7080){\makebox(0,0){\strut{}$127 \cdot 10^3$ cells/core}}%
    }%
    \gplgaddtomacro\gplbacktext{%
      \csname LTb\endcsname%
      \put(917,688){\makebox(0,0)[r]{\strut{}$0$}}%
      \csname LTb\endcsname%
      \put(917,1075){\makebox(0,0)[r]{\strut{}$2$}}%
      \csname LTb\endcsname%
      \put(917,1462){\makebox(0,0)[r]{\strut{}$4$}}%
      \csname LTb\endcsname%
      \put(917,1850){\makebox(0,0)[r]{\strut{}$6$}}%
      \csname LTb\endcsname%
      \put(917,2237){\makebox(0,0)[r]{\strut{}$8$}}%
      \csname LTb\endcsname%
      \put(917,2624){\makebox(0,0)[r]{\strut{}$10$}}%
      \csname LTb\endcsname%
      \put(917,3011){\makebox(0,0)[r]{\strut{}$12$}}%
      \csname LTb\endcsname%
      \put(1122,409){\makebox(0,0){\strut{}$2^{8}$}}%
      \csname LTb\endcsname%
      \put(1841,409){\makebox(0,0){\strut{}$2^{10}$}}%
      \csname LTb\endcsname%
      \put(2561,409){\makebox(0,0){\strut{}$2^{12}$}}%
      \csname LTb\endcsname%
      \put(3281,409){\makebox(0,0){\strut{}$2^{14}$}}%
      \csname LTb\endcsname%
      \put(4000,409){\makebox(0,0){\strut{}$2^{16}$}}%
      \csname LTb\endcsname%
      \put(4720,409){\makebox(0,0){\strut{}$2^{18}$}}%
      \csname LTb\endcsname%
      \put(7382,1251){\makebox(0,0)[l]{\strut{}(\gls{sfc} -- Morton order)}}%
    }%
    \gplgaddtomacro\gplfronttext{%
      \csname LTb\endcsname%
      \put(518,1849){\rotatebox{-270}{\makebox(0,0){\strut{}time in sec.}}}%
      \csname LTb\endcsname%
      \put(3084,130){\makebox(0,0){\strut{}cores}}%
      \csname LTb\endcsname%
      \put(3084,3290){\makebox(0,0){\strut{}$429 \cdot 10^3$ cells/core}}%
      \csname LTb\endcsname%
      \put(7385,2193){\makebox(0,0)[l]{\strut{}entire \gls{amr} cycle}}%
      \csname LTb\endcsname%
      \put(7385,1870){\makebox(0,0)[l]{\strut{}data migration}}%
      \csname LTb\endcsname%
      \put(7385,1547){\makebox(0,0)[l]{\strut{}dynamic load balancing}}%
    }%
    \gplbacktext
    \put(0,0){\includegraphics{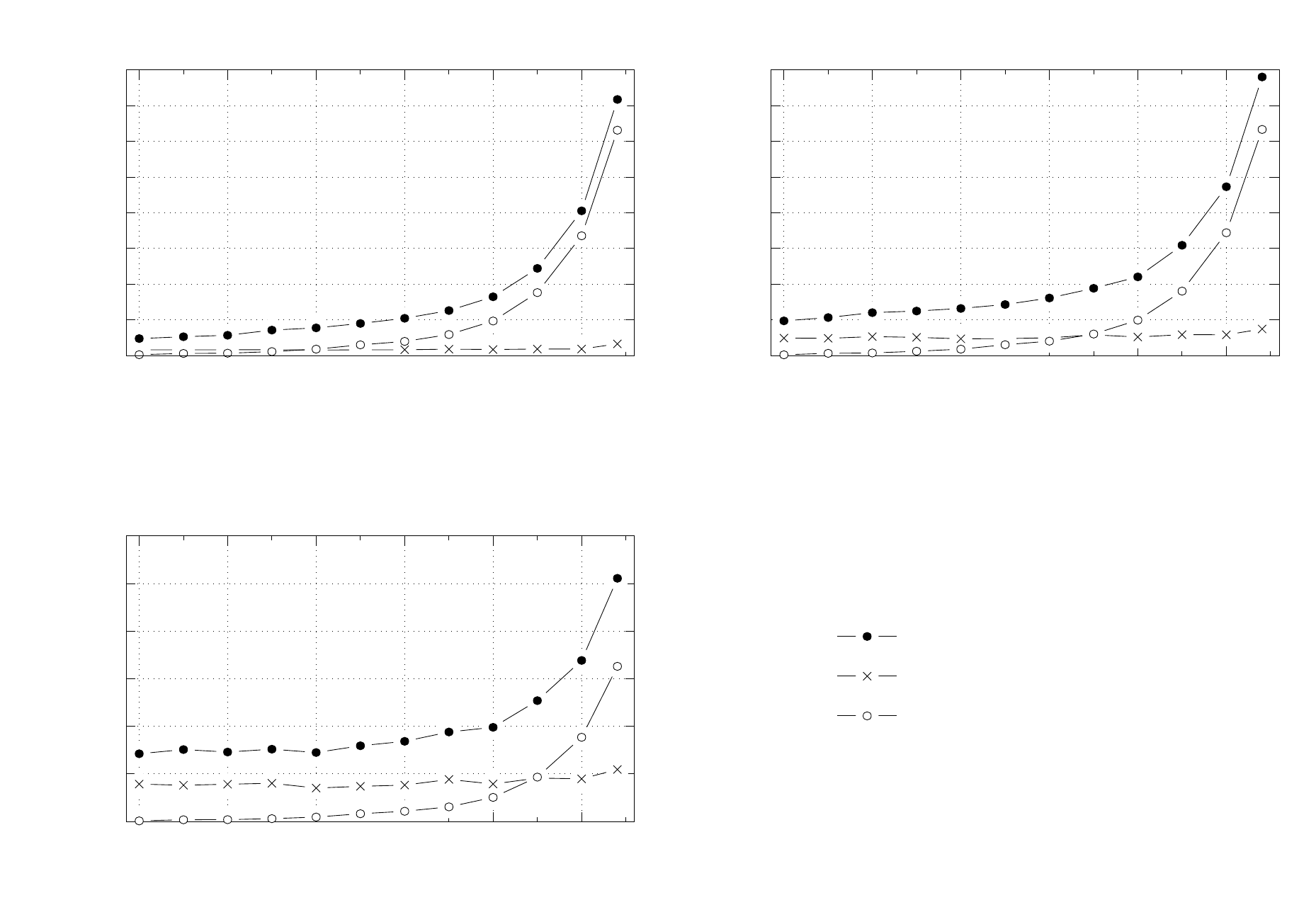}}%
    \gplfronttext
  \end{picture}%
\endgroup

%% file: jdifit_.tex
\begingroup
  \makeatletter
  \providecommand\color[2][]{%
    \GenericError{(gnuplot) \space\space\space\@spaces}{%
      Package color not loaded in conjunction with
      terminal option `colourtext'%
    }{See the gnuplot documentation for explanation.%
    }{Either use 'blacktext' in gnuplot or load the package
      color.sty in LaTeX.}%
    \renewcommand\color[2][]{}%
  }%
  \providecommand\includegraphics[2][]{%
    \GenericError{(gnuplot) \space\space\space\@spaces}{%
      Package graphicx or graphics not loaded%
    }{See the gnuplot documentation for explanation.%
    }{The gnuplot epslatex terminal needs graphicx.sty or graphics.sty.}%
    \renewcommand\includegraphics[2][]{}%
  }%
  \providecommand\rotatebox[2]{#2}%
  \@ifundefined{ifGPcolor}{%
    \newif\ifGPcolor
    \GPcolortrue
  }{}%
  \@ifundefined{ifGPblacktext}{%
    \newif\ifGPblacktext
    \GPblacktexttrue
  }{}%
  \let\gplgaddtomacro\g@addto@macro
  \gdef\gplbacktext{}%
  \gdef\gplfronttext{}%
  \makeatother
  \ifGPblacktext
    \def\colorrgb#1{}%
    \def\colorgray#1{}%
  \else
    \ifGPcolor
      \def\colorrgb#1{\color[rgb]{#1}}%
      \def\colorgray#1{\color[gray]{#1}}%
      \expandafter\def\csname LTw\endcsname{\color{white}}%
      \expandafter\def\csname LTb\endcsname{\color{black}}%
      \expandafter\def\csname LTa\endcsname{\color{black}}%
      \expandafter\def\csname LT0\endcsname{\color[rgb]{1,0,0}}%
      \expandafter\def\csname LT1\endcsname{\color[rgb]{0,1,0}}%
      \expandafter\def\csname LT2\endcsname{\color[rgb]{0,0,1}}%
      \expandafter\def\csname LT3\endcsname{\color[rgb]{1,0,1}}%
      \expandafter\def\csname LT4\endcsname{\color[rgb]{0,1,1}}%
      \expandafter\def\csname LT5\endcsname{\color[rgb]{1,1,0}}%
      \expandafter\def\csname LT6\endcsname{\color[rgb]{0,0,0}}%
      \expandafter\def\csname LT7\endcsname{\color[rgb]{1,0.3,0}}%
      \expandafter\def\csname LT8\endcsname{\color[rgb]{0.5,0.5,0.5}}%
    \else
      \def\colorrgb#1{\color{black}}%
      \def\colorgray#1{\color[gray]{#1}}%
      \expandafter\def\csname LTw\endcsname{\color{white}}%
      \expandafter\def\csname LTb\endcsname{\color{black}}%
      \expandafter\def\csname LTa\endcsname{\color{black}}%
      \expandafter\def\csname LT0\endcsname{\color{black}}%
      \expandafter\def\csname LT1\endcsname{\color{black}}%
      \expandafter\def\csname LT2\endcsname{\color{black}}%
      \expandafter\def\csname LT3\endcsname{\color{black}}%
      \expandafter\def\csname LT4\endcsname{\color{black}}%
      \expandafter\def\csname LT5\endcsname{\color{black}}%
      \expandafter\def\csname LT6\endcsname{\color{black}}%
      \expandafter\def\csname LT7\endcsname{\color{black}}%
      \expandafter\def\csname LT8\endcsname{\color{black}}%
    \fi
  \fi
  \setlength{\unitlength}{0.0500bp}%
  \begin{picture}(9060.00,3220.00)%
    \gplgaddtomacro\gplbacktext{%
      \csname LTb\endcsname%
      \put(645,688){\makebox(0,0)[r]{\strut{} 0}}%
      \csname LTb\endcsname%
      \put(645,1076){\makebox(0,0)[r]{\strut{} 2}}%
      \csname LTb\endcsname%
      \put(645,1464){\makebox(0,0)[r]{\strut{} 4}}%
      \csname LTb\endcsname%
      \put(645,1852){\makebox(0,0)[r]{\strut{} 6}}%
      \csname LTb\endcsname%
      \put(645,2239){\makebox(0,0)[r]{\strut{} 8}}%
      \csname LTb\endcsname%
      \put(645,2627){\makebox(0,0)[r]{\strut{} 10}}%
      \csname LTb\endcsname%
      \put(645,3015){\makebox(0,0)[r]{\strut{} 12}}%
      \csname LTb\endcsname%
      \put(1239,409){\makebox(0,0){\strut{}$2^{8}$}}%
      \csname LTb\endcsname%
      \put(1732,409){\makebox(0,0){\strut{}$2^{9}$}}%
      \csname LTb\endcsname%
      \put(2224,409){\makebox(0,0){\strut{}$2^{10}$}}%
      \csname LTb\endcsname%
      \put(2717,409){\makebox(0,0){\strut{}$2^{11}$}}%
      \csname LTb\endcsname%
      \put(3209,409){\makebox(0,0){\strut{}$2^{12}$}}%
      \csname LTb\endcsname%
      \put(3702,409){\makebox(0,0){\strut{}$2^{13}$}}%
      \csname LTb\endcsname%
      \put(4194,409){\makebox(0,0){\strut{}$2^{14}$}}%
      \csname LTb\endcsname%
      \put(4687,409){\makebox(0,0){\strut{}$2^{15}$}}%
      \csname LTb\endcsname%
      \put(5179,409){\makebox(0,0){\strut{}$2^{16}$}}%
      \csname LTb\endcsname%
      \put(5672,409){\makebox(0,0){\strut{}$2^{17}$}}%
      \csname LTb\endcsname%
      \put(6164,409){\makebox(0,0){\strut{}$2^{18}$}}%
      \csname LTb\endcsname%
      \put(6657,520){\rotatebox{-45}{\makebox(0,0)[l]{\strut{}\small 458752}}}%
      \csname LTb\endcsname%
      \put(144,1851){\rotatebox{-270}{\makebox(0,0){\strut{}iterations}}}%
      \csname LTb\endcsname%
      \put(3948,130){\makebox(0,0){\strut{}cores}}%
    }%
    \gplgaddtomacro\gplfronttext{%
      \csname LTb\endcsname%
      \put(8039,2013){\makebox(0,0)[l]{\strut{}push}}%
      \csname LTb\endcsname%
      \put(8039,1690){\makebox(0,0)[l]{\strut{}push/pull}}%
    }%
    \gplbacktext
    \put(0,0){\includegraphics{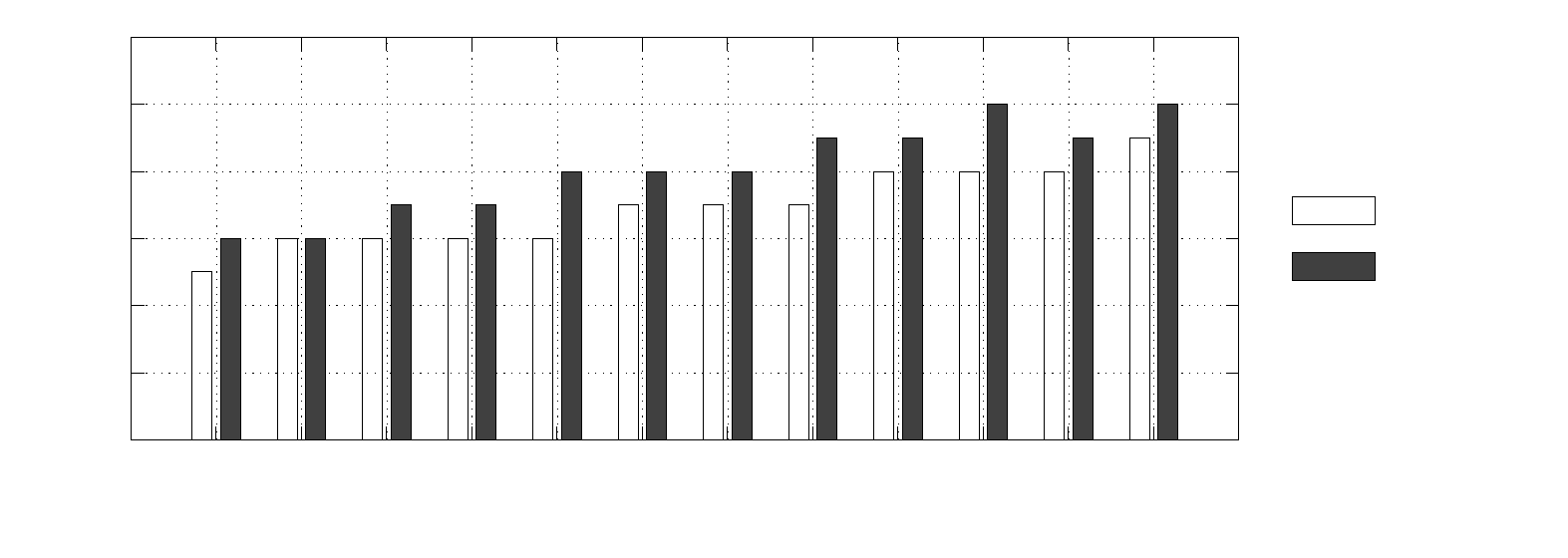}}%
    \gplfronttext
  \end{picture}%
\endgroup

%% file: jdiff_.tex
\begingroup
  \makeatletter
  \providecommand\color[2][]{%
    \GenericError{(gnuplot) \space\space\space\@spaces}{%
      Package color not loaded in conjunction with
      terminal option `colourtext'%
    }{See the gnuplot documentation for explanation.%
    }{Either use 'blacktext' in gnuplot or load the package
      color.sty in LaTeX.}%
    \renewcommand\color[2][]{}%
  }%
  \providecommand\includegraphics[2][]{%
    \GenericError{(gnuplot) \space\space\space\@spaces}{%
      Package graphicx or graphics not loaded%
    }{See the gnuplot documentation for explanation.%
    }{The gnuplot epslatex terminal needs graphicx.sty or graphics.sty.}%
    \renewcommand\includegraphics[2][]{}%
  }%
  \providecommand\rotatebox[2]{#2}%
  \@ifundefined{ifGPcolor}{%
    \newif\ifGPcolor
    \GPcolortrue
  }{}%
  \@ifundefined{ifGPblacktext}{%
    \newif\ifGPblacktext
    \GPblacktexttrue
  }{}%
  \let\gplgaddtomacro\g@addto@macro
  \gdef\gplbacktext{}%
  \gdef\gplfronttext{}%
  \makeatother
  \ifGPblacktext
    \def\colorrgb#1{}%
    \def\colorgray#1{}%
  \else
    \ifGPcolor
      \def\colorrgb#1{\color[rgb]{#1}}%
      \def\colorgray#1{\color[gray]{#1}}%
      \expandafter\def\csname LTw\endcsname{\color{white}}%
      \expandafter\def\csname LTb\endcsname{\color{black}}%
      \expandafter\def\csname LTa\endcsname{\color{black}}%
      \expandafter\def\csname LT0\endcsname{\color[rgb]{1,0,0}}%
      \expandafter\def\csname LT1\endcsname{\color[rgb]{0,1,0}}%
      \expandafter\def\csname LT2\endcsname{\color[rgb]{0,0,1}}%
      \expandafter\def\csname LT3\endcsname{\color[rgb]{1,0,1}}%
      \expandafter\def\csname LT4\endcsname{\color[rgb]{0,1,1}}%
      \expandafter\def\csname LT5\endcsname{\color[rgb]{1,1,0}}%
      \expandafter\def\csname LT6\endcsname{\color[rgb]{0,0,0}}%
      \expandafter\def\csname LT7\endcsname{\color[rgb]{1,0.3,0}}%
      \expandafter\def\csname LT8\endcsname{\color[rgb]{0.5,0.5,0.5}}%
    \else
      \def\colorrgb#1{\color{black}}%
      \def\colorgray#1{\color[gray]{#1}}%
      \expandafter\def\csname LTw\endcsname{\color{white}}%
      \expandafter\def\csname LTb\endcsname{\color{black}}%
      \expandafter\def\csname LTa\endcsname{\color{black}}%
      \expandafter\def\csname LT0\endcsname{\color{black}}%
      \expandafter\def\csname LT1\endcsname{\color{black}}%
      \expandafter\def\csname LT2\endcsname{\color{black}}%
      \expandafter\def\csname LT3\endcsname{\color{black}}%
      \expandafter\def\csname LT4\endcsname{\color{black}}%
      \expandafter\def\csname LT5\endcsname{\color{black}}%
      \expandafter\def\csname LT6\endcsname{\color{black}}%
      \expandafter\def\csname LT7\endcsname{\color{black}}%
      \expandafter\def\csname LT8\endcsname{\color{black}}%
    \fi
  \fi
    \setlength{\unitlength}{0.0500bp}%
    \ifx\gptboxheight\undefined%
      \newlength{\gptboxheight}%
      \newlength{\gptboxwidth}%
      \newsavebox{\gptboxtext}%
    \fi%
    \setlength{\fboxrule}{0.5pt}%
    \setlength{\fboxsep}{1pt}%
\begin{picture}(10700.00,7360.00)%
    \gplgaddtomacro\gplbacktext{%
      \csname LTb\endcsname%
      \put(1054,4478){\makebox(0,0)[r]{\strut{}$0$}}%
      \csname LTb\endcsname%
      \put(1054,5059){\makebox(0,0)[r]{\strut{}$0.5$}}%
      \csname LTb\endcsname%
      \put(1054,5640){\makebox(0,0)[r]{\strut{}$1$}}%
      \csname LTb\endcsname%
      \put(1054,6220){\makebox(0,0)[r]{\strut{}$1.5$}}%
      \csname LTb\endcsname%
      \put(1054,6801){\makebox(0,0)[r]{\strut{}$2$}}%
      \csname LTb\endcsname%
      \put(1255,4199){\makebox(0,0){\strut{}$2^{8}$}}%
      \csname LTb\endcsname%
      \put(1951,4199){\makebox(0,0){\strut{}$2^{10}$}}%
      \csname LTb\endcsname%
      \put(2647,4199){\makebox(0,0){\strut{}$2^{12}$}}%
      \csname LTb\endcsname%
      \put(3343,4199){\makebox(0,0){\strut{}$2^{14}$}}%
      \csname LTb\endcsname%
      \put(4039,4199){\makebox(0,0){\strut{}$2^{16}$}}%
      \csname LTb\endcsname%
      \put(4734,4199){\makebox(0,0){\strut{}$2^{18}$}}%
    }%
    \gplgaddtomacro\gplfronttext{%
      \csname LTb\endcsname%
      \put(518,5639){\rotatebox{-270}{\makebox(0,0){\strut{}time in sec.}}}%
      \csname LTb\endcsname%
      \put(3153,3920){\makebox(0,0){\strut{}cores}}%
      \csname LTb\endcsname%
      \put(3153,7080){\makebox(0,0){\strut{}$31.1 \cdot 10^3$ cells/core}}%
    }%
    \gplgaddtomacro\gplbacktext{%
      \csname LTb\endcsname%
      \put(6297,4478){\makebox(0,0)[r]{\strut{}$0$}}%
      \csname LTb\endcsname%
      \put(6297,4943){\makebox(0,0)[r]{\strut{}$0.5$}}%
      \csname LTb\endcsname%
      \put(6297,5407){\makebox(0,0)[r]{\strut{}$1$}}%
      \csname LTb\endcsname%
      \put(6297,5872){\makebox(0,0)[r]{\strut{}$1.5$}}%
      \csname LTb\endcsname%
      \put(6297,6336){\makebox(0,0)[r]{\strut{}$2$}}%
      \csname LTb\endcsname%
      \put(6297,6801){\makebox(0,0)[r]{\strut{}$2.5$}}%
      \csname LTb\endcsname%
      \put(6498,4199){\makebox(0,0){\strut{}$2^{8}$}}%
      \csname LTb\endcsname%
      \put(7194,4199){\makebox(0,0){\strut{}$2^{10}$}}%
      \csname LTb\endcsname%
      \put(7890,4199){\makebox(0,0){\strut{}$2^{12}$}}%
      \csname LTb\endcsname%
      \put(8586,4199){\makebox(0,0){\strut{}$2^{14}$}}%
      \csname LTb\endcsname%
      \put(9282,4199){\makebox(0,0){\strut{}$2^{16}$}}%
      \csname LTb\endcsname%
      \put(9977,4199){\makebox(0,0){\strut{}$2^{18}$}}%
    }%
    \gplgaddtomacro\gplfronttext{%
      \csname LTb\endcsname%
      \put(5761,5639){\rotatebox{-270}{\makebox(0,0){\strut{}time in sec.}}}%
      \csname LTb\endcsname%
      \put(8396,3920){\makebox(0,0){\strut{}cores}}%
      \csname LTb\endcsname%
      \put(8396,7080){\makebox(0,0){\strut{}$127 \cdot 10^3$ cells/core}}%
    }%
    \gplgaddtomacro\gplbacktext{%
      \csname LTb\endcsname%
      \put(1070,688){\makebox(0,0)[r]{\strut{}$0$}}%
      \csname LTb\endcsname%
      \put(1070,1153){\makebox(0,0)[r]{\strut{}$1$}}%
      \csname LTb\endcsname%
      \put(1070,1617){\makebox(0,0)[r]{\strut{}$2$}}%
      \csname LTb\endcsname%
      \put(1070,2082){\makebox(0,0)[r]{\strut{}$3$}}%
      \csname LTb\endcsname%
      \put(1070,2546){\makebox(0,0)[r]{\strut{}$4$}}%
      \csname LTb\endcsname%
      \put(1070,3011){\makebox(0,0)[r]{\strut{}$5$}}%
      \csname LTb\endcsname%
      \put(1271,409){\makebox(0,0){\strut{}$2^{8}$}}%
      \csname LTb\endcsname%
      \put(1964,409){\makebox(0,0){\strut{}$2^{10}$}}%
      \csname LTb\endcsname%
      \put(2657,409){\makebox(0,0){\strut{}$2^{12}$}}%
      \csname LTb\endcsname%
      \put(3350,409){\makebox(0,0){\strut{}$2^{14}$}}%
      \csname LTb\endcsname%
      \put(4043,409){\makebox(0,0){\strut{}$2^{16}$}}%
      \csname LTb\endcsname%
      \put(4736,409){\makebox(0,0){\strut{}$2^{18}$}}%
      \csname LTb\endcsname%
      \put(7382,1251){\makebox(0,0)[l]{\strut{}(diffusion -- push/pull)}}%
    }%
    \gplgaddtomacro\gplfronttext{%
      \csname LTb\endcsname%
      \put(518,1849){\rotatebox{-270}{\makebox(0,0){\strut{}time in sec.}}}%
      \csname LTb\endcsname%
      \put(3161,130){\makebox(0,0){\strut{}cores}}%
      \csname LTb\endcsname%
      \put(3161,3290){\makebox(0,0){\strut{}$429 \cdot 10^3$ cells/core}}%
      \csname LTb\endcsname%
      \put(7385,2193){\makebox(0,0)[l]{\strut{}entire \gls{amr} cycle}}%
      \csname LTb\endcsname%
      \put(7385,1870){\makebox(0,0)[l]{\strut{}data migration}}%
      \csname LTb\endcsname%
      \put(7385,1547){\makebox(0,0)[l]{\strut{}dynamic load balancing}}%
    }%
    \gplbacktext
    \put(0,0){\includegraphics{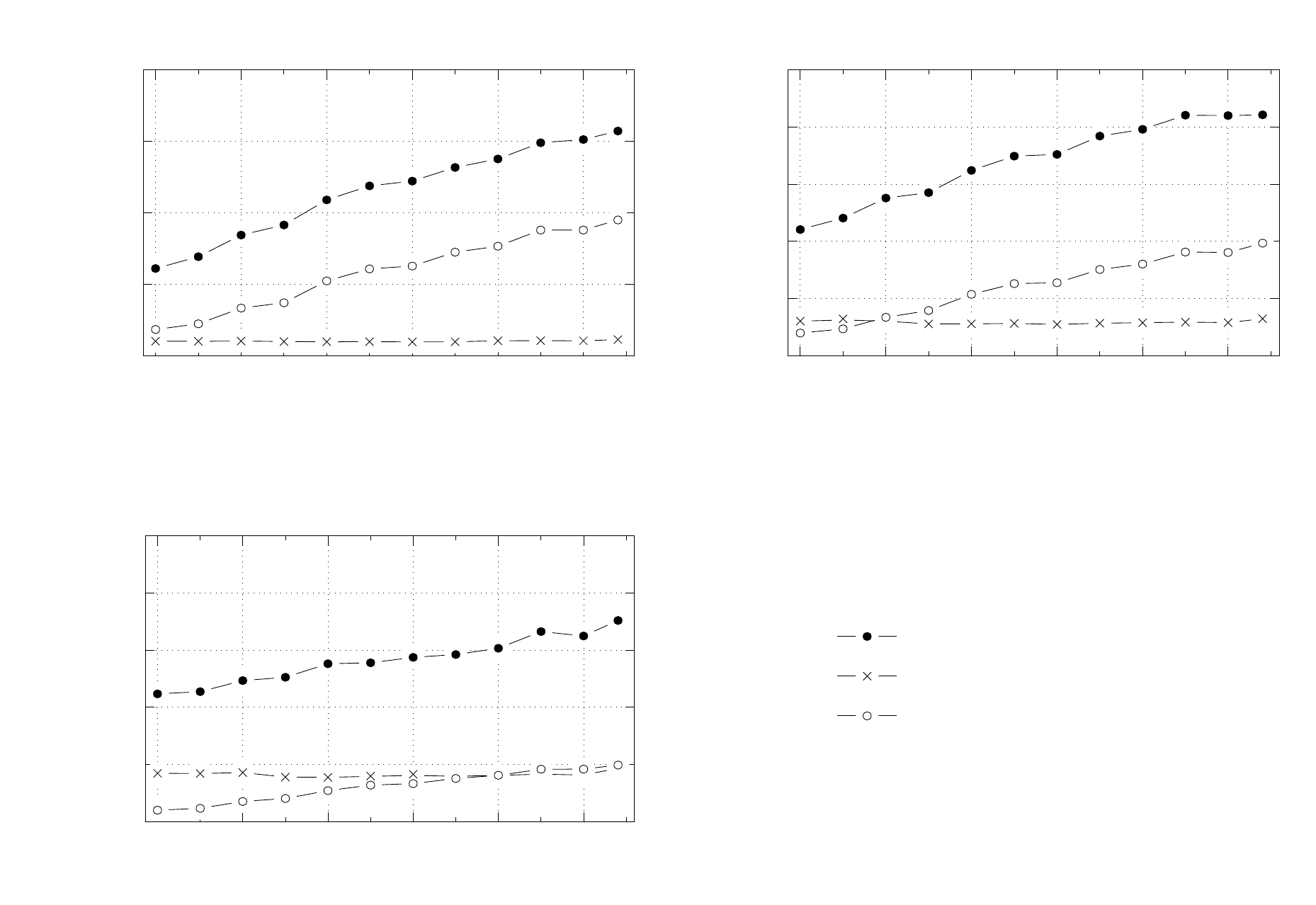}}%
    \gplfronttext
  \end{picture}%
\endgroup

%% file: jamr_.tex
\begingroup
  \makeatletter
  \providecommand\color[2][]{%
    \GenericError{(gnuplot) \space\space\space\@spaces}{%
      Package color not loaded in conjunction with
      terminal option `colourtext'%
    }{See the gnuplot documentation for explanation.%
    }{Either use 'blacktext' in gnuplot or load the package
      color.sty in LaTeX.}%
    \renewcommand\color[2][]{}%
  }%
  \providecommand\includegraphics[2][]{%
    \GenericError{(gnuplot) \space\space\space\@spaces}{%
      Package graphicx or graphics not loaded%
    }{See the gnuplot documentation for explanation.%
    }{The gnuplot epslatex terminal needs graphicx.sty or graphics.sty.}%
    \renewcommand\includegraphics[2][]{}%
  }%
  \providecommand\rotatebox[2]{#2}%
  \@ifundefined{ifGPcolor}{%
    \newif\ifGPcolor
    \GPcolortrue
  }{}%
  \@ifundefined{ifGPblacktext}{%
    \newif\ifGPblacktext
    \GPblacktexttrue
  }{}%
  \let\gplgaddtomacro\g@addto@macro
  \gdef\gplbacktext{}%
  \gdef\gplfronttext{}%
  \makeatother
  \ifGPblacktext
    \def\colorrgb#1{}%
    \def\colorgray#1{}%
  \else
    \ifGPcolor
      \def\colorrgb#1{\color[rgb]{#1}}%
      \def\colorgray#1{\color[gray]{#1}}%
      \expandafter\def\csname LTw\endcsname{\color{white}}%
      \expandafter\def\csname LTb\endcsname{\color{black}}%
      \expandafter\def\csname LTa\endcsname{\color{black}}%
      \expandafter\def\csname LT0\endcsname{\color[rgb]{1,0,0}}%
      \expandafter\def\csname LT1\endcsname{\color[rgb]{0,1,0}}%
      \expandafter\def\csname LT2\endcsname{\color[rgb]{0,0,1}}%
      \expandafter\def\csname LT3\endcsname{\color[rgb]{1,0,1}}%
      \expandafter\def\csname LT4\endcsname{\color[rgb]{0,1,1}}%
      \expandafter\def\csname LT5\endcsname{\color[rgb]{1,1,0}}%
      \expandafter\def\csname LT6\endcsname{\color[rgb]{0,0,0}}%
      \expandafter\def\csname LT7\endcsname{\color[rgb]{1,0.3,0}}%
      \expandafter\def\csname LT8\endcsname{\color[rgb]{0.5,0.5,0.5}}%
    \else
      \def\colorrgb#1{\color{black}}%
      \def\colorgray#1{\color[gray]{#1}}%
      \expandafter\def\csname LTw\endcsname{\color{white}}%
      \expandafter\def\csname LTb\endcsname{\color{black}}%
      \expandafter\def\csname LTa\endcsname{\color{black}}%
      \expandafter\def\csname LT0\endcsname{\color{black}}%
      \expandafter\def\csname LT1\endcsname{\color{black}}%
      \expandafter\def\csname LT2\endcsname{\color{black}}%
      \expandafter\def\csname LT3\endcsname{\color{black}}%
      \expandafter\def\csname LT4\endcsname{\color{black}}%
      \expandafter\def\csname LT5\endcsname{\color{black}}%
      \expandafter\def\csname LT6\endcsname{\color{black}}%
      \expandafter\def\csname LT7\endcsname{\color{black}}%
      \expandafter\def\csname LT8\endcsname{\color{black}}%
    \fi
  \fi
    \setlength{\unitlength}{0.0500bp}%
    \ifx\gptboxheight\undefined%
      \newlength{\gptboxheight}%
      \newlength{\gptboxwidth}%
      \newsavebox{\gptboxtext}%
    \fi%
    \setlength{\fboxrule}{0.5pt}%
    \setlength{\fboxsep}{1pt}%
\begin{picture}(10700.00,7360.00)%
    \gplgaddtomacro\gplbacktext{%
      \csname LTb\endcsname%
      \put(968,4478){\makebox(0,0)[r]{\strut{}$0$}}%
      \csname LTb\endcsname%
      \put(968,4943){\makebox(0,0)[r]{\strut{}$4$}}%
      \csname LTb\endcsname%
      \put(968,5407){\makebox(0,0)[r]{\strut{}$8$}}%
      \csname LTb\endcsname%
      \put(968,5872){\makebox(0,0)[r]{\strut{}$12$}}%
      \csname LTb\endcsname%
      \put(968,6336){\makebox(0,0)[r]{\strut{}$16$}}%
      \csname LTb\endcsname%
      \put(968,6801){\makebox(0,0)[r]{\strut{}$20$}}%
      \csname LTb\endcsname%
      \put(1172,4199){\makebox(0,0){\strut{}$2^{8}$}}%
      \csname LTb\endcsname%
      \put(1882,4199){\makebox(0,0){\strut{}$2^{10}$}}%
      \csname LTb\endcsname%
      \put(2593,4199){\makebox(0,0){\strut{}$2^{12}$}}%
      \csname LTb\endcsname%
      \put(3304,4199){\makebox(0,0){\strut{}$2^{14}$}}%
      \csname LTb\endcsname%
      \put(4015,4199){\makebox(0,0){\strut{}$2^{16}$}}%
      \csname LTb\endcsname%
      \put(4725,4199){\makebox(0,0){\strut{}$2^{18}$}}%
    }%
    \gplgaddtomacro\gplfronttext{%
      \csname LTb\endcsname%
      \put(518,5639){\rotatebox{-270}{\makebox(0,0){\strut{}time in sec.}}}%
      \csname LTb\endcsname%
      \put(3110,3920){\makebox(0,0){\strut{}cores}}%
      \csname LTb\endcsname%
      \put(3110,7080){\makebox(0,0){\strut{}$31.1 \cdot 10^3$ cells/core}}%
    }%
    \gplgaddtomacro\gplbacktext{%
      \csname LTb\endcsname%
      \put(6211,4478){\makebox(0,0)[r]{\strut{}$0$}}%
      \csname LTb\endcsname%
      \put(6211,4943){\makebox(0,0)[r]{\strut{}$4$}}%
      \csname LTb\endcsname%
      \put(6211,5407){\makebox(0,0)[r]{\strut{}$8$}}%
      \csname LTb\endcsname%
      \put(6211,5872){\makebox(0,0)[r]{\strut{}$12$}}%
      \csname LTb\endcsname%
      \put(6211,6336){\makebox(0,0)[r]{\strut{}$16$}}%
      \csname LTb\endcsname%
      \put(6211,6801){\makebox(0,0)[r]{\strut{}$20$}}%
      \csname LTb\endcsname%
      \put(6415,4199){\makebox(0,0){\strut{}$2^{8}$}}%
      \csname LTb\endcsname%
      \put(7125,4199){\makebox(0,0){\strut{}$2^{10}$}}%
      \csname LTb\endcsname%
      \put(7836,4199){\makebox(0,0){\strut{}$2^{12}$}}%
      \csname LTb\endcsname%
      \put(8547,4199){\makebox(0,0){\strut{}$2^{14}$}}%
      \csname LTb\endcsname%
      \put(9258,4199){\makebox(0,0){\strut{}$2^{16}$}}%
      \csname LTb\endcsname%
      \put(9968,4199){\makebox(0,0){\strut{}$2^{18}$}}%
    }%
    \gplgaddtomacro\gplfronttext{%
      \csname LTb\endcsname%
      \put(5761,5639){\rotatebox{-270}{\makebox(0,0){\strut{}time in sec.}}}%
      \csname LTb\endcsname%
      \put(8353,3920){\makebox(0,0){\strut{}cores}}%
      \csname LTb\endcsname%
      \put(8353,7080){\makebox(0,0){\strut{}$127 \cdot 10^3$ cells/core}}%
    }%
    \gplgaddtomacro\gplbacktext{%
      \csname LTb\endcsname%
      \put(968,688){\makebox(0,0)[r]{\strut{}$0$}}%
      \csname LTb\endcsname%
      \put(968,1153){\makebox(0,0)[r]{\strut{}$3$}}%
      \csname LTb\endcsname%
      \put(968,1617){\makebox(0,0)[r]{\strut{}$6$}}%
      \csname LTb\endcsname%
      \put(968,2082){\makebox(0,0)[r]{\strut{}$9$}}%
      \csname LTb\endcsname%
      \put(968,2546){\makebox(0,0)[r]{\strut{}$12$}}%
      \csname LTb\endcsname%
      \put(968,3011){\makebox(0,0)[r]{\strut{}$15$}}%
      \csname LTb\endcsname%
      \put(1172,409){\makebox(0,0){\strut{}$2^{8}$}}%
      \csname LTb\endcsname%
      \put(1882,409){\makebox(0,0){\strut{}$2^{10}$}}%
      \csname LTb\endcsname%
      \put(2593,409){\makebox(0,0){\strut{}$2^{12}$}}%
      \csname LTb\endcsname%
      \put(3304,409){\makebox(0,0){\strut{}$2^{14}$}}%
      \csname LTb\endcsname%
      \put(4015,409){\makebox(0,0){\strut{}$2^{16}$}}%
      \csname LTb\endcsname%
      \put(4725,409){\makebox(0,0){\strut{}$2^{18}$}}%
    }%
    \gplgaddtomacro\gplfronttext{%
      \csname LTb\endcsname%
      \put(518,1849){\rotatebox{-270}{\makebox(0,0){\strut{}time in sec.}}}%
      \csname LTb\endcsname%
      \put(3110,130){\makebox(0,0){\strut{}cores}}%
      \csname LTb\endcsname%
      \put(3110,3290){\makebox(0,0){\strut{}$429 \cdot 10^3$ cells/core}}%
      \csname LTb\endcsname%
      \put(6788,1178){\makebox(0,0)[l]{\strut{}\shortstack[l]{diffusion -- push/pull\\(\mpi{}+OpenMP $\rightarrow$ 4 threads/proc.)}}}%
      \csname LTb\endcsname%
      \put(6788,1825){\makebox(0,0)[l]{\strut{}\shortstack[l]{\gls{sfc} -- Morton order\\(\mpi{}+OpenMP $\rightarrow$ 8 threads/proc.)}}}%
      \csname LTb\endcsname%
      \put(6788,2472){\makebox(0,0)[l]{\strut{}\shortstack[l]{\gls{sfc} -- Morton order\\(\mpi{} only)}}}%
    }%
    \gplbacktext
    \put(0,0){\includegraphics{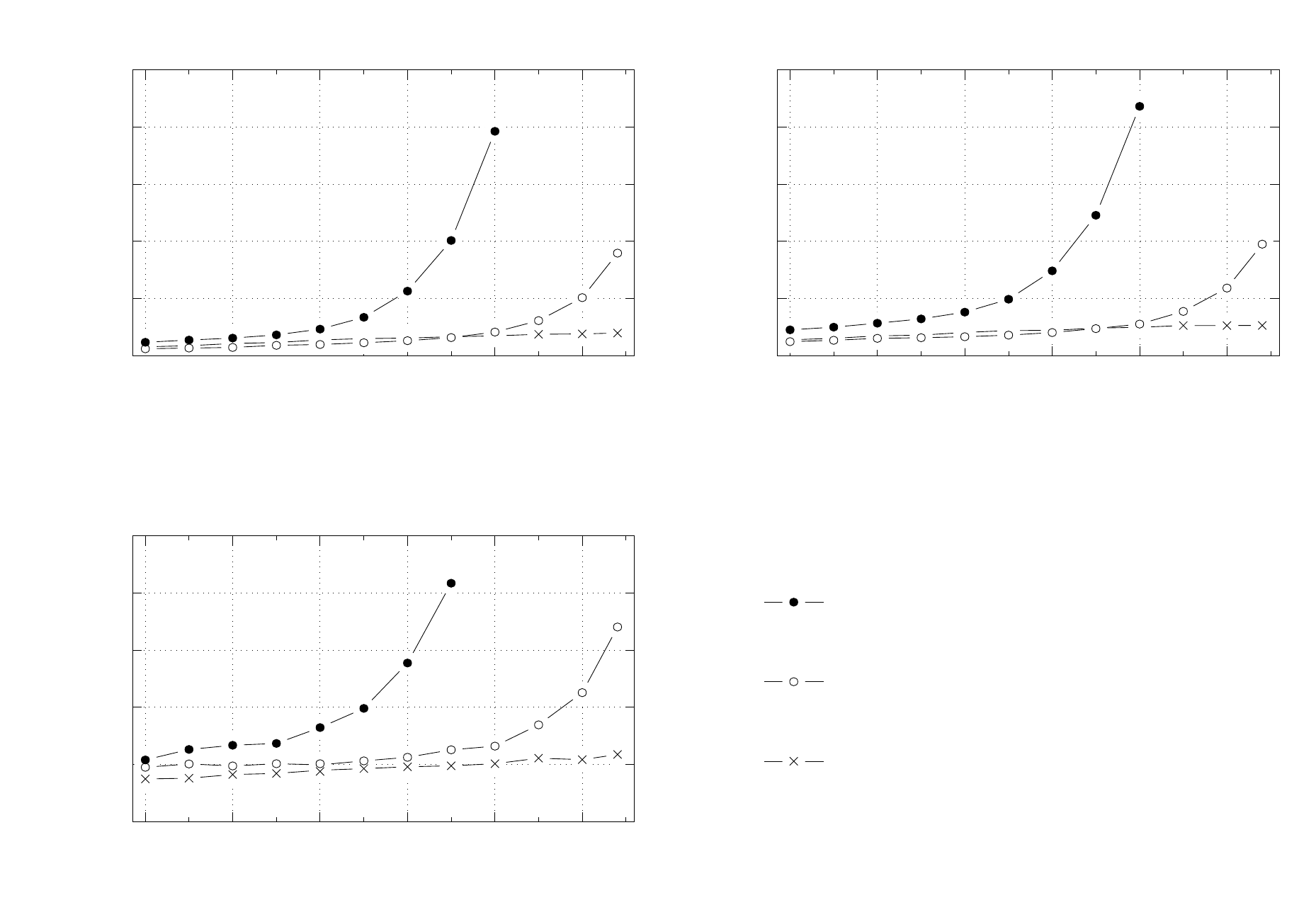}}%
    \gplfronttext
  \end{picture}%
\endgroup

%% file: ssfc_.tex
\begingroup
  \makeatletter
  \providecommand\color[2][]{%
    \GenericError{(gnuplot) \space\space\space\@spaces}{%
      Package color not loaded in conjunction with
      terminal option `colourtext'%
    }{See the gnuplot documentation for explanation.%
    }{Either use 'blacktext' in gnuplot or load the package
      color.sty in LaTeX.}%
    \renewcommand\color[2][]{}%
  }%
  \providecommand\includegraphics[2][]{%
    \GenericError{(gnuplot) \space\space\space\@spaces}{%
      Package graphicx or graphics not loaded%
    }{See the gnuplot documentation for explanation.%
    }{The gnuplot epslatex terminal needs graphicx.sty or graphics.sty.}%
    \renewcommand\includegraphics[2][]{}%
  }%
  \providecommand\rotatebox[2]{#2}%
  \@ifundefined{ifGPcolor}{%
    \newif\ifGPcolor
    \GPcolortrue
  }{}%
  \@ifundefined{ifGPblacktext}{%
    \newif\ifGPblacktext
    \GPblacktexttrue
  }{}%
  \let\gplgaddtomacro\g@addto@macro
  \gdef\gplbacktext{}%
  \gdef\gplfronttext{}%
  \makeatother
  \ifGPblacktext
    \def\colorrgb#1{}%
    \def\colorgray#1{}%
  \else
    \ifGPcolor
      \def\colorrgb#1{\color[rgb]{#1}}%
      \def\colorgray#1{\color[gray]{#1}}%
      \expandafter\def\csname LTw\endcsname{\color{white}}%
      \expandafter\def\csname LTb\endcsname{\color{black}}%
      \expandafter\def\csname LTa\endcsname{\color{black}}%
      \expandafter\def\csname LT0\endcsname{\color[rgb]{1,0,0}}%
      \expandafter\def\csname LT1\endcsname{\color[rgb]{0,1,0}}%
      \expandafter\def\csname LT2\endcsname{\color[rgb]{0,0,1}}%
      \expandafter\def\csname LT3\endcsname{\color[rgb]{1,0,1}}%
      \expandafter\def\csname LT4\endcsname{\color[rgb]{0,1,1}}%
      \expandafter\def\csname LT5\endcsname{\color[rgb]{1,1,0}}%
      \expandafter\def\csname LT6\endcsname{\color[rgb]{0,0,0}}%
      \expandafter\def\csname LT7\endcsname{\color[rgb]{1,0.3,0}}%
      \expandafter\def\csname LT8\endcsname{\color[rgb]{0.5,0.5,0.5}}%
    \else
      \def\colorrgb#1{\color{black}}%
      \def\colorgray#1{\color[gray]{#1}}%
      \expandafter\def\csname LTw\endcsname{\color{white}}%
      \expandafter\def\csname LTb\endcsname{\color{black}}%
      \expandafter\def\csname LTa\endcsname{\color{black}}%
      \expandafter\def\csname LT0\endcsname{\color{black}}%
      \expandafter\def\csname LT1\endcsname{\color{black}}%
      \expandafter\def\csname LT2\endcsname{\color{black}}%
      \expandafter\def\csname LT3\endcsname{\color{black}}%
      \expandafter\def\csname LT4\endcsname{\color{black}}%
      \expandafter\def\csname LT5\endcsname{\color{black}}%
      \expandafter\def\csname LT6\endcsname{\color{black}}%
      \expandafter\def\csname LT7\endcsname{\color{black}}%
      \expandafter\def\csname LT8\endcsname{\color{black}}%
    \fi
  \fi
    \setlength{\unitlength}{0.0500bp}%
    \ifx\gptboxheight\undefined%
      \newlength{\gptboxheight}%
      \newlength{\gptboxwidth}%
      \newsavebox{\gptboxtext}%
    \fi%
    \setlength{\fboxrule}{0.5pt}%
    \setlength{\fboxsep}{1pt}%
\begin{picture}(10700.00,7360.00)%
    \gplgaddtomacro\gplbacktext{%
      \csname LTb\endcsname%
      \put(1019,4478){\makebox(0,0)[r]{\strut{}$0$}}%
      \csname LTb\endcsname%
      \put(1019,4943){\makebox(0,0)[r]{\strut{}$0.1$}}%
      \csname LTb\endcsname%
      \put(1019,5407){\makebox(0,0)[r]{\strut{}$0.2$}}%
      \csname LTb\endcsname%
      \put(1019,5872){\makebox(0,0)[r]{\strut{}$0.3$}}%
      \csname LTb\endcsname%
      \put(1019,6336){\makebox(0,0)[r]{\strut{}$0.4$}}%
      \csname LTb\endcsname%
      \put(1019,6801){\makebox(0,0)[r]{\strut{}$0.5$}}%
      \csname LTb\endcsname%
      \put(1240,4199){\makebox(0,0){\strut{}$2^{9}$}}%
      \csname LTb\endcsname%
      \put(1783,4199){\makebox(0,0){\strut{}$2^{10}$}}%
      \csname LTb\endcsname%
      \put(2327,4199){\makebox(0,0){\strut{}$2^{11}$}}%
      \csname LTb\endcsname%
      \put(2870,4199){\makebox(0,0){\strut{}$2^{12}$}}%
      \csname LTb\endcsname%
      \put(3414,4199){\makebox(0,0){\strut{}$2^{13}$}}%
      \csname LTb\endcsname%
      \put(3957,4199){\makebox(0,0){\strut{}$2^{14}$}}%
      \csname LTb\endcsname%
      \put(4501,4199){\makebox(0,0){\strut{}$2^{15}$}}%
      \csname LTb\endcsname%
      \put(5044,4199){\makebox(0,0){\strut{}$2^{16}$}}%
    }%
    \gplgaddtomacro\gplfronttext{%
      \csname LTb\endcsname%
      \put(518,5639){\rotatebox{-270}{\makebox(0,0){\strut{}time in sec.}}}%
      \csname LTb\endcsname%
      \put(3135,3920){\makebox(0,0){\strut{}cores}}%
      \csname LTb\endcsname%
      \put(3135,7080){\makebox(0,0){\strut{}$62.1 \cdot 10^3$ cells/core}}%
    }%
    \gplgaddtomacro\gplbacktext{%
      \csname LTb\endcsname%
      \put(6262,4478){\makebox(0,0)[r]{\strut{}$0$}}%
      \csname LTb\endcsname%
      \put(6262,4943){\makebox(0,0)[r]{\strut{}$0.2$}}%
      \csname LTb\endcsname%
      \put(6262,5407){\makebox(0,0)[r]{\strut{}$0.4$}}%
      \csname LTb\endcsname%
      \put(6262,5872){\makebox(0,0)[r]{\strut{}$0.6$}}%
      \csname LTb\endcsname%
      \put(6262,6336){\makebox(0,0)[r]{\strut{}$0.8$}}%
      \csname LTb\endcsname%
      \put(6262,6801){\makebox(0,0)[r]{\strut{}$1$}}%
      \csname LTb\endcsname%
      \put(6483,4199){\makebox(0,0){\strut{}$2^{9}$}}%
      \csname LTb\endcsname%
      \put(7026,4199){\makebox(0,0){\strut{}$2^{10}$}}%
      \csname LTb\endcsname%
      \put(7570,4199){\makebox(0,0){\strut{}$2^{11}$}}%
      \csname LTb\endcsname%
      \put(8113,4199){\makebox(0,0){\strut{}$2^{12}$}}%
      \csname LTb\endcsname%
      \put(8657,4199){\makebox(0,0){\strut{}$2^{13}$}}%
      \csname LTb\endcsname%
      \put(9200,4199){\makebox(0,0){\strut{}$2^{14}$}}%
      \csname LTb\endcsname%
      \put(9744,4199){\makebox(0,0){\strut{}$2^{15}$}}%
      \csname LTb\endcsname%
      \put(10287,4199){\makebox(0,0){\strut{}$2^{16}$}}%
    }%
    \gplgaddtomacro\gplfronttext{%
      \csname LTb\endcsname%
      \put(5761,5639){\rotatebox{-270}{\makebox(0,0){\strut{}time in sec.}}}%
      \csname LTb\endcsname%
      \put(8378,3920){\makebox(0,0){\strut{}cores}}%
      \csname LTb\endcsname%
      \put(8378,7080){\makebox(0,0){\strut{}$210 \cdot 10^3$ cells/core}}%
    }%
    \gplgaddtomacro\gplbacktext{%
      \csname LTb\endcsname%
      \put(1019,688){\makebox(0,0)[r]{\strut{}$0$}}%
      \csname LTb\endcsname%
      \put(1019,978){\makebox(0,0)[r]{\strut{}$0.5$}}%
      \csname LTb\endcsname%
      \put(1019,1269){\makebox(0,0)[r]{\strut{}$1$}}%
      \csname LTb\endcsname%
      \put(1019,1559){\makebox(0,0)[r]{\strut{}$1.5$}}%
      \csname LTb\endcsname%
      \put(1019,1850){\makebox(0,0)[r]{\strut{}$2$}}%
      \csname LTb\endcsname%
      \put(1019,2140){\makebox(0,0)[r]{\strut{}$2.5$}}%
      \csname LTb\endcsname%
      \put(1019,2430){\makebox(0,0)[r]{\strut{}$3$}}%
      \csname LTb\endcsname%
      \put(1019,2721){\makebox(0,0)[r]{\strut{}$3.5$}}%
      \csname LTb\endcsname%
      \put(1019,3011){\makebox(0,0)[r]{\strut{}$4$}}%
      \csname LTb\endcsname%
      \put(1240,409){\makebox(0,0){\strut{}$2^{9}$}}%
      \csname LTb\endcsname%
      \put(1783,409){\makebox(0,0){\strut{}$2^{10}$}}%
      \csname LTb\endcsname%
      \put(2327,409){\makebox(0,0){\strut{}$2^{11}$}}%
      \csname LTb\endcsname%
      \put(2870,409){\makebox(0,0){\strut{}$2^{12}$}}%
      \csname LTb\endcsname%
      \put(3414,409){\makebox(0,0){\strut{}$2^{13}$}}%
      \csname LTb\endcsname%
      \put(3957,409){\makebox(0,0){\strut{}$2^{14}$}}%
      \csname LTb\endcsname%
      \put(4501,409){\makebox(0,0){\strut{}$2^{15}$}}%
      \csname LTb\endcsname%
      \put(5044,409){\makebox(0,0){\strut{}$2^{16}$}}%
      \csname LTb\endcsname%
      \put(7382,1251){\makebox(0,0)[l]{\strut{}(\gls{sfc} -- Morton order)}}%
    }%
    \gplgaddtomacro\gplfronttext{%
      \csname LTb\endcsname%
      \put(518,1849){\rotatebox{-270}{\makebox(0,0){\strut{}time in sec.}}}%
      \csname LTb\endcsname%
      \put(3135,130){\makebox(0,0){\strut{}cores}}%
      \csname LTb\endcsname%
      \put(3135,3290){\makebox(0,0){\strut{}$971 \cdot 10^3$ cells/core}}%
      \csname LTb\endcsname%
      \put(7385,2193){\makebox(0,0)[l]{\strut{}entire \gls{amr} cycle}}%
      \csname LTb\endcsname%
      \put(7385,1870){\makebox(0,0)[l]{\strut{}data migration}}%
      \csname LTb\endcsname%
      \put(7385,1547){\makebox(0,0)[l]{\strut{}dynamic load balancing}}%
    }%
    \gplbacktext
    \put(0,0){\includegraphics{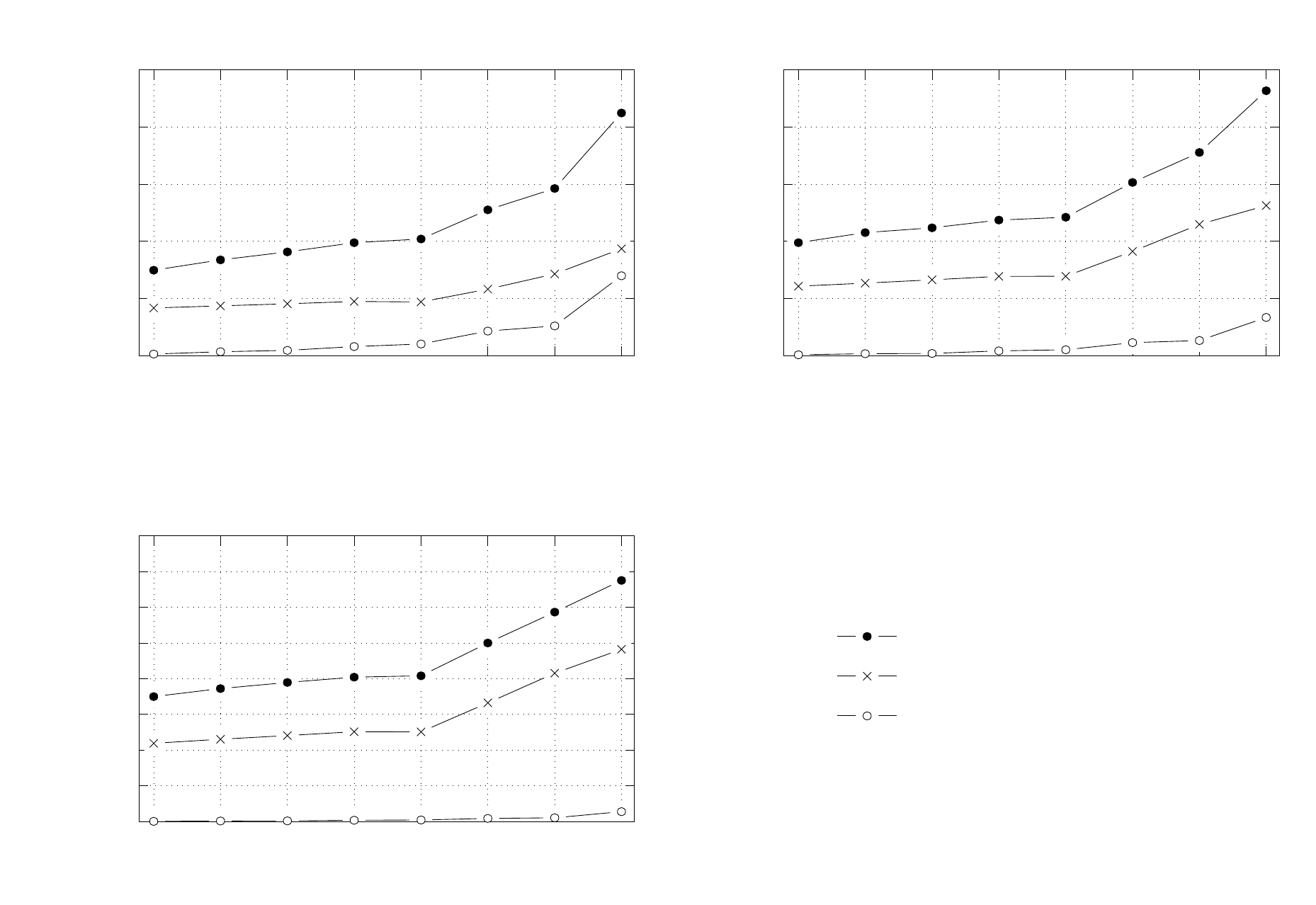}}%
    \gplfronttext
  \end{picture}%
\endgroup

%% file: sdifit_.tex
\begingroup
  \makeatletter
  \providecommand\color[2][]{%
    \GenericError{(gnuplot) \space\space\space\@spaces}{%
      Package color not loaded in conjunction with
      terminal option `colourtext'%
    }{See the gnuplot documentation for explanation.%
    }{Either use 'blacktext' in gnuplot or load the package
      color.sty in LaTeX.}%
    \renewcommand\color[2][]{}%
  }%
  \providecommand\includegraphics[2][]{%
    \GenericError{(gnuplot) \space\space\space\@spaces}{%
      Package graphicx or graphics not loaded%
    }{See the gnuplot documentation for explanation.%
    }{The gnuplot epslatex terminal needs graphicx.sty or graphics.sty.}%
    \renewcommand\includegraphics[2][]{}%
  }%
  \providecommand\rotatebox[2]{#2}%
  \@ifundefined{ifGPcolor}{%
    \newif\ifGPcolor
    \GPcolortrue
  }{}%
  \@ifundefined{ifGPblacktext}{%
    \newif\ifGPblacktext
    \GPblacktexttrue
  }{}%
  \let\gplgaddtomacro\g@addto@macro
  \gdef\gplbacktext{}%
  \gdef\gplfronttext{}%
  \makeatother
  \ifGPblacktext
    \def\colorrgb#1{}%
    \def\colorgray#1{}%
  \else
    \ifGPcolor
      \def\colorrgb#1{\color[rgb]{#1}}%
      \def\colorgray#1{\color[gray]{#1}}%
      \expandafter\def\csname LTw\endcsname{\color{white}}%
      \expandafter\def\csname LTb\endcsname{\color{black}}%
      \expandafter\def\csname LTa\endcsname{\color{black}}%
      \expandafter\def\csname LT0\endcsname{\color[rgb]{1,0,0}}%
      \expandafter\def\csname LT1\endcsname{\color[rgb]{0,1,0}}%
      \expandafter\def\csname LT2\endcsname{\color[rgb]{0,0,1}}%
      \expandafter\def\csname LT3\endcsname{\color[rgb]{1,0,1}}%
      \expandafter\def\csname LT4\endcsname{\color[rgb]{0,1,1}}%
      \expandafter\def\csname LT5\endcsname{\color[rgb]{1,1,0}}%
      \expandafter\def\csname LT6\endcsname{\color[rgb]{0,0,0}}%
      \expandafter\def\csname LT7\endcsname{\color[rgb]{1,0.3,0}}%
      \expandafter\def\csname LT8\endcsname{\color[rgb]{0.5,0.5,0.5}}%
    \else
      \def\colorrgb#1{\color{black}}%
      \def\colorgray#1{\color[gray]{#1}}%
      \expandafter\def\csname LTw\endcsname{\color{white}}%
      \expandafter\def\csname LTb\endcsname{\color{black}}%
      \expandafter\def\csname LTa\endcsname{\color{black}}%
      \expandafter\def\csname LT0\endcsname{\color{black}}%
      \expandafter\def\csname LT1\endcsname{\color{black}}%
      \expandafter\def\csname LT2\endcsname{\color{black}}%
      \expandafter\def\csname LT3\endcsname{\color{black}}%
      \expandafter\def\csname LT4\endcsname{\color{black}}%
      \expandafter\def\csname LT5\endcsname{\color{black}}%
      \expandafter\def\csname LT6\endcsname{\color{black}}%
      \expandafter\def\csname LT7\endcsname{\color{black}}%
      \expandafter\def\csname LT8\endcsname{\color{black}}%
    \fi
  \fi
  \setlength{\unitlength}{0.0500bp}%
  \begin{picture}(7080.00,3220.00)%
    \gplgaddtomacro\gplbacktext{%
      \csname LTb\endcsname%
      \put(645,688){\makebox(0,0)[r]{\strut{} 0}}%
      \csname LTb\endcsname%
      \put(645,1153){\makebox(0,0)[r]{\strut{} 2}}%
      \csname LTb\endcsname%
      \put(645,1619){\makebox(0,0)[r]{\strut{} 4}}%
      \csname LTb\endcsname%
      \put(645,2084){\makebox(0,0)[r]{\strut{} 6}}%
      \csname LTb\endcsname%
      \put(645,2550){\makebox(0,0)[r]{\strut{} 8}}%
      \csname LTb\endcsname%
      \put(645,3015){\makebox(0,0)[r]{\strut{} 10}}%
      \csname LTb\endcsname%
      \put(1238,409){\makebox(0,0){\strut{}$2^{9}$}}%
      \csname LTb\endcsname%
      \put(1730,409){\makebox(0,0){\strut{}$2^{10}$}}%
      \csname LTb\endcsname%
      \put(2221,409){\makebox(0,0){\strut{}$2^{11}$}}%
      \csname LTb\endcsname%
      \put(2712,409){\makebox(0,0){\strut{}$2^{12}$}}%
      \csname LTb\endcsname%
      \put(3204,409){\makebox(0,0){\strut{}$2^{13}$}}%
      \csname LTb\endcsname%
      \put(3695,409){\makebox(0,0){\strut{}$2^{14}$}}%
      \csname LTb\endcsname%
      \put(4186,409){\makebox(0,0){\strut{}$2^{15}$}}%
      \csname LTb\endcsname%
      \put(4678,409){\makebox(0,0){\strut{}$2^{16}$}}%
      \csname LTb\endcsname%
      \put(144,1851){\rotatebox{-270}{\makebox(0,0){\strut{}iterations}}}%
      \csname LTb\endcsname%
      \put(2958,130){\makebox(0,0){\strut{}cores}}%
    }%
    \gplgaddtomacro\gplfronttext{%
      \csname LTb\endcsname%
      \put(6059,2013){\makebox(0,0)[l]{\strut{}push}}%
      \csname LTb\endcsname%
      \put(6059,1690){\makebox(0,0)[l]{\strut{}push/pull}}%
    }%
    \gplbacktext
    \put(0,0){\includegraphics{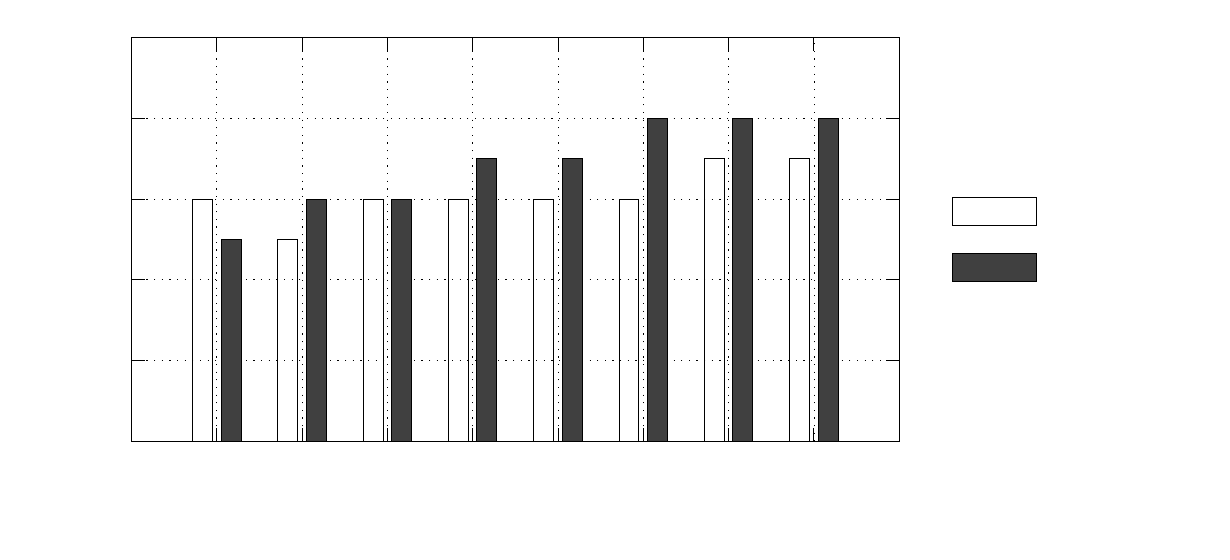}}%
    \gplfronttext
  \end{picture}%
\endgroup

%% file: sdiff_.tex
\begingroup
  \makeatletter
  \providecommand\color[2][]{%
    \GenericError{(gnuplot) \space\space\space\@spaces}{%
      Package color not loaded in conjunction with
      terminal option `colourtext'%
    }{See the gnuplot documentation for explanation.%
    }{Either use 'blacktext' in gnuplot or load the package
      color.sty in LaTeX.}%
    \renewcommand\color[2][]{}%
  }%
  \providecommand\includegraphics[2][]{%
    \GenericError{(gnuplot) \space\space\space\@spaces}{%
      Package graphicx or graphics not loaded%
    }{See the gnuplot documentation for explanation.%
    }{The gnuplot epslatex terminal needs graphicx.sty or graphics.sty.}%
    \renewcommand\includegraphics[2][]{}%
  }%
  \providecommand\rotatebox[2]{#2}%
  \@ifundefined{ifGPcolor}{%
    \newif\ifGPcolor
    \GPcolortrue
  }{}%
  \@ifundefined{ifGPblacktext}{%
    \newif\ifGPblacktext
    \GPblacktexttrue
  }{}%
  \let\gplgaddtomacro\g@addto@macro
  \gdef\gplbacktext{}%
  \gdef\gplfronttext{}%
  \makeatother
  \ifGPblacktext
    \def\colorrgb#1{}%
    \def\colorgray#1{}%
  \else
    \ifGPcolor
      \def\colorrgb#1{\color[rgb]{#1}}%
      \def\colorgray#1{\color[gray]{#1}}%
      \expandafter\def\csname LTw\endcsname{\color{white}}%
      \expandafter\def\csname LTb\endcsname{\color{black}}%
      \expandafter\def\csname LTa\endcsname{\color{black}}%
      \expandafter\def\csname LT0\endcsname{\color[rgb]{1,0,0}}%
      \expandafter\def\csname LT1\endcsname{\color[rgb]{0,1,0}}%
      \expandafter\def\csname LT2\endcsname{\color[rgb]{0,0,1}}%
      \expandafter\def\csname LT3\endcsname{\color[rgb]{1,0,1}}%
      \expandafter\def\csname LT4\endcsname{\color[rgb]{0,1,1}}%
      \expandafter\def\csname LT5\endcsname{\color[rgb]{1,1,0}}%
      \expandafter\def\csname LT6\endcsname{\color[rgb]{0,0,0}}%
      \expandafter\def\csname LT7\endcsname{\color[rgb]{1,0.3,0}}%
      \expandafter\def\csname LT8\endcsname{\color[rgb]{0.5,0.5,0.5}}%
    \else
      \def\colorrgb#1{\color{black}}%
      \def\colorgray#1{\color[gray]{#1}}%
      \expandafter\def\csname LTw\endcsname{\color{white}}%
      \expandafter\def\csname LTb\endcsname{\color{black}}%
      \expandafter\def\csname LTa\endcsname{\color{black}}%
      \expandafter\def\csname LT0\endcsname{\color{black}}%
      \expandafter\def\csname LT1\endcsname{\color{black}}%
      \expandafter\def\csname LT2\endcsname{\color{black}}%
      \expandafter\def\csname LT3\endcsname{\color{black}}%
      \expandafter\def\csname LT4\endcsname{\color{black}}%
      \expandafter\def\csname LT5\endcsname{\color{black}}%
      \expandafter\def\csname LT6\endcsname{\color{black}}%
      \expandafter\def\csname LT7\endcsname{\color{black}}%
      \expandafter\def\csname LT8\endcsname{\color{black}}%
    \fi
  \fi
    \setlength{\unitlength}{0.0500bp}%
    \ifx\gptboxheight\undefined%
      \newlength{\gptboxheight}%
      \newlength{\gptboxwidth}%
      \newsavebox{\gptboxtext}%
    \fi%
    \setlength{\fboxrule}{0.5pt}%
    \setlength{\fboxsep}{1pt}%
\begin{picture}(10700.00,7360.00)%
    \gplgaddtomacro\gplbacktext{%
      \csname LTb\endcsname%
      \put(1070,4478){\makebox(0,0)[r]{\strut{}$0$}}%
      \csname LTb\endcsname%
      \put(1070,5059){\makebox(0,0)[r]{\strut{}$0.1$}}%
      \csname LTb\endcsname%
      \put(1070,5640){\makebox(0,0)[r]{\strut{}$0.2$}}%
      \csname LTb\endcsname%
      \put(1070,6220){\makebox(0,0)[r]{\strut{}$0.3$}}%
      \csname LTb\endcsname%
      \put(1070,6801){\makebox(0,0)[r]{\strut{}$0.4$}}%
      \csname LTb\endcsname%
      \put(1289,4199){\makebox(0,0){\strut{}$2^{9}$}}%
      \csname LTb\endcsname%
      \put(1826,4199){\makebox(0,0){\strut{}$2^{10}$}}%
      \csname LTb\endcsname%
      \put(2363,4199){\makebox(0,0){\strut{}$2^{11}$}}%
      \csname LTb\endcsname%
      \put(2899,4199){\makebox(0,0){\strut{}$2^{12}$}}%
      \csname LTb\endcsname%
      \put(3436,4199){\makebox(0,0){\strut{}$2^{13}$}}%
      \csname LTb\endcsname%
      \put(3972,4199){\makebox(0,0){\strut{}$2^{14}$}}%
      \csname LTb\endcsname%
      \put(4509,4199){\makebox(0,0){\strut{}$2^{15}$}}%
      \csname LTb\endcsname%
      \put(5046,4199){\makebox(0,0){\strut{}$2^{16}$}}%
    }%
    \gplgaddtomacro\gplfronttext{%
      \csname LTb\endcsname%
      \put(518,5639){\rotatebox{-270}{\makebox(0,0){\strut{}time in sec.}}}%
      \csname LTb\endcsname%
      \put(3161,3920){\makebox(0,0){\strut{}cores}}%
      \csname LTb\endcsname%
      \put(3161,7080){\makebox(0,0){\strut{}$62.1 \cdot 10^3$ cells/core}}%
    }%
    \gplgaddtomacro\gplbacktext{%
      \csname LTb\endcsname%
      \put(6313,4478){\makebox(0,0)[r]{\strut{}$0$}}%
      \csname LTb\endcsname%
      \put(6313,5059){\makebox(0,0)[r]{\strut{}$0.2$}}%
      \csname LTb\endcsname%
      \put(6313,5640){\makebox(0,0)[r]{\strut{}$0.4$}}%
      \csname LTb\endcsname%
      \put(6313,6220){\makebox(0,0)[r]{\strut{}$0.6$}}%
      \csname LTb\endcsname%
      \put(6313,6801){\makebox(0,0)[r]{\strut{}$0.8$}}%
      \csname LTb\endcsname%
      \put(6532,4199){\makebox(0,0){\strut{}$2^{9}$}}%
      \csname LTb\endcsname%
      \put(7069,4199){\makebox(0,0){\strut{}$2^{10}$}}%
      \csname LTb\endcsname%
      \put(7606,4199){\makebox(0,0){\strut{}$2^{11}$}}%
      \csname LTb\endcsname%
      \put(8142,4199){\makebox(0,0){\strut{}$2^{12}$}}%
      \csname LTb\endcsname%
      \put(8679,4199){\makebox(0,0){\strut{}$2^{13}$}}%
      \csname LTb\endcsname%
      \put(9215,4199){\makebox(0,0){\strut{}$2^{14}$}}%
      \csname LTb\endcsname%
      \put(9752,4199){\makebox(0,0){\strut{}$2^{15}$}}%
      \csname LTb\endcsname%
      \put(10289,4199){\makebox(0,0){\strut{}$2^{16}$}}%
    }%
    \gplgaddtomacro\gplfronttext{%
      \csname LTb\endcsname%
      \put(5761,5639){\rotatebox{-270}{\makebox(0,0){\strut{}time in sec.}}}%
      \csname LTb\endcsname%
      \put(8404,3920){\makebox(0,0){\strut{}cores}}%
      \csname LTb\endcsname%
      \put(8404,7080){\makebox(0,0){\strut{}$210 \cdot 10^3$ cells/core}}%
    }%
    \gplgaddtomacro\gplbacktext{%
      \csname LTb\endcsname%
      \put(1070,688){\makebox(0,0)[r]{\strut{}$0$}}%
      \csname LTb\endcsname%
      \put(1070,1153){\makebox(0,0)[r]{\strut{}$0.5$}}%
      \csname LTb\endcsname%
      \put(1070,1617){\makebox(0,0)[r]{\strut{}$1$}}%
      \csname LTb\endcsname%
      \put(1070,2082){\makebox(0,0)[r]{\strut{}$1.5$}}%
      \csname LTb\endcsname%
      \put(1070,2546){\makebox(0,0)[r]{\strut{}$2$}}%
      \csname LTb\endcsname%
      \put(1070,3011){\makebox(0,0)[r]{\strut{}$2.5$}}%
      \csname LTb\endcsname%
      \put(1289,409){\makebox(0,0){\strut{}$2^{9}$}}%
      \csname LTb\endcsname%
      \put(1826,409){\makebox(0,0){\strut{}$2^{10}$}}%
      \csname LTb\endcsname%
      \put(2363,409){\makebox(0,0){\strut{}$2^{11}$}}%
      \csname LTb\endcsname%
      \put(2899,409){\makebox(0,0){\strut{}$2^{12}$}}%
      \csname LTb\endcsname%
      \put(3436,409){\makebox(0,0){\strut{}$2^{13}$}}%
      \csname LTb\endcsname%
      \put(3972,409){\makebox(0,0){\strut{}$2^{14}$}}%
      \csname LTb\endcsname%
      \put(4509,409){\makebox(0,0){\strut{}$2^{15}$}}%
      \csname LTb\endcsname%
      \put(5046,409){\makebox(0,0){\strut{}$2^{16}$}}%
      \csname LTb\endcsname%
      \put(7382,1251){\makebox(0,0)[l]{\strut{}(diffusion -- push/pull)}}%
    }%
    \gplgaddtomacro\gplfronttext{%
      \csname LTb\endcsname%
      \put(518,1849){\rotatebox{-270}{\makebox(0,0){\strut{}time in sec.}}}%
      \csname LTb\endcsname%
      \put(3161,130){\makebox(0,0){\strut{}cores}}%
      \csname LTb\endcsname%
      \put(3161,3290){\makebox(0,0){\strut{}$971 \cdot 10^3$ cells/core}}%
      \csname LTb\endcsname%
      \put(7385,2193){\makebox(0,0)[l]{\strut{}entire \gls{amr} cycle}}%
      \csname LTb\endcsname%
      \put(7385,1870){\makebox(0,0)[l]{\strut{}data migration}}%
      \csname LTb\endcsname%
      \put(7385,1547){\makebox(0,0)[l]{\strut{}dynamic load balancing}}%
    }%
    \gplbacktext
    \put(0,0){\includegraphics{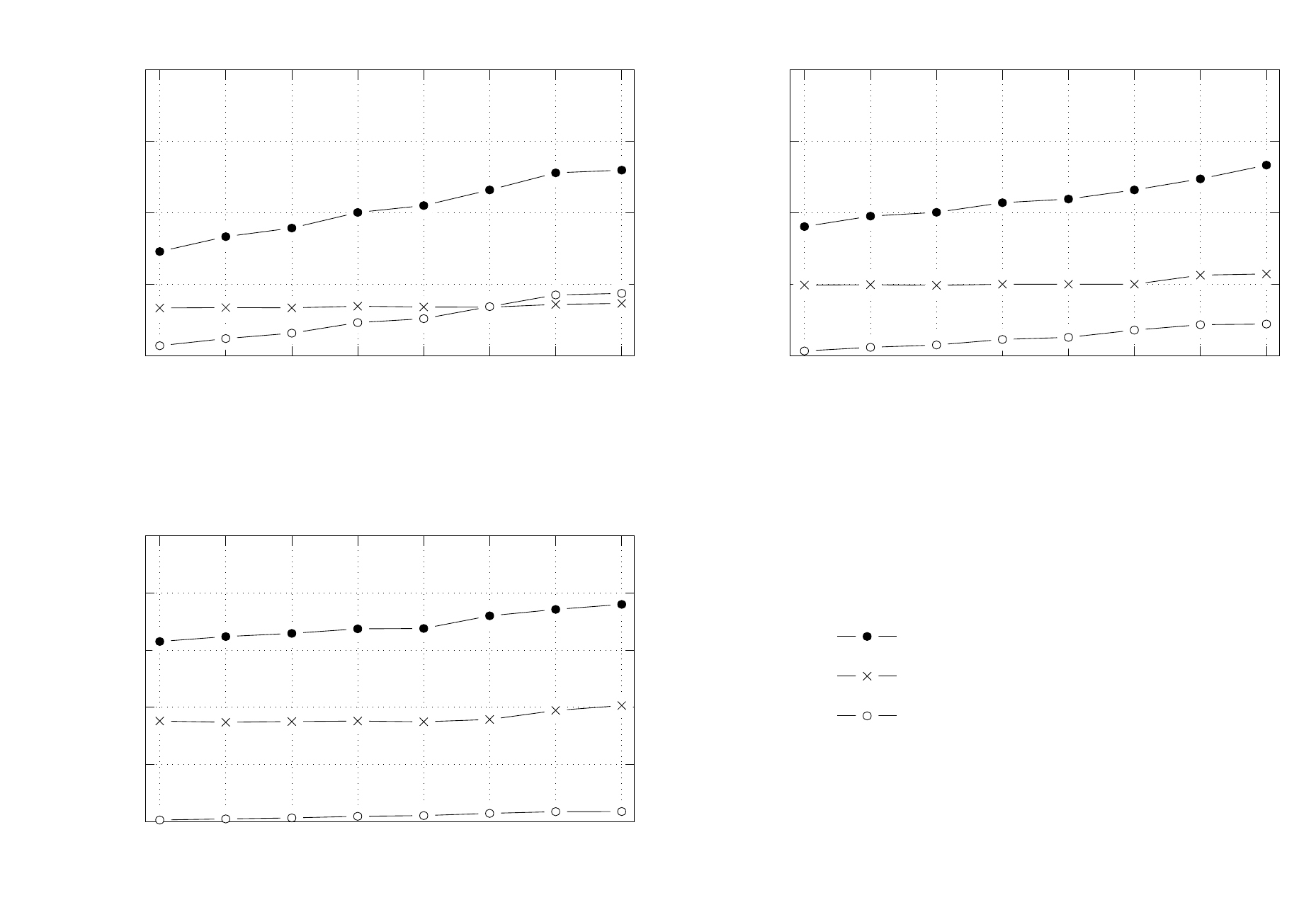}}%
    \gplfronttext
  \end{picture}%
\endgroup